\documentclass[fleqn,usenatbib]{mnras}

\usepackage{newtxtext,newtxmath}

\usepackage[T1]{fontenc}

\DeclareRobustCommand{\VAN}[3]{#2}
\let\VANthebibliography\thebibliography
\def\thebibliography{\DeclareRobustCommand{\VAN}[3]{##3}\VANthebibliography}

\usepackage{graphicx}	
\usepackage{amsmath}	
\usepackage{rotating}
\usepackage[abs]{overpic}
\usepackage{float}
\usepackage{color}

\usepackage[abs]{overpic}
\usepackage{float}
\usepackage{color}

\title[]{Gas permeability and mechanical properties of dust grain aggregates at hyper- and zero-gravity}

\author[H.L. Capelo et al.]{Holly L. Capelo,$^{1}$\thanks{E-mail: holly.capelo@space.unibe.ch} Jean-David Bod\'enan,$^{2}$ Martin Jutzi,$^{1}$  Jonas K\"uhn,$^{1}$  Romain Cerubini,$^{1}$  Bernhard Jost,$^{1}$ \newauthor Linus St\"ockli ,$^{1}$  Stefano Spadaccia,$^{1}$  Clemence Herny,$^{1}$  Bastian Gundlach ,$^{3}$  G\"unter Kargl,$^{4}$ Cl\'ement Surville,$^{5}$ \newauthor Lucio Mayer, $^{5}$  Maria Sch\"onb\"achler, $^{2}$ Nicolas Thomas $^{1}$ and Antoine Pommerol $^{1}$\\
$^{1}$Space Research and Planetary Sciences, Physics institute, University of Bern, Sidlerstrasse 5, CH-3012 Bern, Switzerland\\
$^{2}$ETH Zurich, Institute of Geochemistry and Petrology, 8092 Zurich, Switzerland\\ 
$^{3}$Institut für Planetologie, Universit\"at M\"unster, Wilhelm-Klemm Str. 10, 48149 M\"unster, Germany\\ 
$^{4}$Space Research Institute, Austrian Academy of Sciences, Schmiedlstrasse 6, A-8042 Graz, Austria \\
$^{5}$Center for Theoretical Astrophysics and Cosmology, Institute for Computational Science, University of Zurich, Winterthurerstrasse 190, CH-8057 Zurich, Switzerland}

\date{Accepted XXX. Received YYY; in original form ZZZ}

\pubyear{2023}

\begin{document}
\label{firstpage}
\pagerange{\pageref{firstpage}--\pageref{lastpage}}
\maketitle
\begin{abstract}
Particle-particle and particle-gas processes significantly impact planetary precursors such as dust aggregates and planetesimals. We investigate gas permeability ($\kappa$) in 12 granular samples, mimicking planetesimal dust regoliths.  Using parabolic flights, this study assesses how gravitational compression -- and lack thereof-- influences gas permeation, impacting the equilibrium state of low-gravity objects. Transitioning between micro- and hyper-gravity induces granular sedimentation dynamics, revealing collective dust-grain aerodynamics. Our experiments measure $\kappa$ across Knudsen number (Kn) ranges, reflecting transitional flow. Using mass and momentum conservation, we derive $\kappa$ and calculate pressure gradients within the granular matrix. Key findings: 1. As confinement pressure increases with gravitational load and mass flow, $\kappa$ and average pore space decrease. This implies that a planetesimal's unique dust-compaction history limits sub-surface volatile outflows. 2. The derived pressure gradient enables tensile strength determination for asteroid regolith simulants with cohesion. This offers a unique approach to studying dust-layer properties when suspended in confinement pressures comparable to the equilibrium state on planetesimals surfaces, which will be valuable for modelling their collisional evolution. 3. We observe a dynamical flow symmetry breaking when granular material moves against the pressure gradient. This occurs even at low Reynolds numbers, suggesting that Stokes numbers for drifting dust aggregates near the Stokes-Epstein transition require a drag force modification based on permeability.
\end{abstract}




\section{Introduction} \label{section:intro}

The early Solar System formed out of a protosolar disc, which contained gas of densities $10^{-13}$kg m$^{-3}$ -- $10^{-6}$kg m$^{-3}$, depending upon location and evolutionary stage. The solids embedded in the gas disc migrated around in this fluid, experiencing aerodynamic forces such as drag \citep{Weidenschilling:1977a} and momentum exchange \citep{Nakagawa1986}. Small solar-system solids exist at various scales, where dust particles can range from nm-mm, compact dust aggregates or pebbles are generally solids that have grown to cm-dm scale, and planetesimals which can be 1-100s of km. By comparison, the mean free path of the gas molecules in the disc can often be comparable to or much larger than dust and pebbles, which means in turn that Epstein drag forces \citep{epstein} act on these particles.  These components interact significantly with the gas during the planetary assembly processes --  such as core formation and accretion \citep[][]{PollackEtal:1996,Lambrechts_2014}, impacts \citep{Jutzi:2017} and cratering -- as well as disassembly processes, such as volatile release \citep{Thomas:2015,Burn:2019,Spadaccia:2021}, erosion  \citep{SchraeplerBulm:2011,krijt,Schaffer:2020,Demirci:2019,Demirci:2020}, or destructive collisions \citep{Paszun,Jutzi:2015,Schwartz}.

The growth pathway of planetesimals tends to make them porous. Despite different evolution and processing histories, the regoliths of both asteroids and comets are studied in order to yield insight into the formation history of planetesimals.  Granular surface layers arise firstly due to the cohesion of primordial dust grains to one another upon contact, which produces dust agglomerates (for a complete review of dust aggregate types see e.g. \cite{Wooden:2008}). Such agglomerates verifiably \citep{Mannel:2016} become incorporated into planetesimals, such as comet Churyumov-Gerasimenko (67P). The pathway to generating a self-gravitating planetesimal is an open field of inquiry, with aero-gravitational instability currently being the favoured scenario \citep{You_good2005,You_jo2007j,Jo_you2007j,Johansen2007}. The formation process does bear upon the compactness of a minor body \citep{WahlbergI,WahlbergII,mohtashim}. See \cite{Simon:2022} for a review of how different formation scenarios correspond to the compaction state of particles. The compressive strength of granular materials is sufficiently high that bodies in the size ranges from meters to 10s of kilometers experience negligible gravitational compression and are naturally loosely bound. For the same reason, the packing density of the soil may assume a wide range of values, with the potential to reach porosities that are much higher than would be permitted under Earth's gravity. As a mathematical principle, it has long been noted that a close random packing of monodisperse solid spheres produces a volumetric filling factor (or total volume occupied by solid material), of $\phi=0.65$ \citep{Scott_1969, zaccone_2022}. For symbol definitions, value, and units, see Table \ref{parameter_table}. Different void fractions can result from the way by which the particles are packed, from non-spherical particle shapes, and from the existence of a particle size distribution \citep{Donev_2004}.  While a random packing is a reasonable starting point for analogy to soil types, additional considerations arise when one considers, for example, the regolith on low-gravity objects such as asteroids, comets, and satellites. The particles are neither spherical, nor are they compressed together by Earth's gravitational acceleration g$_{\rm Earth}$. 

For example, \cite{wada} consider values of $0.1 \lesssim \phi \lesssim 0.6$ to be a reasonable estimate for the filling factor of Asteroid Ryugu, implying that the actual porosity is unknown and may vary as a function of scale, but can reach up to 90\% voids. It was also found by \cite{Sakatani2021} that anomalously high porosities could exist in the boulders on the surface of Ryugu. The bulk porosity of 67P was reported by \cite{patzold2016} to be 74\%. However, \citep{Brouet_2015} interpreted data from the MIRO instrument aboard the Rosetta spacecraft to suggest local variations in porosity in dust mantle of the comet nucleus. It is expected that the thermal skin depth of the dust mantle is only a few centimeters deep, and that gases which flow through the dust regolith are released from this depth \citep{thomas_2008}. Hence, we here concern ourselves with the voided space at the scale of the particles in such a mantle.

Again due to the low gravitational binding energy, planetesimals and their morphological surface features do not require high tensile or shear strengths in order to remain coherent. For instance, the size scale at which overhangs on 67P have been observed to collapse suggests a tensile strength of $<$ 150 Pa \citep{groussin:2015,Attree:2018}, which is an order of magnitude lower than the strength measured for cm-dm-sized dust aggregates in laboratories, though we acknowledge that strength may decrease with size scale. There is increasing evidence that the surfaces of small asteroids, such as the recently visited Ryugu and Bennu are very weak \citep[e.g.][]{Arakawa2020,Perry2022}. The Hayabusa2 spacecraft \citep{Arakawa2020} sent a projectile into the surface of the asteroid Ryugu and from the crater and ejecta analysis found that the cohesion of the surface is smaller than $\sim$1 Pa \citep{Jutzi:2022}. Our experiments, conducted in microgravity (henceforce $0g$), with low-pressure gas exerting a force on the granular samples, start to approach these low-strength surface properties. 

Porous planetary precursors experience both a static and dynamic pressure due to differential motion with the gas.  Examples include the sub-Keplerian headwind of drifting particles in a disc, or dust comae in comets. Gas drag, originating from sub-surface volatile sublimation flows, leads to the liberation of grains from the comets nucleus. However, the calculation of drag in a particle-rich environment is challenging mathematically and computationally, and experiments testing the build-up of pressure gradients in the presence of opposing forces (gravity, cohesion) are therefore important to understand comet activity and more broadly forces on particle collectives (regolith matrix, aggregates). 

In the most general sense, the equilibrium state of a planetesimal embedded within a gas disc is determined by the confining gas-pressure gradient on a given layer of granular media. Gravity and pressure are typically the two opposing forces, and for solid layers, the equillibrium balance may also depend upon the compactness and cohesiveness of the material. Therefore, modeling of collisions between planetesimals, which can be addressed according to the response of the porous bodies to compressive forces \citep{JUTZI20153}, can also be informed by measurements involving granular media that is suspended in a low-density-gas pressure gradient. Due to lack of buoyancy when the particles are solid and the fluid is under vacuum, such suspension can only be achieved in the absence of gravity.

Our experiments are designed to monitor simultaneously the compression state of granular matter and the media's resistance to low-pressure fluid flow. We determine gas permeability coefficients in the Knudsen regime, by measuring the pressure differential across a bed of granular material, through which gas of known flow rate passes. A related set of experiments were conducted by \cite{Schweighart:2021}. Our experiments build on this work by adding the capability to conduct the measurements aboard parabolic flights. We continue our measurement throughout the entire variation in gravitational load, $0 < g < 2 g_{\rm Earth}$. In this way, we study a wide range of granular packing fractions, which may differ from the values that we measure on Earth. The permeability coefficient serves as a proxy for the porosity, as it is a complex function of the particle shape and size, which impacts the mean inter-particle spacing. By utilising permeability and confining pressure measurements, we proceed to calculate the pressure gradient in selected samples for which cohesion is observed, which yields an estimated limit on dust layer tensile strength. In addition to exploring the equilibrium conditions of the samples at fixed gravity level, we also make observations during the dynamic phase of the parabola, namely the transition from zero- to hyper-gravity(henceforth $0g$ and $hg$, respectively), as the particles sediment and actively compress against the pressure gradient. This enables us to study inertial effects of the particle population upon the gas, as in a collective drag effect which can break the symmetry of the flow.

Flow symmetry breaking caused by objects in a fluid is quite well illustrated at moderate and high Reynolds numbers \citep{vanDyke:1982}. However, for flows where the Reynolds number is low due to the extremely low gas density, and when the aerodynamic drag is unrelated to viscosity as in the Epstein regime, it remains an open question as to whether symmetry breaking can be expected to occur. If so, it would have implications for the drag force on dust aggregates and hence impact, for example, the Stokes numbers of aggregates in protoplanetary discs or the production of dust tails from comets.

We summarise the aims of the experiments as such: i) determine permeability in planetesimal-surface analogues as function of Knudsen number and filling factor; ii) investigate the relationship between sample material properties such as cohesiveness and compactness of the regolith layer under the application of very small forces, as in gentle collisions of planetesimals; iii) establish the role of inertia in causing pressure gradients in granular media and low-pressure gas, in analogy to a collective particle drag effect. 
We address all three aims in the present work. Note however that we do not analyse collective particle inertia in very low particle density and high mass loading, and reserve such analysis for a followup work that is oriented towards the conditions in protoplanetary discs. 

The rest of this document follows the outline given here. Section \ref{section:methods} explains our experimental methodology; in \ref{subsection:experiments} we introduce the experimental facility, in \ref{subsection:definitions} we define relevant equations and parameters, in \ref{subsection:flightseq} we describe the flight sequences, justify the choice of sample materials, and report the basic extracted data products from the measurements. Section \ref{section:analysis} contains the data analysis and scientific results; in \ref{subsection:compression} and \ref{subsection:expansion} we assess the effect of gravitational load on the measured pressure differential caused by the granular sample, in \ref{subsection:coeffs} we report permeability coefficients and fit the Klinkenberg correction for flow conditions which span the transition from continuum flow to free-molecular flow, in \ref{subsection:overpressure} we show the relationship between the confining pressure around the sample and the mean particle separation, in \ref{subsection:cohesion} we illustrate the presence of particle cohesion for a very narrow set of conditions and calculate the shear and tensile strength of the granular matrix. We conclude in section \ref{section:conclusions}; A brief technical summary of all results can be found in \ref{subsection:summary}, in \ref{subsection:synthesis} we conclude from the combined evidence of all the prior analysis that we have detected a dynamical contributor to the measured pressure gradient for specific samples and conditions, and finally \ref{subsection:significance} contains a discussion of the significance and applications of the findings. 

\section{Methods and Measurements} \label{section:methods}
\subsection{The TEMPusVoLA experiment} \label{subsection:experiments}
In a previous publication we describe thoroughly the purpose and capabilities of the Timed Epstein Multi-pressure vessel at Low Accelerations (TEMPusVoLA; \cite{Capelo:2022} ). There are three separate experiments in the TEMPusVoLA facility, the first of which is dedicated to the direct measurement of gas permeability coefficients. The results from this experiment form the subject of the present work. In addition to the variations in gravitational loading, a unique aspect of our experiment compared to previous gas permeability experiments is the design of the container shape, which has been optimised to control the pressure gradient in the sample. The purpose of this is to test the strength of the sample, when converting the pressure gradient to a force profile, and comparing to the time-dependent gravitational force. Moreover, we are interested in the modification of drag laws in the instance of a many-particle population, and so we exploit the relationship between pressure gradient and drag force to study the dynamical influence of the particle population acting upon the gas. 

In these experiments, we measure the differential pressure caused by gas flow through a granular bed. Two pressure transducers, each placed either above or below the granular sample, monitor the net pressure gradient, while the gas mass flow rate remains controlled and fixed, but the gravity level varies. The variation in gravity results from the maneuvers of the aircraft which passes repeatedly through the phases: steady flight (g$_{\rm Earth}$, 9.81 m s$^{-1}$), hyper-gravity ($hg$, $ \sim$ 1.8 $\times$g$_{\rm Earth}$), and either zero-gravity ($0g$, 0$\times$g$_{\rm Earth}$) or reduced gravity (Martian or Lunar, at 0.4 $\times$g$_{\rm Earth}$ or 0.1 $\times$g$_{\rm Earth}$, respectively). 

Intended to study particle-gas dynamics in the Epstein aerodynamic drag regime \citep{epstein,Sharipov,Lasseux:2017} and under zero- or reduced-gravity conditions, the TEMPus VoLA facility has been utilised aboard three parabolic flight campaigns (PFC; 4th Swiss PFC, ESA 75th PFC and ESA 78th PFC). The first campaign was a pilot flight, and consisted of 16 parabolas executed on a single day. The second two campaigns each consisted of three flight days, each day providing a sequence of 31 parabolas. Each parabola provided 23s of reduced gravity. Several features of our experiment enabled us to fully exploit the large number of parabolas such that we could vary system parameters such as Knudsen number (gas flow regime), and sample type. First, we directly control the mass flow of gas into the system with a flow regulator. This allows us to vary the pressure of the downstream gas flow and therefore adjust the flow conditions to cross the transition between continuum and free molecular flow regimes. Second, we installed redundant chambers, so that we could pre-load multiple different sample types; we use an electromagnetic valve system to close off unused chambers and divert the gas stream through the chamber of choice. The interior chamber walls are treated to prevent the buildup of static electricity.

The camera system for this experiment was described in \cite{Capelo:2022}. However, an upgrade to increase the speed and image quality was made for use in the 78th ESA PFC. We selected a high resolution, monochromatic Kiralux CS126MU camera for this purpose. The camera has been carefully configured to minimize frame drops while maximizing the resolution. In order to capture the highest detail, we did not use any binning of the pixels. Only the center of the sensor was read out, in order to avoid geometric distortion effects close to the edge of the frame. Gain was set to 0 and the exposure time is set to 39999 microseconds. The frames were individually saved as raw .tif files with a resolution of 3080x2256, which corresponds to 6.9 MP to provide the most brightness information.
For a quick view, they were converted into compressed movies with a frame rate of 24 fps. Examples of these movies are provided as supperlementary Online Material.

\subsection{Definitions}\label{subsection:definitions}
\subsubsection{Permeability: mass and momentum balance}
A granular bed will cause a pressure gradient when gas flows through it. For flow with velocity $u$ in a constant mean direction through a granular bed, the difference in pressure at a height $h$ is given by:

\begin{equation}\label{equ:pressure_drop}
\frac{dP}{dh}=-\frac{\eta}{\kappa}u.
\end{equation}
The dynamic viscosity $\eta$ is an intrinsic property of the gas, and it is related to the kinematic viscosity $\mu$ through the gas density: 
\begin{equation}\label{equ:kinematic}
\mu=\frac{\eta}{\rho}.
\end{equation}
 We emphasise that the units of $\kappa$ are m$^{2}$, indicating the open channel space available for flow passage, and so this quantity is directly connected to the porosity, which is an important point, to which we will return several times in this work. The average size of the pores in the granular matrix mediates the speed of the gas at all heights in the bed, resulting in a pressure gradient within the sample. The gas velocity can be calculated due to the conservation of gas mass flow rate $Q$ from the lower end of the granular bed through all layers to the upper-most layer of the granular bed, as 
\begin{equation}\label{equ:flux}
Q=u_{l}a_{l}\rho_{l}=u_{h}a_{h}\rho_{h} = u_{u}a_{u}\rho_{u}. 
\end{equation}
At the lower end of the granular bed, the three quantities linear gas speed $u$, container cross-sectional area $a$, and density of the gas $\rho$ are denoted with subscript $l$, and at the upper end of the bed with subscript $u$. All three quantities vary as a function of height, yet their product remains constant and is always equal to the controlled massflow rate Q. In a cylindrical container, integrating Equation \ref{equ:pressure_drop}, which is the momentum equation, and applying gas mass conservation expressed by Equation \ref{equ:flux} results in an expression for the permeability:
\begin{equation} \label{equ:kappa}
    \kappa= \frac{\mu\delta z}{|\delta P | \frac{P_{u}+P_{l}}{2}} \frac{Q}{a_{l}}\frac{RT}{M}.
\end{equation}
The quantities $P_{u}$ and $P_{l}$ correspond to the pressure at the upper and lower boundaries of the container, respectively. And $\delta P$ = $P_{u} - P_{l}$ is the net pressure difference.  The quantities $R$, $M$, and $T$ are the gas constant, mean molecular mass, and constant temperature, the numerical values for which are given in \ref{parameter_table}. We assume an isothermal gas, and do not expect temperature gradients in the sample since the experiment is conducted at ambient conditions and the vacuum line draws air from the cabin of the aircraft.
In our apparatus, the ends of the sample containers are conical (see Figure \ref{fig:experiment_illustration}), and so an extra function, $\psi(\beta)$ which accounts for the variation in cross-sectional area with height must be included. This takes the form:
\begin{equation}\label{equ:shape}
\psi(\beta)= \frac{1}{\pi~tan \beta~(r-\delta z~tan \beta)} - \frac{1}{\pi~r~tan\beta},
\end{equation}
with $r$ being the radius of the narrowest part of the cone. In this case, the opening angle $\beta=15^{\circ}$. The differential in Equation \ref{equ:pressure_drop} can be expressed as $\frac{\delta P}{\delta z}$. Moveover, all parameters in Equations \ref{equ:kappa} and \ref{equ:shape} are constant when Q is constant. Therefore, with a known granular bed height $\delta z$ and assuming mass flow conservation, the permeability coefficient $\kappa$, can be extracted simply from the measurement of $\delta P$. For this reason, we will report $\delta P$ as the first quantity of interest in the results below. The quantities $\delta P$ and $\delta z$ can also be considered as the confinement pressure and confinement height. By taking the difference between $\delta P$ and the equivalent gravitational pressure, we will study how the total confinement on the sample affects the pore size that is reflected in $\kappa$. 

\subsubsection{Porosity: particle and pore properties}
Variations in the pressure differential at different gravitational accelerations are expected to be caused by the possible compression and expansion of the granular matrix. We notice by inspection of the momentum balance in Equation \ref{equ:pressure_drop} that the value of $\kappa$ regulates the pressure gradient. However, $\kappa$ itself depends upon the size, shape, and generally the spacing of the grains. For example, a version of the law expressed in Equation \ref{equ:kappa} is known as the Carman-Kozenzy correlation \citep{CARMAN,kozeny1924kapillaren,richardson}. For particles of diameter d$_{\rm p}$ , 
\begin{equation}
 \kappa={K}_{\rm Koz} d_{p}^{2}(1-c)^{3}/c^{2}, 
 \end{equation} 
 with c being the porosity, and related to $\phi$, the filling factor, $c+\phi=1$ see also eqn. 7.36 and fig 7.5 of \cite{feder_flekkøy_hansen_2022}. For a random-packed bed composed of equal-size spheres, $\phi$=0.64 and ${K}_{\rm Koz}$=2/90. The value of ${K}_{\rm Koz}$ for irregular particles is not known and parameters such as capillary diameter, and tortuosity can change the coefficient. Regardless, it is clear that a particular value of $\kappa$ results from a particular packing state of particles. Therefore, by measuring $\kappa$ at different loadings, we can check how the flow conditions and pressure differential are impacted for different particle types.

\subsubsection{Pressure gradient: sample strength}
The pressure gradient at each height in the sample can also be calculated directly. In \cite{Capelo:2022} we derived and experimentally verified the vertical pressure profile using Equations \ref{equ:pressure_drop}, \ref{equ:flux}, and \ref{equ:shape}, i.e. 
\begin{equation}\label{equ:profile}
P(h)=\sqrt{P_{l}^{2}-\left(\frac{2\eta u_{l}P_{l}}{\kappa}\right)h}.
\end{equation}
At the top of the sample, $h = \delta z$ and $P(\delta z)=P_{u}$. Together, the product of the pressure profile and the height-dependent cross-sectional area, returns a vertical force profile. Due to the geometry of this system, the force profile maximizes close to the surface of the sample, as opposed to the bottom of the sample, as would be the case for a cylindrical container. By observing the breaking point of the sample, and calculating the force at this point in the profile, we obtain a direct measurement of the sample's cohesive force, if there is any. 

Although the two concepts are often used inter-changeably, shear strength differs from tensile strength in that the former refers to a force applied perpendicular to the surface, and the latter refers to a force applied parallel to the surface. During the compression stage, when gas is flowing through the sample, we are probing the tensile strength. 
The shear strength can be assessed when the aggregate is airborne during $0g$ and experiences a cross-flow. We use the formalism provided by \cite{Schaffer:2020}, for an object experiencing a cross-wind in a proto-planetary disc, with the shear force, $\tau_{w}$, expressed as:
\begin{equation}\label{equ:shear}
    \tau_{w}=\frac{\rho u^{2}}{\rm{Re}^{1/2}}.
\end{equation}
The parameter in the denominator is the Reynolds number, defined:
 
\begin{equation} \label{equ:reynolds}
{\rm Re}= u L/\mu,
\end{equation} 
with L the length scale under consideration, such as particle or container size.

\subsubsection{Flow regimes: drag}
There are several parameters which determine the properties and behaviour of a hydrodynamical system. For objects embedded in a flow, one considers the response of the fluid to the object according to whether perturbations caused by such an object will result in turbulence. Furthermore, it is important to consider how the boundary conditions near such an object should be handled, in light of the fluid pressure. Because of qualitatively different flow behaviour for high- versus low- pressure fluids, it is common to divide flow regimes according to Knudsen number,
\begin{equation}
Kn = \lambda/d_{p},
\end{equation}
which compares the mean free path of the gas $\lambda$ to, in our case, the particle size. As a general guide, the following values of $Kn$ belong to the various flow regimes that are listed:
\begin{multline}
Kn \lesssim 0.1:\; \; {\rm continuum}\\
0.1 \lesssim Kn \lesssim 10:\;\; {\rm transition}\\
10 \lesssim Kn: \;\; {\rm free-molecular}\\.
\end{multline}
These guidelines by no means represent sharp transitions between regimes. It is a common convention to already use the free molecular formulation in the transition region, for example, when calculating the drag force in the equation of motion for dust grains. In general, such approximation is not detrimental to an understanding of the relevant flow dynamics. First, because the transition region corresponds to a narrow range of pressures when dealing with small particles due to the inverse relationship between $\lambda$ and P; for micrometer-sized particles, differences in pressure in multiples of 10$^{-6}$ result in the same factor of change in $Kn$. More importantly, flows in either the transition or free-molecular flow drag regime generally both allow for the drag force term in the momentum equation to be expressed in linear form. This condition is set by the Reynolds number, which is defined by Equation \ref{equ:reynolds} above. When Re$<1$, the linear drag formulation applies, which has the special property of being symmetric in time and space.  $Re$ decreases with decreasing gas pressure, according to Equation \ref{equ:kinematic}, and so for systems possessing both small-scale L and low-density gas, the Carman-Kozenzy correlation can predict the pressure differential caused by a bed of spheres. For porous samples in general, Equation \ref{equ:pressure_drop} directly applies. We notice, however, that this expression appears impervious to flow regime. On the other hand, $\kappa$ becomes pressure dependent for sufficiently low pressure. We define $\kappa_{\infty}$ as the special case of Equation \ref{equ:kappa}, where essentially continuum flow conditions apply (P$_{l}$ and $P_{u}$ are both relatively high), but for which Re is still less than 1.  \cite{Klinkenberg2012ThePO} originally predicted that the permeability coefficient will be modified as mean pressure, $\delta P/2$ decreases:
\begin{equation}\label{klinkenberg}
\kappa_{\rm K}=\kappa_{\infty}\left(1+\frac{2b}{\delta P} \right). 
\end{equation}
This expression employs a parameter often expressed as a decreasing exponential: $b= X \kappa ^{-Y}$. However, X and Y can vary by orders of magnitude, depending upon how they are derived \citep{Lasseux:2017}. The important feature to notice, is that $\kappa_{\infty}$ provides the baseline value, which takes a pressure-dependent correction. For vanishing pressure differential, which results concurrently with vanishing mass flux rates, $\kappa_{\rm K}$ increases asymptotically. This function is therefore useful to bridge between the continuum regime and, upon the asymptotic branch, the regime where pure molecular diffusion applies. While Equation \ref{equ:kappa} is sufficient to extract $\kappa$ and already reflects the pressure-dependence, we will compare the results to Equation \ref{klinkenberg}, to remind ourselves where on the flow spectrum the system is, and how this may be modified under different compression states. For the same reason, we obtain measurements of $\delta P$ at initially higher pressures -- to represent $\kappa_{\infty}$ and decrease mass flux rate to yield smaller $\delta P$.

The force responsible for liberating particles from a surface is aerodynamic drag, which is true regardless of the flow regime. However, the formulation of the correct drag law can be challenging, since simple expressions such as either the Stokes or Epstein laws apply for the case of relatively isolated spheres. In a particle-rich environment, the boundary conditions around each particle are affected by the presence of other particles. \cite{Brinkman} recognized this fact, but also provided the insight, that the local conditions around a particle in a swarm could be predicted by using Darcy’s law, which is an early empirical version of Equation \ref{equ:pressure_drop}. Brinkman derived a modification to Stokes drag for particles traveling in a swarm through a fluid, reasoning that in the reference frame between particle and flow, a solid bed of grains confined by a pressure gradient is equivalent to a group of particles with relative motion with respect to a gas. Such modified drag-laws have been used to model two-phase flow. Here, we are concerned with the additional consequence of this formulation, which is that a pressure differential should develop up and down stream of such a particle swarm, and that this is directly modulated by the relative velocity between the granular matrix and the gas. Such a consequence is curious, for the case of a dust aggregate experiencing relative motion in a gas, since it would imply a breaking of the symmetry of the surrounding flow, which is contrary to the expectation from calculating the Reynolds number of the macro-scale of the agglomerate. We therefore pay attention to the pressure differential during the sedimentation of particles, when the relative particle-gas velocity is increased and particle packing is increased due to the entry into $hg$ after the $0g$ suspension.

\subsection{Flight Sequence Overview}\label{subsection:flightseq}

\subsubsection{Data preparation} \label{subsubsection:dataprep}
We separate the pressure measurements into intervals corresponding to the duration of a parabola, including the steady-flight phase preceding entry into, duration of, and pull out of the parabola. The acceleration is recorded with an accelerometer and we then use this signal as a reference for the external force experienced by the sample material at any given time. In a single flight, we vary the mass flow rate $Q$ upon repeated parabolas, and so we cover a wide range of $Kn$ for each sample. Below we describe the materials used over four separate flights, three samples per flight, resulting in a study of twelve different analogs. For each chamber, we plot the differential pressure to show that the pressure varies systematically according to g-level and $Q$.  

\begin{figure}
\includegraphics[angle=270,width=0.48\textwidth]{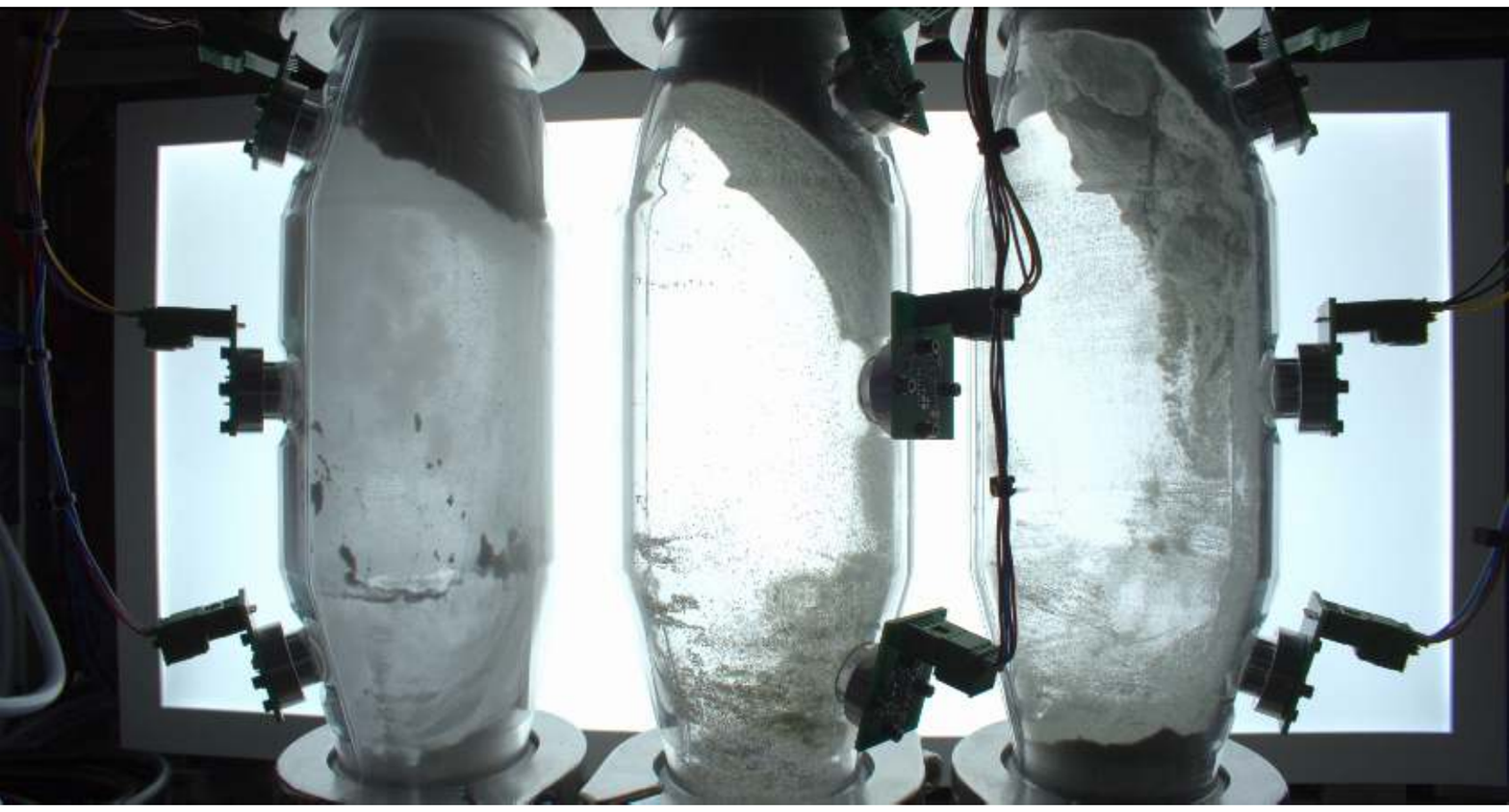}\\
\includegraphics[angle=270,width=0.48\textwidth]{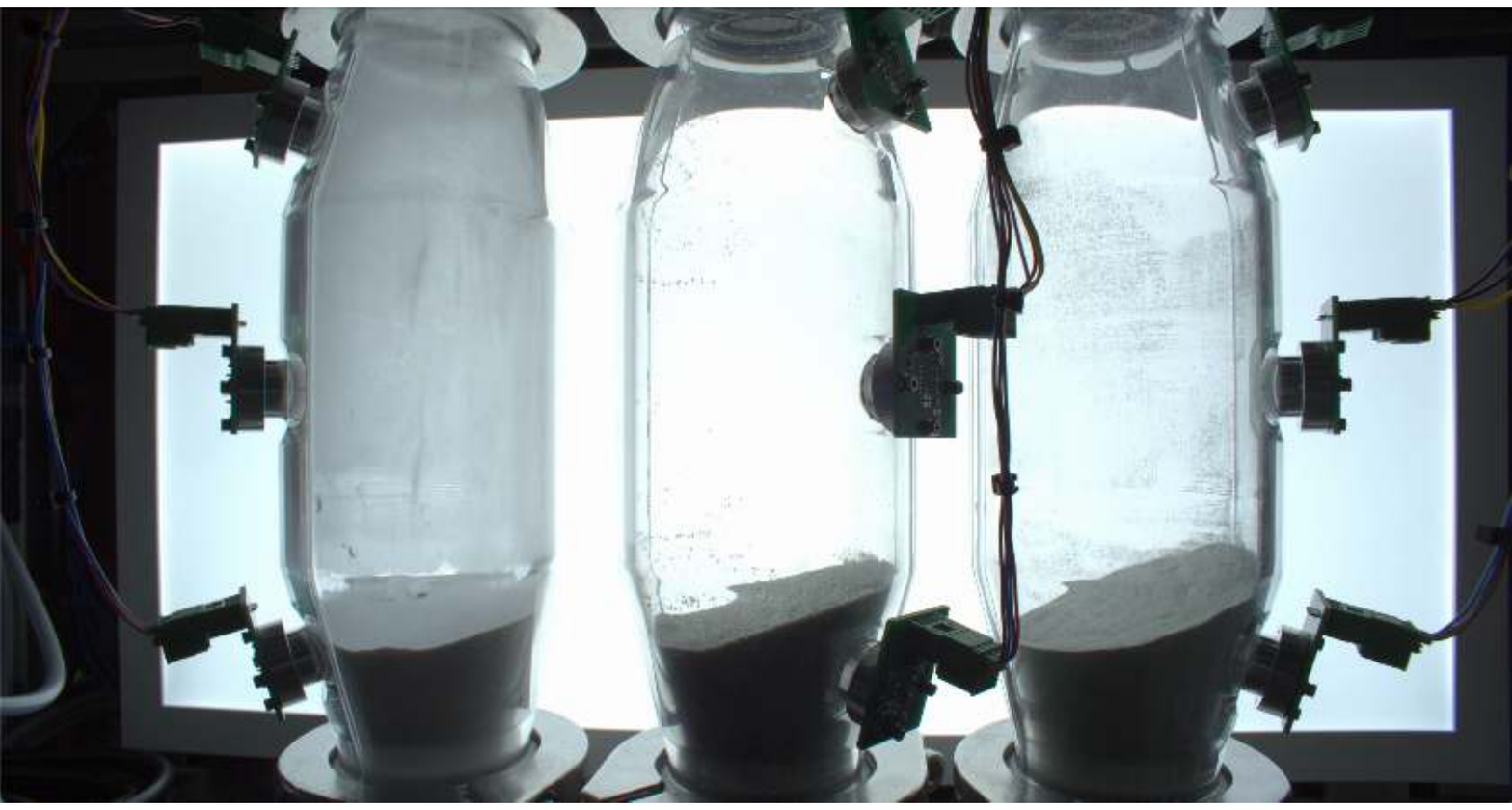}\\
\caption{\label{fig:experiment_illustration} From left to right, chambers 1, 2, and 3, containing Microgrit WCA-40, olivine coarse, and olivine fine. Top: $0g$; Bottom, $g_{\rm Earth}$. The total height of the chambers is 24 cm and maximum inner diameter is 8.24 cm. Obtained during flight 2, parabola number 22, during which time chamber 1 was active. See Table \ref{exp_table2} for particle sizes and mass flux rate. }
\end{figure}

We also obtain imaging data using a high-resolution camera, to use a cross reference to interpret the pressure gradients created by the granular media. Figure \ref{fig:experiment_illustration} shows all three chambers, each containing a different sample (see data corresponding to flight 2 below). Henceforth, we refer to the chambers as  1,2,3, from left to right, respectively. In this image sequence, chamber 1 is the `active' chamber, meaning that the gas line is diverted to this chamber and we will study the pressure gradient in this chamber alone for the duration of an entire parabola, at fixed mass flux rate. The direction of the flow is in opposition to the direction of gravity (from bottom to top). Therefore the higher pressure is measured continuously upstream of (below) the sample, and the reduced pressure is measured downstream of (above) the sample.  The upper panel of the figure shows the situation in $0g$, when most of the material packs into the upper portion of the chambers. The lower panel shows the situation at Earth's gravity, where the particles are settled in the lower portion of the chamber. 

\subsubsection{Samples}\label{subsubsection:samples}
To operate the experimental apparatus, we switch between three identical chambers which are each loaded with a different particle type. This allows us to study more than one kind of granular medium, without changing the samples during flight. As can be seen in the definitions above, some of the determinants of the measured values of $\delta P$ are the sample height of a granular bed, the particle size, the fluid density, and the particle packing density. To make a fair comparison between samples, we try to hold many of these properties constant and vary the parameters systematically. For each of the four flights that we report on, our samples were chosen with a slightly different focus in mind. We initially worked with relatively large spherical particles, so that we could apply simple continuum equations to model the flow and verify the performance \citep{Capelo:2022}. We then experimented with particles of different sizes, shapes, and mass densities. A summary of dust analog materials used, along with their general properties such as composition, shape, porosity, etc. can be found in Table \ref{sample_table}.
\begin{figure*}
\centering
\includegraphics[width=0.8\textwidth]{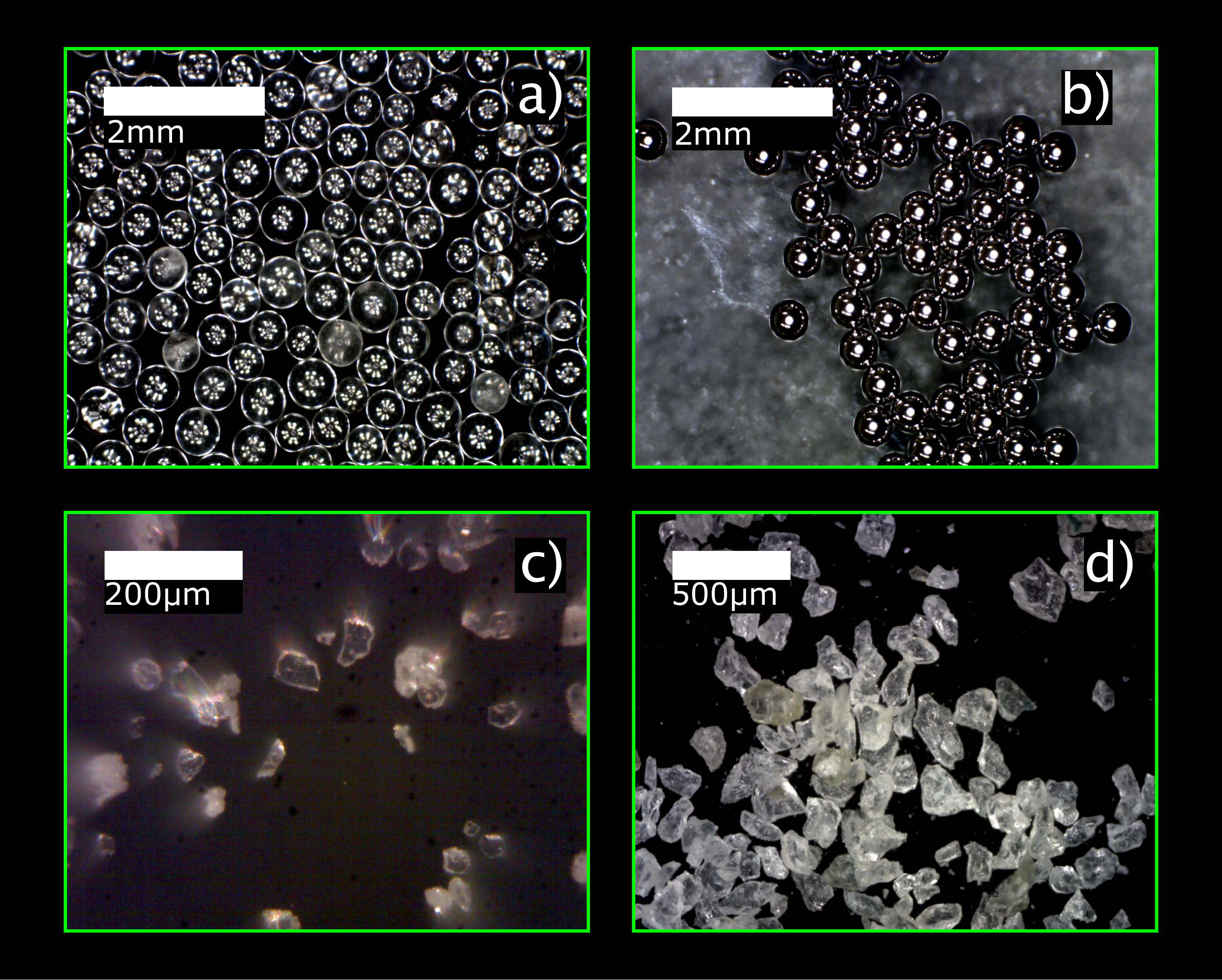}
\caption{\label{Microscope_images} Examples of idealized spherical and more realistic irregular sample materials ; a) glass beads, b) coarse steel beads, c) Microgrit WCA 40, d) Olivine fine. See Tables\ref{exp_table1},  \ref{exp_table2}, \ref{exp_table3}, \ref{exp_table4} for descriptions of the particle size and Table \ref{app:sample_tab} for additional properties.}  \end{figure*}
In Figure \ref{Microscope_images}, we show microscope images of some of our samples, to demonstrate the difference between the spherical samples such as glass (a) and steel (b) beads with the more irregular particles such as Microgrit (c) and olivine sand (d). 

In Flight 1, we were interested in verifying the sensitivity of our instruments using the simple laws stated in Equations \ref{equ:pressure_drop} and \ref{klinkenberg}. So we used particles that were relatively large -- on the order of 0.5 mm --  to result in small $Kn$. The three types of particles, corresponding to chamber 1, 2, and 3 respectively, were soda-lime glass beads, olivine sand, and coarse steel beads. The order in which we collected the data is shown in Table \ref{exp_table1}. The height of each granular bed under$g_{\rm Earth}$ is the same, at $z=5$ cm. The main difference between these samples is therefore the degree of particle sphericity. The coarse steel particles are exactly spherical and uniform in diameter, since they are made for use in high-precision machinery. The glass beads are also spherical, but have a wider dispersion in size. The olivine sand, being naturally occurring, is rough and more irregular in shape, also with a natural variation in size. 

\begin{table*}
\caption{Parameters of the data collected during flight 1. D$_{\rm f}$ is the Feret diameter, Q is the mass flow rate expressed as a percentage of the maximum rate of 1$\times10^{-5}$ kg s$^{-1}$, $\delta P_{\rm min}$ is the mean pressure gradient during the reduced gravity phase, $\delta P_{\rm hg}$ is the mean pressure gradient during the $hg$ phase, and $\delta P_{\rm Earth-g}$ is the mean pressure gradient at $g_{\rm Earth}$, given in hPa. The column g level indicates whether the aircraft entered a $0g$, Martian, or Lunar gravity parabola.}
\label{exp_table1}
\centering
\footnotesize
\begin{tabular}{c c c c c c c c c }  
\hline\hline  
chamber&Parabola Number& Sample type & D$_{\rm f}$ ($\mu$m)& Q (\% 1$\times10^{-5}$ kg s$^{-1}$) &$\delta$P$_{\rm min}$ (hPa)
&$\delta$P$_{\rm hg}$ (hPa)  & $\delta$P$_{\rm Earth-g}$ (hPa) & g level \\
\hline
 1& 1 	& Glass	beads	    	& 552$\pm$ 69  	& 0.5	&0.50& 0.57& 0.58 &0 \\
2 & 2 	& Olivine coarse	& 629$\pm$ 133 	& 0.5	&0.55 &  0.69& 0.70 & 0 \\
3 & 3 	& Steel	beads	    	& 542$\pm$ 76   	& 0.5 	& 0.43&0.54 & 0.54 &0 \\
1 & 5 	& Glass	beads	    	& 552$\pm$ 69  	& 0.8 	&0.67 & 0.76 & 0.77 &0 \\
2 & 6 	& Olivine coarse	& 629$\pm$ 133 	& 0.8 	&0.74 &0.94 & 0.94 & 0\\
3 & 7 	& Steel	beads	    	& 542$\pm$ 76   	& 0.8 	&0.62 &0.75 &0.78  &0 \\
1 & 9 	& Glass	beads	    	& 552$\pm$ 69  	& 0.8	&3.18 & 3.47&3.46 &0 \\
2 & 10 	& Olivine coarse	& 629$\pm$ 133 	& 0.8	&2.75 & 3.86&3.93 & 0\\
3 & 11 	& Steel beads	    	& 542$\pm$ 76  	& 0.8	&2.94 &3.38 &3.36 &0 \\

\hline\hline  

2 & 13 	& Olivine coarse	& 629$\pm$ 133 	& 5	    &3.82 &3.82 & 3.82&  Martian\\
2 & 14 	& Olivine coarse	& 629$\pm$ 133 	& 5	    & 3.81&3.82 & 3.83&  Martian\\
2 & 15 	& Olivine coarse	& 629$\pm$ 133 	& 5	    &3.82 &3.89 & 3.88&  Lunar\\
2 & 16 	& Olivine coarse	& 629$\pm$ 133 	& 5	    &2.85 &3.92 & 3.93&  0\\
\hline
\end{tabular}
\end{table*}

In Flight 2, we again worked with olivine sand in chamber 2, to check the repeatability of the performance between measurement campaigns. We added finer olivine sand with diameter $\sim$ 200 $\mu$m to chamber 3. In chamber 1, we added a new component composed of aluminum oxide Al$_{2}$O$_{3}$ referred to as microgrit WCA 30. This product is smaller in size than the olivine samples, and the shape is rather that of a platelet, as opposed to any degree of sphericity. The order in which we collected the data is shown in Table \ref{exp_table2}. As in flight 1, the height of the granular beds under $g_{\rm Earth}$ was fixed at 5 cm. 

\begin{table*}
\caption{Parameters of the data collected during flight 2. We report measured values of P and $\delta$P. D$_{\rm f}$ is the Feret diameter, Q is the mass flow rate expressed as a percentage of the maximum rate of 1$\times10^{-5}$ kg s$^{-1}$, $\delta P_{\rm min}$ is the mean pressure gradient during the reduced gravity phase, $\delta P_{\rm hg}$ is the mean pressure gradient during the $hg$ phase, and $\delta P_{\rm Earth-g}$ is the mean pressure gradient at $g_{\rm Earth}$, given in hPa.}
\label{exp_table2}
\centering
\footnotesize
\begin{tabular}{c c c c c c c c c }  

\hline\hline  
chamber&Parabola Number& Sample type & D$_{\rm f}$ ($\mu$m)& Q (\% 1$\times10^{-5}$ kg s$^{-1}$) &$\delta$P$_{\rm min}$ (hPa)
&$\delta$P$_{\rm hg}$ (hPa)  & $\delta$P$_{\rm Earth-g}$ (hPa) & g level \\
\hline

1	&2	& Microgrit WCA-40	& 51 $\pm$ 17  	& 0.5	&0.86 & 4.12 &4.74 &  0\\
2	&3	& Olivine coarse	& 629$\pm$ 133 	& 0.5	&0.67 &0.97 &1.00 &  0\\
3	&4	& Olivine fine		& 205$\pm$ 62  	& 0.5	& 0.84&1.86 &1.76 &  0\\

1	&7	& Microgrit WCA-40	& 51 $\pm$ 17  	& 0.8	&1.08 &4.79 &4.71 &  0\\
2	&8	& Olivine coarse	& 629$\pm$ 133 	& 0.8	&0.95 &1.27 &1.17 &  0\\
3	&9	& Olivine fine		& 205$\pm$ 62  	& 0.8 	&1.00 &2.24 &2.18  & 0\\

1	&12 	& Microgrit WCA-40	& 51 $\pm$ 17  	& 1	&1.94 &5.72 &4.71  & 0\\
2	&13 	& Olivine coarse	& 629$\pm$ 133 	& 1	&0.94 &1.19 &1.18  & 0\\
3	&14 	& Olivine fine		& 205$\pm$ 62  	& 1	&0.98 &2.52 &2.18  & 0\\

1	&17 	& Microgrit WCA-40	& 51 $\pm$ 17  	& 2	&2.71 &9.42 &5.32 &  0\\
2	&18 	& Olivine coarse	& 629$\pm$ 133 	& 2	&1.33 &1.99 &1.15 &  0\\
3	&19 	& Olivine fine		& 205$\pm$ 62  	& 2	&1.84 & 4.07&2.56 &  0\\

1	&22 	& Microgrit WCA-40 	& 51 $\pm$ 17  	& 3	& 2.91&10.76 &9.19 &  0\\
2	&23 	& Olivine coarse 	& 629$\pm$ 133 	& 3	&2.36 &2.53 &1.99 &  0\\
3	&24 	& Olivine fine		& 205$\pm$ 62  	& 3	&2.12 &5.53 &3.99 &  0\\

1	&27 	& Microgrit WCA-40	& 51 $\pm$ 17  	& 5	&3.7 & 13.86&8.1 &  0\\
2	&28 	& Olivine coarse	& 629$\pm$ 133 	& 5	&2.88 &3.86 & 2.19&  0\\
3	&29 	& Olivine fine		& 205$\pm$ 62  	& 5	&3.68 &7.95 & 5.46&  0\\

\hline
\end{tabular}
\end{table*}

In flight 3, we worked with particles in the 50$\mu$m size range. Our focus here is to explore the impact of mass density on the resulting pressure gradient. Since the previous measurement campaigns showed that the 5 cm bed height was more than sufficient to create a pressure drop, we reduced the bed height to achieve lower filling factors during the 0g phase. This also results in lower average pressures in the chambers because there is less granular material to impede the flux of gas. Chamber 1, contained fine steel beads in the size range 40-60 $\mu$m. Chambers 2 and 3 both contained olivine sand in the 40-60 $\mu$m range. In one of the olivine-laden chambers (2) we used an equivalent total particle mass as the fine steel reference sample, of 64 g measured under $g_{\rm Earth}$. In the other olivine-laden chamber (3) we used the equivalent volume of particles as the fine steel reference chamber, in the quantity 8.1 ml.

\begin{table*}
\caption{Parameters of the data collected during flight 3. D$_{\rm f}$ is the diameter of particles that were sieved into the indicated size range, Q is the mass flow rate expressed as a percentage of the maximum rate of 1$\times10^{-5}$ kg s$^{-1}$, $\delta P_{\rm min}$ is the mean pressure gradient during the reduced gravity phase, $\delta P_{\rm hg}$ is the mean pressure gradient during the $hg$ phase, and $\delta P_{\rm Earth-g}$ is the mean pressure gradient at $g_{\rm Earth}$, given in hPa}.
\label{exp_table3}
\centering
\footnotesize
\begin{tabular}{c c c c c c c c c }  
\hline\hline  
chamber&Parabola Number& Sample type & D$_{\rm f}$ ($\mu$m)& Q (\% 1$\times10^{-5}$ kg s$^{-1}$) &$\delta$P$_{\rm min}$ (hPa)
&$\delta$P$_{\rm hg}$ (hPa)  & $\delta$P$_{\rm Earth-g}$ (hPa) & g level \\
\hline

1	&2	& Steel	& 40-80  	& 0.5	&0.45 &0.75 & 0.70&  0\\
2	&3	& Olivine	& 40-80  	& 0.5	&0.60 &0.97 & 0.97&  0\\
3	&4	& Olivine	& 40-80  	& 0.5	& 0.45&0.68 &0.68 &  0\\

1	&7	& Steel	& 40-80   & 0.8	&0.62 &0.87 &0.87 &  0\\
2	&8	& Olivine	& 40-80  	& 0.8	&0.65 &1.27 & 1.27&  0\\
3	&9	& Olivine		& 40-80  	& 0.8 	&0.56 &0.90 & 0.90&  0\\

1	&12 	& Steel	& 40-80  	& 1	&0.64 &1.03 & 1.02&  0\\
2	&13 	& Olivine	& 40-80  	& 1	&0.69 &1.43 & 1.51&  0\\
3	&14 	& Olivine		& 40-80  	& 1	&0.68 &1.06 &1.00 &  0\\

1	&17 	& Steel	& 40-80  	& 2	&1.16 &1.83 &1.96 &  0\\
2	&18 	& Olivine	& 40-80  	& 2	&1.22 &2.40 &2.33 &  0\\
3	&19 	& Olivine		& 40-80  	& 2	&1.21 &1.77 &1.73 &  0\\

1	&22 	& Steel 	& 40-80  	& 3	&2.42 &2.53 & 2.73&  0\\
2	&23 	& Olivine 	& 40-80  	& 3	&2.14 & 3.68& 3.68 &  0\\
3	&24 	& Olivine		& 40-80  	& 3	&1.79 & 2.28&2.43 &  0\\

1	&27 	& Steel	& 40-80  	& 5	&2.77 &4.14 &6.06 &  0\\
2	&28 	& Olivine	& 40-80   & 5	&3.97 &5.19 &5.05 &  0\\
3	&29 	& Olivine	& 40-80  & 5	&2.79&3.79 &3.52 &  0\\

\hline
\end{tabular}
\end{table*}

During flight 4, we used sample material that has been engineered to mimic the mineral composition of the CR class of carbonaceous chondrite meteorites. The product CR Asteroid Regolith Simulant from the supplier Exolith Labs (https://exolithsimulants.com) has been well-characterized in \cite{Britt:2019} and \cite{Schweighart:2021}. In particular, the size distribution of the grains is measured in \cite{Schweighart:2021}. We have dry sieved the CR asteroid regolith simulant into three different size ranges: 40-80 $\mu$m (chamber 1), 80-200 $\mu$m (chamber 2), 200-800 $\mu$m (chamber 3).
 It is worth to note, however that the smallest particles in this simulant can easily agglomerate and subsequently break apart, and so particles which belong to a particular size range while sieving may disintegrate while handling the samples, thus adding smaller particles to the population. 

\begin{figure*}
\includegraphics[width=0.7\textwidth,angle=270]{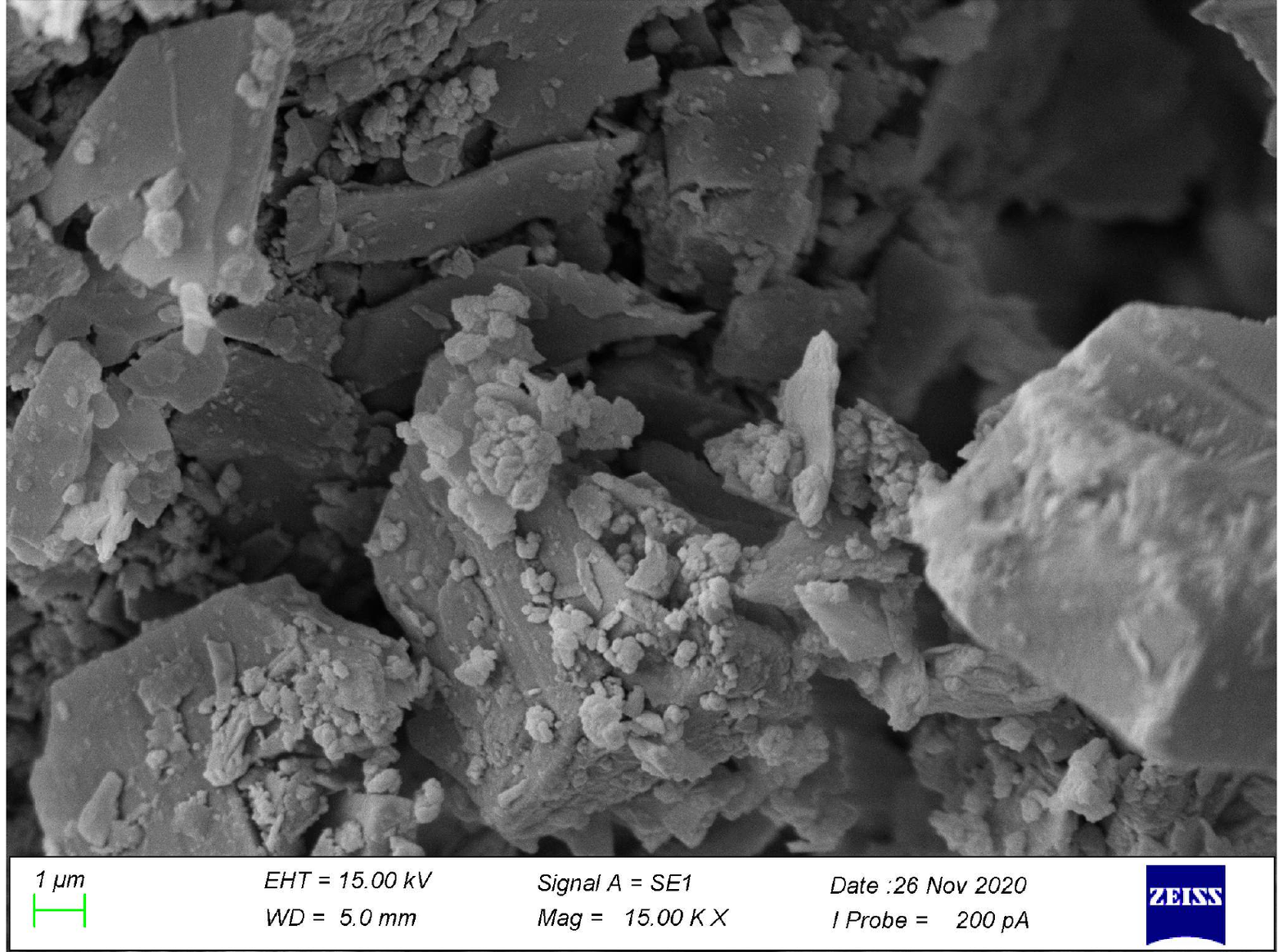}
\caption{\label{SEM_image} Scanning Electron Microscope image of CR asteroid regolith simulant from Exolith labs. Scale bar represents 1 $\mu$m. Irregular shape, as well as broad range of sizes can be identified.}
\end{figure*}

\begin{table*}
\caption{Parameters of the data collected during flight 4. D$_{\rm f}$ is the diameter of particles that were sieved into the indicated size range, Q is the mass flow rate expressed as a percentage of the maximum rate of 1$\times10^{-5}$ kg s$^{-1}$, $\delta P_{\rm min}$ is the mean pressure gradient during the reduced gravity phase, $\delta P_{\rm hg}$ is the mean pressure gradient during the $hg$ phase, and $\delta P_{\rm Earth-g}$ is the mean pressure gradient at $g_{\rm Earth}$, given in hPa}.
\label{exp_table4}
\centering
\footnotesize
\begin{tabular}{c c c c c c c c c}  
\hline\hline  
chamber&Parabola Number& Sample type & D$_{\rm f}$ ($\mu$m)& Q (\% 1$\times10^{-5}$ kg s$^{-1}$) &$\delta$P$_{\rm min}$ (hPa) &$\delta$P$_{\rm hg}$ (hPa)  & $\delta$P$_{\rm Earth-g}$ (hPa) & g level \\
\hline

1	&2	& CR simulant	& 40-80   	& 0.5	& 0.50 &1.2 &1.25 &  0\\
2	&3	& CR simulant	& 80-200  	& 0.5	& 0.54&1.78 &1.91 &  0\\
3	&4	& CR simulant	& 200-800   & 0.5	&0.52 &0.54 &0.54 &  0\\

1	&7	& CR simulant	& 40-80   		& 0.8	&0.62 &1.17 &1.13  & 0\\
2	&8	& CR simulant	& 80-200  		& 0.8	& 0.69&2.45 &2.27  & 0\\
3	&9	& CR simulant	& 200-800  	& 0.8 & 0.68 &0.76 &0.75  & 0\\

1	&12 	& CR simulant	& 40-80   	& 1	& 0.74&1.52 &1.46  & 0\\
2	&13 	& CR simulant	& 80-200  		& 1	&0.80 & 2.95&2.37  & 0\\
3	&14 	& CR simulant	& 200-800   & 1	& 0.78 &0.82 &0.82  & 0\\

1	&17 	& CR simulant	& 40-80   		& 2	&1.18 &2.78 &2.05  & 0\\
2	&18 	& CR simulant	& 80-200  		& 2	&1.54 & 5.09&3.95  & 0\\
3	&19 	& CR simulant	& 200-800   & 2 &1.27 &1.36 &1.34  & 0\\

1	&22 	& CR simulant 	& 40-80   		& 3	& 2.04 &3.80 &2.90  & 0\\
2	&23 	& CR simulant 	& 80-200  		& 3	&1.83 &6.03 &4.46  & 0\\
3	&24 	& CR simulant	& 200-800  	& 3	&1.83 &1.84 &1.82  & 0\\

1	&27 	& CR simulant	& 40-80   		& 5	&2.83 &4.52 &4.53  & 0\\
2	&28 	& CR simulant	& 80-200  		& 5	&3.05 &7.23 &7.23  & 0\\
3	&29 	& CR simulant	& 200-800  	& 5	&2.84 &2.84 &2.84  & 0\\

\hline
\end{tabular}
\end{table*}

\subsubsection{Pressure Gradients}\label{subsubsection:pressure_gradient_overview}

In Tables \ref{exp_table1}-\ref{exp_table4}, we give an overview of the flight sequence indicating which chamber was operated during subsequent parabolas. Each chamber corresponds to a particular sample type, defined by the particle size and material composition. We systematically alternate through chambers 1, 2, and 3, and adjust the mass flow rate $Q$ after a measurement has been obtained for each chamber. We refer to a set of measurements at fixed mass flow rate as a series. For each measurement, we report a representative average value of the pressure gradient, corresponding to a small range in gravitational load. The mean of pressure gradients measured during $0g$, $\delta P_{\rm min}$ correspond to when the gravitational load g is in the range 0 < g < 0.1 g$_{\rm Earth}$; the mean of the pressure gradient measured during steady flight, $\delta P_{\rm Earth-g}$, is in the range $0.9 g_{\rm Earth}$< g < 1.1 g$_{\rm Earth}$; the mean of pressure gradients measured during the $hg$ pull-up and pull-out phases of the parabola, $\delta P_{\rm hg}$, correspond to the range 1.1 g$_{\rm Earth}$< $g$ < 2.0 g$_{\rm Earth}$. Exceptions to these data ranges are the three partial -- Lunar or Martian-- gravity measurements, shown in Table \ref{exp_table1}, in which case $\delta P_{\rm min}$ corresponds to 0 g$_{\rm Earth}$<  $g$ < 0.1g$_{\rm Earth}$.

\begin{figure}
\includegraphics[width=0.45\textwidth]{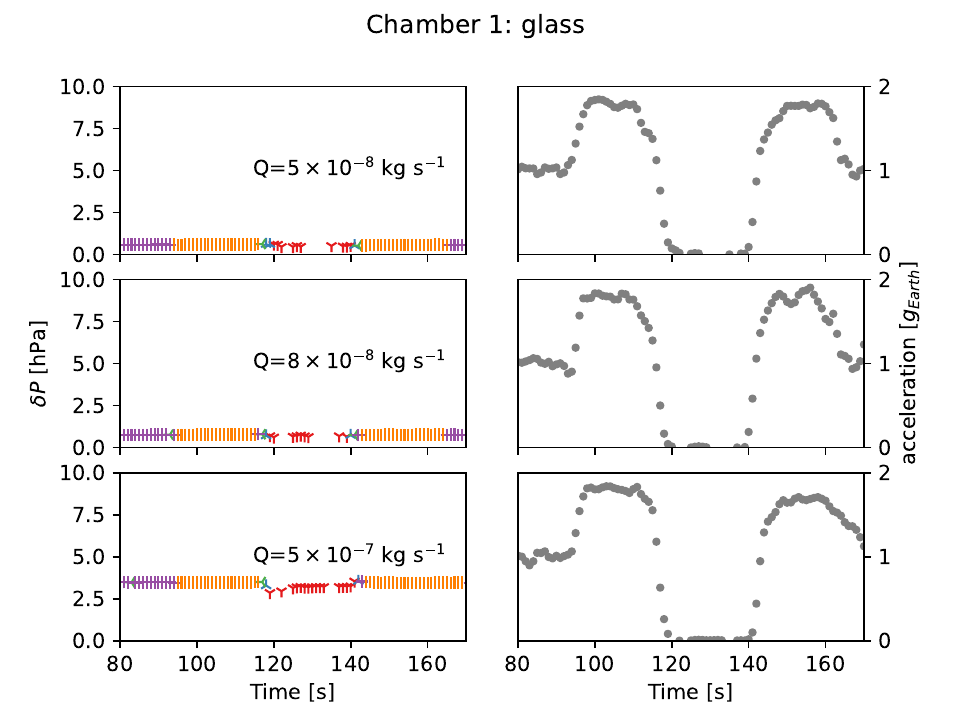}\\
\includegraphics[width=0.45\textwidth]{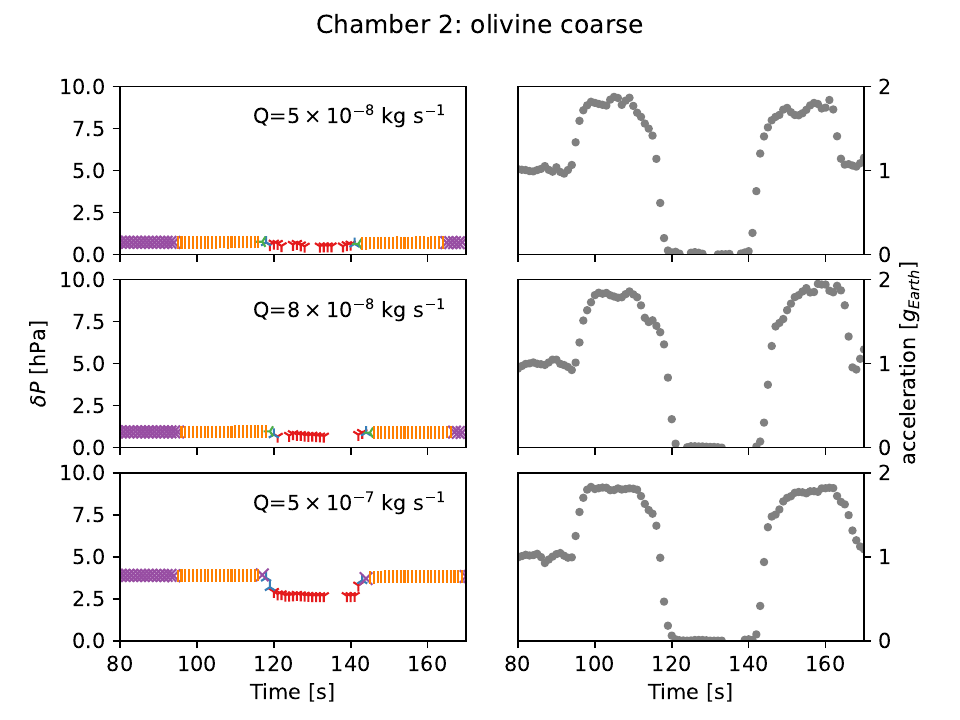}\\
\includegraphics[width=0.45\textwidth]{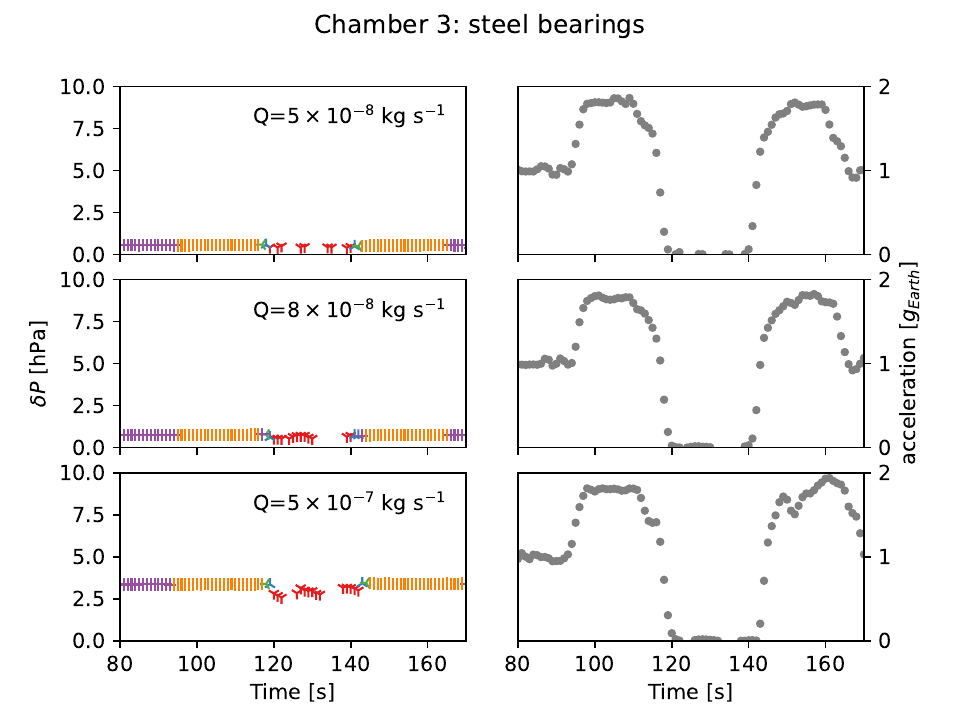}

\caption{\label{flight1_1} $\delta P$ vs. time and acceleration vs time, for each of the three samples in flight 1.  All samples consist of particles $\sim$50 mm. Purple plus signs correspond to steady-flight at $g_{\rm Earth}$, orange |'s correspond to the $hg$ phases; red down-arrows correspond to the $0g$ phase, whereas the green left arrows and blue up arrows correspond to intermediate phases, transitioning into and out of the $0g$ phase. See also Table \ref{exp_table1}.}
\end{figure}

\begin{figure}
\includegraphics[width=0.45\textwidth]{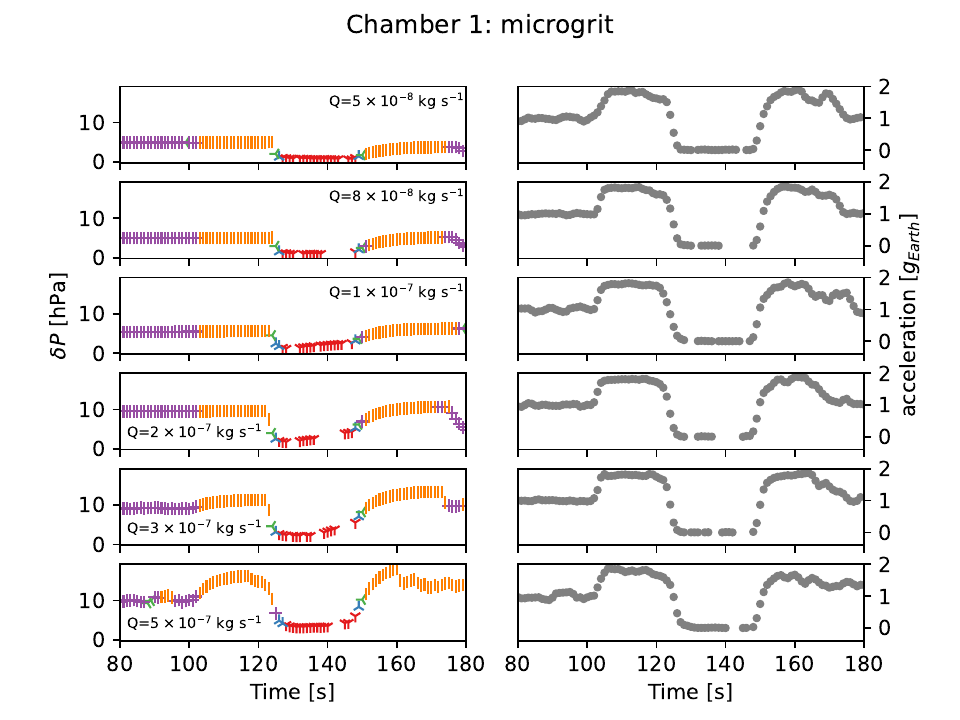}\\
\includegraphics[width=0.45\textwidth]{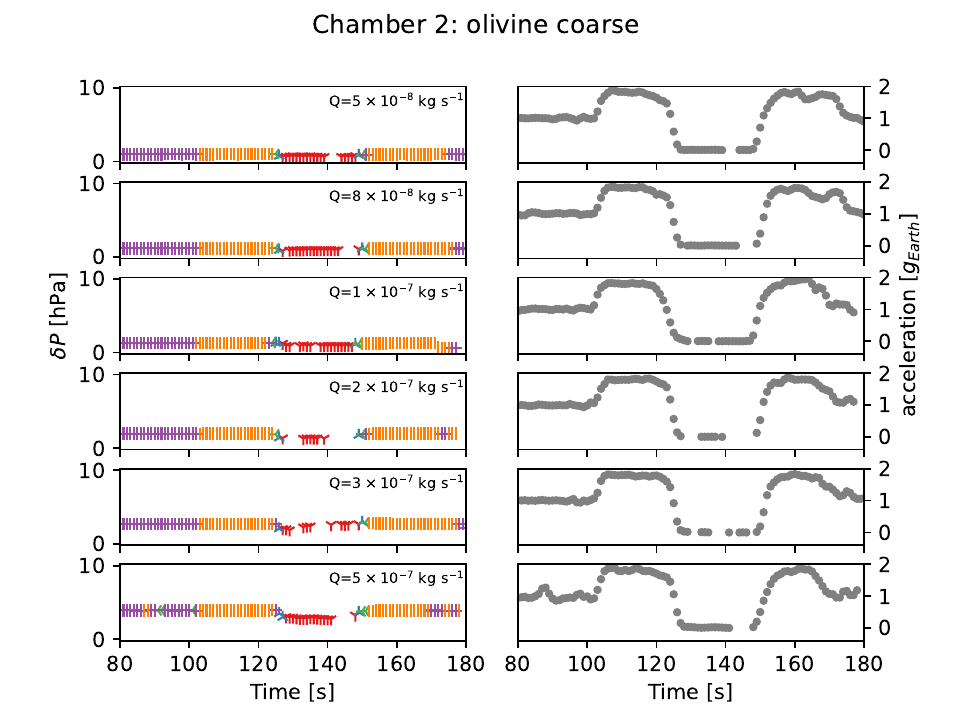}\\
\includegraphics[width=0.45\textwidth]{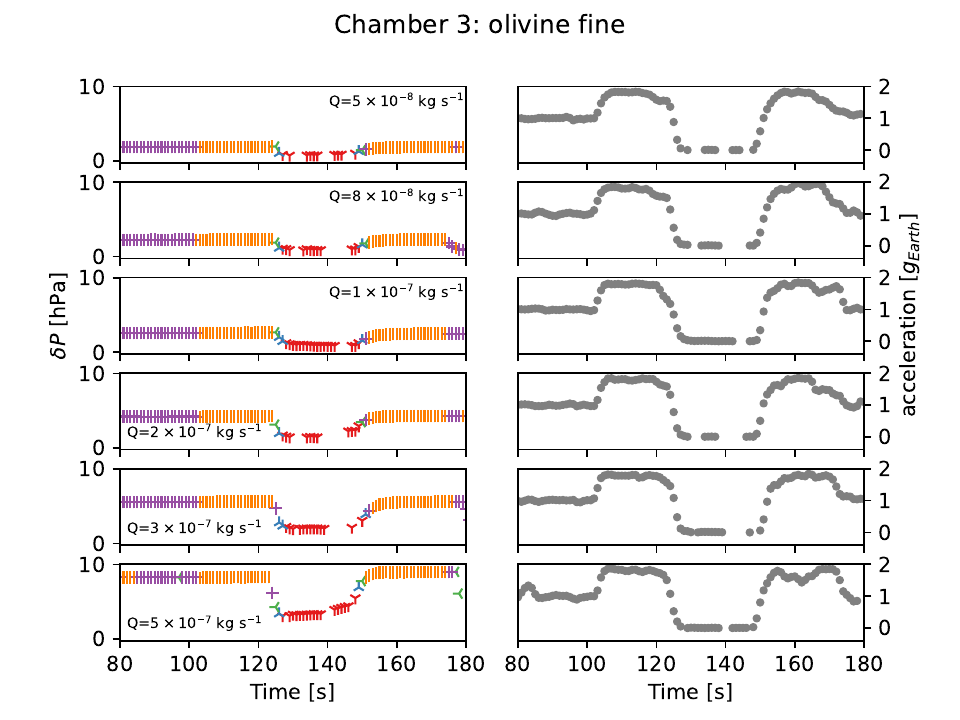}
\caption{\label{flight1_2} $\delta P$ vs. time and acceleration vs time, for each of the three samples in flight 2. Coloring and symbols are the same as for Figure \ref{flight1_1}. See also Table \ref{exp_table2}.}
\end{figure}

\begin{figure}
\includegraphics[width=0.45\textwidth]{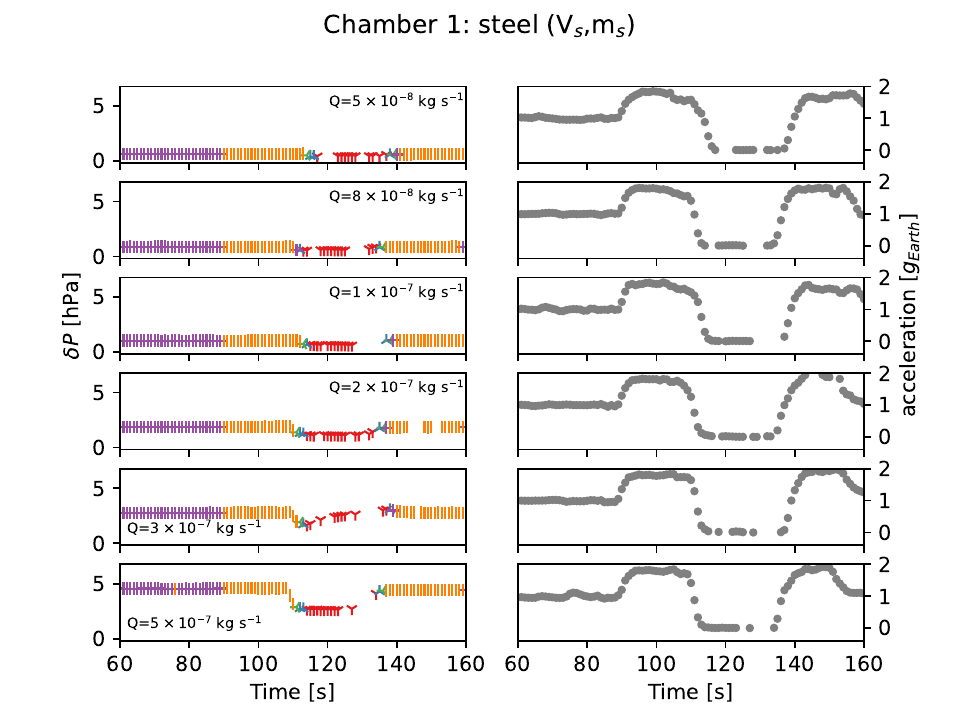}\\
\includegraphics[width=0.45\textwidth]{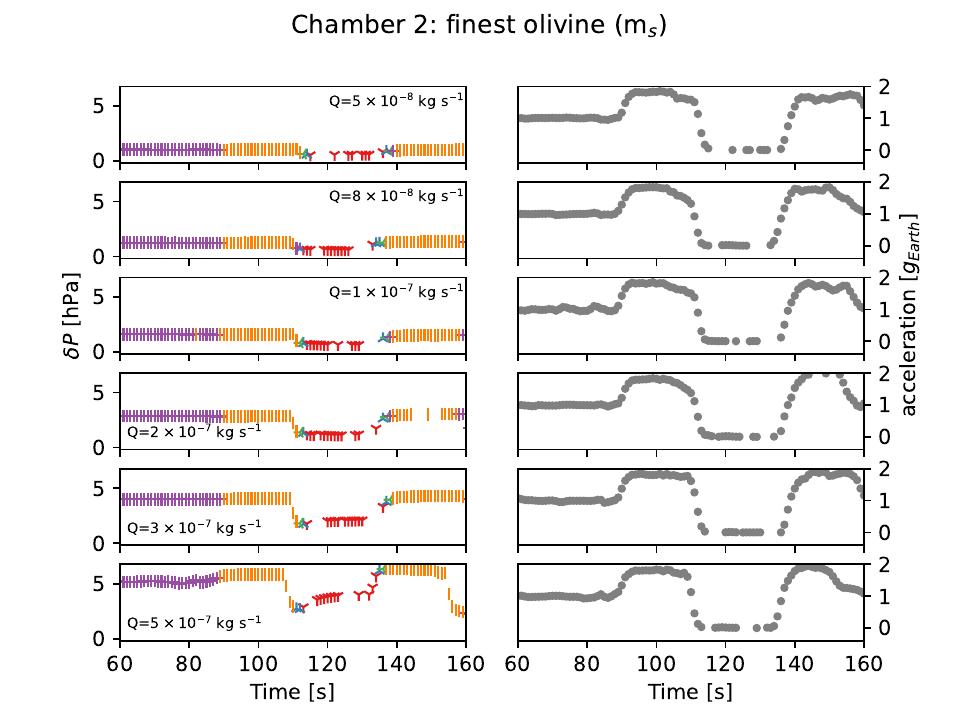}\\
\includegraphics[width=0.45\textwidth]{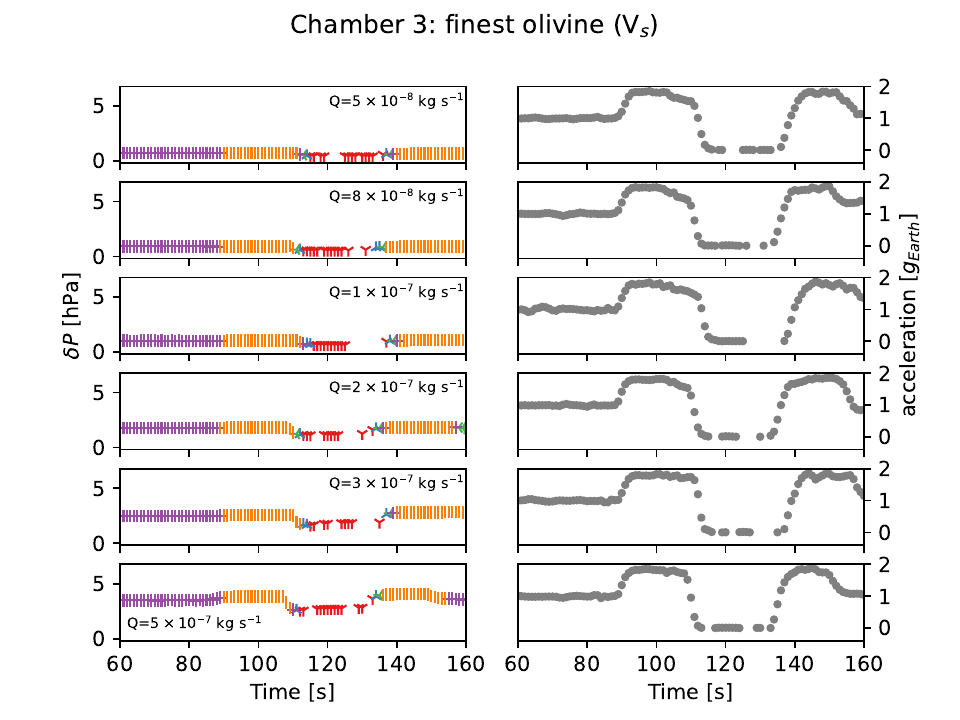}
\caption{\label{flight1_3} $\delta P$ vs. time and acceleration vs time, for each of the three samples in flight 3. All particles have 40-80 micrometer diameter and the purpose is to compare mass density effects. Coloring and symbols are the same as for Figure \ref{flight1_1}. See also Table \ref{exp_table3}.} 
\end{figure}

\begin{figure}
\includegraphics[width=0.45\textwidth]{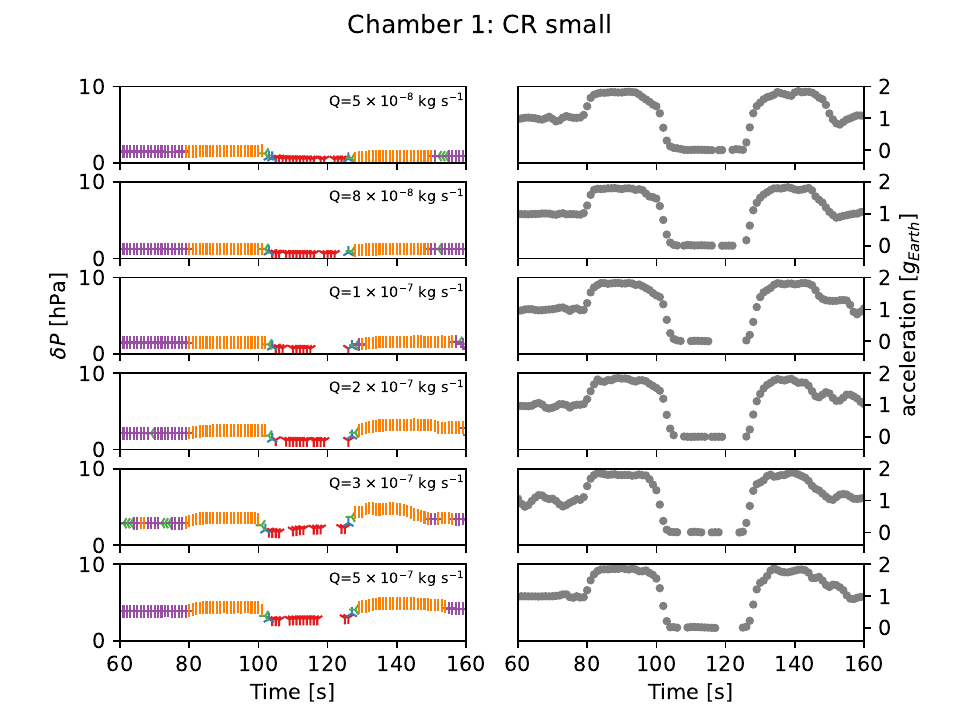}\\
\includegraphics[width=0.45\textwidth]{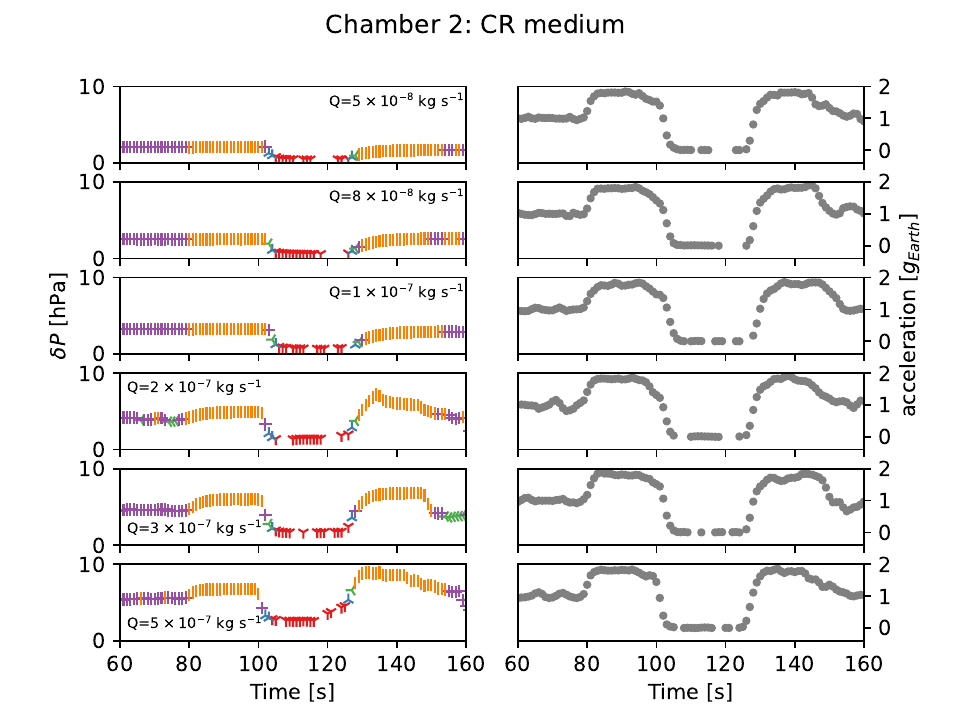}\\
\includegraphics[width=0.45\textwidth]{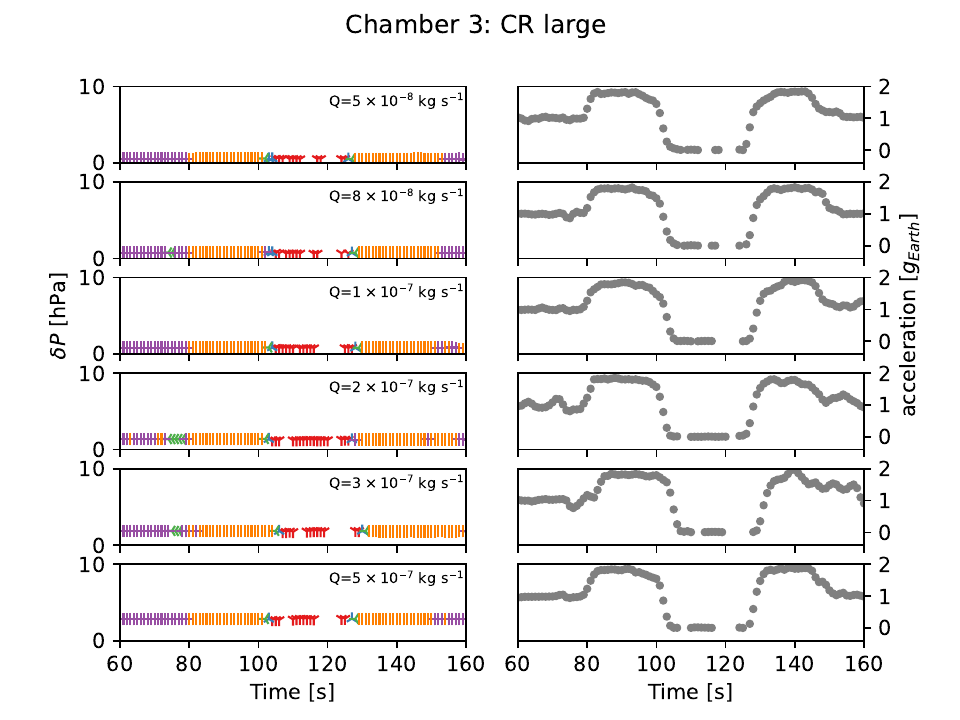}
\caption{\label{flight1_4} $\delta P$ vs. time and acceleration vs. time, for each of the three samples in flight 4. Coloring and symbols are the same as for Figure \ref{flight1_1}. See also Table \ref{exp_table4}. } 
\end{figure}

The mean pressure gradients reported in Tables \ref{exp_table1}-\ref{exp_table4} provide insights into fundamental behavior, but they do not fully encompass the intriguing aspects of the pressure gradient's temporal evolution, which is influenced by gravity and pressure changes. In Figures \ref{flight1_1}-\ref{flight1_4} we show the temporal variation of $\delta P$, according to gravitational load and mass flow rate. We do not include any errors on the reported pressure and time values because they are around 2\%, and therefore would not appear on the graphs. We have divided the parabolas into phases according to gravitational load, and shown them with different colors, as described in the caption of Figure \ref{flight1_1}.  For clarity, the $\delta P$ values reported in the tables are the averages of the points of a particular color corresponding to range in gravitational load. The mass flow rate $Q$ in Figures \ref{flight1_1}-\ref{flight1_4} (all panels) increases from top to bottom, and so the pressure of the gas coming into the chamber is progressively higher. The associated change in the normalisation of the value $\delta P$ with increasing $Q$ reflects the expectation that the magnitude of the pressure gradient decreases with increasing Knudsen number.  A notable feature at $0g$ is the variation around the minimum $\delta$P. This is due to the jitter of the aircraft (small fluctuations resulting from outside air turbulence, which can be seen as variability in the acceleration profiles plotted in the right side panels of Figures \ref{flight1_1}-\ref{flight1_2}) moving the experimental apparatus and causing fluctuations around the $0g$ level of $\sim \pm 0.003$ g. 

The minimum is an important value, however, because it represents the situation in which all of the material is packed into the upper conical reservoir of the chamber.  One would expect, due to symmetry, that the pressure difference would return to its steady-flight value if the packing volume is the same when the particles reach the opposite side of the chamber. Instead, the pressure difference is at all times significantly smaller at $0g$ than at $g_{\rm Earth}$. This is an indicator of the reduced compression and thus increased porosity, of the granular bed when gravity is not acting upon it.

In Figure \ref{flight1_1}, we plot $\delta P$ vs. time and acceleration vs. time, for each sample of flight 1, and for the relative mass flow rates listed in Table \ref{exp_table1}. Chamber 1 refers to the glass beads, Chamber 2 refers to the olivine sand, and Chamber 3 refers to the coarse steel beads. For all these samples with particles in the size close to 0.5 mm, and same initial sample height, there is very little difference between the measured pressure differentials. The coarse steel and glass are practically identical. Olivine is slightly more effective at creating a pressure differential at $g_{\rm Earth}$ and has a deeper dip in pressure differential at $0g$. A feature to notice is that the $\delta P$ curves before and after the 0$g$ phase are relatively flat and symmetric. The angle of repose of the settled material can be seen to change sometimes when the aircraft pulls out of the parabola, but apparently this does not significantly affect $\delta P$.  

In Figure \ref{flight1_2}, we plot $\delta P$ vs. time and acceleration vs. time, for each sample of flight 2, and for the mass flow rates listed in Table \ref{exp_table2}. Chamber 1 refers to the microgrit samples, Chamber 2 refers to the coarse olivine sand, and Chamber 3 refers to the fine olivine sand. The sample shown in the middle panel of Figure \ref{flight1_2} is the same as in the middle panel of Figure \ref{flight1_1}, showing experimental repeatability. The extracted averages of $\delta$P reported in tables \ref{exp_table1} and \ref{exp_table2} agree to about $20\%$. We do not expect exact one-to-one correspondence since the unavoidable variability of the aircraft acceleration induces variability in the pressure gradient. The microgrit sample shows a strong asymmetry in the pull-up vs. pullout phases and a very enhanced pressure differential in $hg$. For clarification, the sample is primarily located in the lower part of the chamber during these phases, however the pull-out phases involve a very brief period of sedimentation that corresponds to the transition in gravity level (e.g. points shown in blue and green in figure \ref{flight1_2}).

In Figure \ref{flight1_3}, we plot $\delta P$ vs. time and acceleration vs. time, for each sample of flight 3, and for the mass flow rates listed in Table \ref{exp_table3}. Chamber 1 refers to fine steel beads, chamber 2 refers to the fine olivine sand with the same total mass as the fine steel in chamber 1, and chamber 3 refers to the fine olivine sand with the same total volume as the fine steel in chamber 1. The purpose of these samples was to compare the relative importance of mass density in creating pressure difference.  Therefore, we notice that when comparing chamber 1 to chamber 2, the baseline value of $\delta P$ at $g_{\rm Earth}$ is at a lower average value than of the fine steel in chamber 1 for low pressure, but that this gap closes at higher pressure. Comparing chamber 1 to chamber 3, we notice that the baseline behavior is very similar, except at the highest mass flow rate. These trends indicate that the more massive material becomes equally effective at generating a pressure gradient only when the momentum diffusivity is low. We analyze these trends further below. 

In figure \ref{flight1_4}, we plot $\delta P$ vs. time and acceleration vs. time, for each sample of flight 4, and for the mass flow rates listed in Table \ref{exp_table4}. All three chambers contained asteroid regolith simulant, in different size ranges. Chamber 1 refers to the smallest size range, chamber 2 refers to the medium size range, and chamber 3 refers to the largest size range.  It is easily noticeable that the samples in chamber 1 and 2 are both very susceptible to pressure gradient variations during the $hg$ phases, whereas the sample in the third chamber is not. The samples in chambers 1 and 2 also have a remarkably similar behavior to that of the microgrit studied in flight 2, (see Table \ref{exp_table2} and Figure \ref{flight1_2}, chamber 1 top panel). In particular, there is a strong asymmetry in the $\delta P$ curve for the two $hg$ phases. For both the microgrit, and the regolith simulant, there is a distinctive pattern of increasing $\delta P$ with increasing mass flow rate during the pull-out phase of the parabola, that changes dramatically over the course of several sequences. We point out that since this pattern occurs in logical order from low- to high mass flow rate, and since it is repeated for three different samples, this asymmetric behavior can be neither random nor systematic. It is also the case that we study samples with overlapping size ranges which do not exhibit this behavior such as the fine olivine sand shown in Figures \ref{flight1_2} and \ref{flight1_3}.  The difference seems to be the irregularity of the particle shapes found in the microgrit and regolith samples. That the compression history should matter in the apparent gas permeability of a sample merits further analysis below.  

\section{Analysis} \label{section:analysis}
\subsection{Compression effects: hyper-g}\label{subsection:compression}
\begin{figure}
\includegraphics[width=0.45\textwidth]{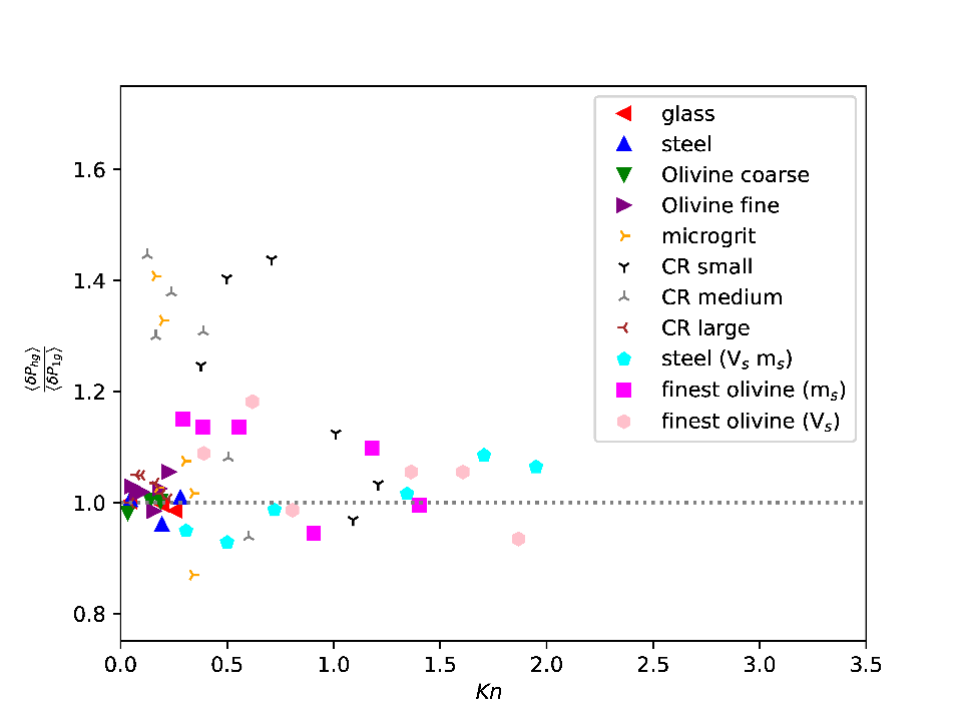}
\caption{\label{fig:compression_Knudsen} Compression factor as a function of $Kn$, for all samples and the complete flight measurement sequences. Values correspond to the $hg$ phase. Glass (red left triangles), steel (blue up triangles), as well as coarse and fine olivine (green down triangles or purple right triangles) show little response to enhanced gravitational load. Of the CR asteroid regolith simulant samples, only the small and medium size bins (black down-arrows and grey up-arrows) show a dramatic response, as opposed to the large size fraction (brown left-arrows). The microgrit (orange right-arrows) behaves similarly to the smaller CR samples. While the responsive samples show a trend with changing mass flow rate, the responsiveness does not strictly depend upon $Kn$, as different samples with similar $Kn$ can have different behaviors. The three samples that were measured at a wide range of $Kn$ (steel as cyan pentagrams, finest olivine as either magenta squares or pink hexagons), fluctuate in their response to gravitational load, with the two types of sample materials showing inverse behavior.}
\end{figure}

We address the effect of compression on the measured pressure gradient, due to enhanced gravitational load during the $hg$ phase. The `compression factor'  $f_{c}=\frac{\left<\delta P_{hg}\right>}{\left<\delta P_{\rm Earth-g}\right>}$, which is the mean pressure gradient at $hg$ vs. the mean pressure gradient at $g_{\rm Earth}$, represents the relative effect of enhanced external force on the sample. In Figure \ref{fig:compression_Knudsen}, we plot $f_{c}$ as a function of Knudsen number $Kn$, for all samples described in Section \ref{subsubsection:samples}, and all sequences indicated in Section \ref{subsubsection:pressure_gradient_overview}. We choose $Kn$ as the x-axis value to make a fair comparison of gas pressure-dependency, while controlling for particle size. For the calculation of $Kn$, we use $\frac{\delta P}{2}$ as the mean gas pressure in the chamber. We convert to mean free path assuming an ideal gas composed of air with molar mass M$=0.028914$ kg mol$^{-1}$, viscosity of 1.8$\times10^{-5}$ kg m$^{-1}$ s$^{-1}$ and temperature 295K.  

It can be seen directly in this figure that the three samples of sizes near 500$\mu$m studied during flight 1 --  Glass, coarse steel beads, and coarse olivine sand -- interact with gas with relatively low $Kn$, because of relatively large mean free paths compared to particle size. Other samples studied during flights 2 and 4 have similar $Kn$ as those studied in flight 1, even though the particle sizes used were much smaller. This is because small particles result in small pore-opening space and hence smaller permeability coefficient: these samples are very effective at blocking the flow and generating a pressure gradient, and so the average pressure in the chamber remains limited at a relatively high pressure, hence also limiting $Kn$. This is the case for the microgrit sample shown with orange right-arrows, as well as the medium size fraction of CR asteroid regolith simulant shown with grey up-arrows. Other samples reach greater $Kn$, such as the coarse steel beads and olivine samples, which all had sizes in the range 40-80$\mu$m; in the experiments of flight 3, the granular bed height $dh$ was greatly reduced, hence creating a smaller pressure differential and allowing lower overall pressures to be achieved.  

A horizontal dotted line at $f_{c}=1$ is overplotted for reference in Figure \ref{fig:compression_Knudsen}, since this value represents the case where there is no change during the $hg$ phase, as compared to the $g_{\rm Earth}$ phase. The most dramatic deviations from $f_{c}=1$ arise with the three samples in particular: microgrit, CR small size fraction, and CR medium size fraction. These samples can have compression factors up to $40-50\%$. Note that this factor involves a mean value at $hg$ including both pull-up and pull-out phases, and so excursions could be even greater.

 Despite averaging, the trend of a growing asymmetry with increasing gas flow rate, which was apparent in Figures \ref{flight1_2} and \ref{flight1_4} can also be recognised here, as the excursions from $f_{c}=1$ grow inversely with $Kn$. This trend is challenging to interpret. First, it is not obvious why the pressure gradient should depend so strongly upon the history of the gravitational loading. Second, it is counter-intuitive that the apparent highly packed state correlates with increasing gas flow rate. One might postulate the opposite: that with greater gas pressure the sample should be supported against gravity and therefore not compress as easily. This behavior can be comprehended best when we revisit the definitions from Section \ref{subsection:definitions}, as applied to our data; we will present the permeability coefficients in Section \ref{section:analysis}, and discuss the meaning of the time-dependent evolution further in section \ref{section:conclusions}. Regardless, it is clear that the common feature of the three samples which exhibit this surprising trend is the highly irregular particle shape. Both the microgrit and CR asteroid regolith simulant are highly non-spherical, and therefore voided space can be filled much more easily and in a flexible manner, compared to the case of hard, non-overlapping spheres. The coarse asteroid regolith simulant, consisting of 200-800 $\mu$m particles behaves similarly to the spherical particles of flight 1, all with particle sizes close to half mm. This underscores the importance of particle size in limiting the extent to which the pore space can be minimized by compression. Since the total voided area depends upon the surface area of the particles, larger particles do not compress to low porosity conditions as readily as small particles do. Therefore, the permeability is similar at all g-levels.

The relatively spherical samples show little or no deviations from $f_{c}=1$. A moderate exception are the three samples studied during fight 3, which are the particles of sizes 60-80$\mu$m, of fine steel and `finest' olivine. Recall that the fine steel sample in chamber 1 possessed an initially fixed volume $V_{s}$ and mass $m_{s}$, whereas the finest olivine sample in chamber 2 possessed fixed $m_{s}$ and the finest olivine sample in chamber 2 possessed fixed $V_{s}$. The samples were prepared in this way in order to eventually study in more detail the different effects of mass and volume upon the pressure gradient. All three of these samples can show deviations from $f_{c}=1$ up to approximately $10-20\%$. The behavior is much different here, however, compared to the three irregularly shaped particles. Interestingly, the steel and olivine samples show opposite behavior: the steel particles are below the $f_{c}=1$ curve for low $Kn$ and increase to values greater than one for increasing $Kn$, whereas both olivine samples start above $f_{c}=1$ and progressively reach or drop below $f_{c}=1$ for increasing $Kn$. This behavior is probably a real effect, that is related to the inversion of the normalization of the curves, as was first noticed in Figure \ref{flight1_3}. In other words, as the baseline value of $\delta P$ at $g_{\rm Earth}$ shifts, so does its relation to $\delta P$ at $hg$. The reasons for this inversion are explored further in a separate work, with respect to the role of mass-loading in generating pressure gradients. We do not consider further the data from flight 3 in the proceeding analysis, since it displays a different class of behavior from the other samples.   

\subsection{Expansion effects: zero-g}\label{subsection:expansion}

\begin{figure}
\includegraphics[width=0.5\textwidth]{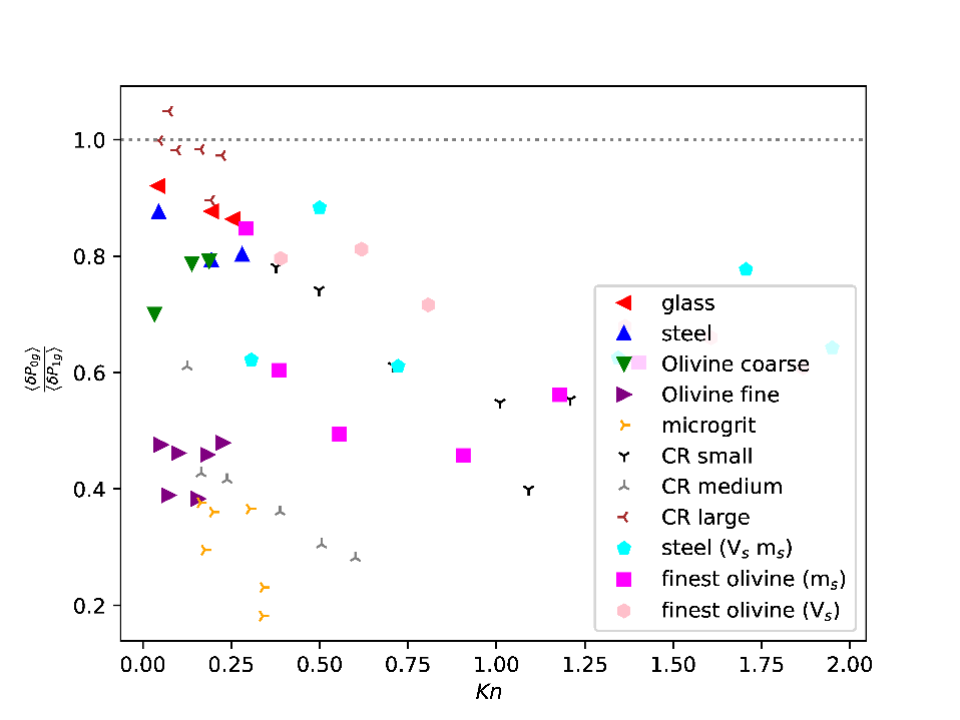}
\caption{\label{fig:expansion_Knudsen} Expansion factor as a function of $Kn$, for all samples and the complete flight measurement sequences. Values correspond to the $0g$ phase.} Symbols and coloring are the same as for Figure \ref{fig:compression_Knudsen}.
\end{figure}

In a similar fashion as we have studied the compression factor, we define the expansion factor as the ratio of the mean pressure differential at $0g$ to that at $g_{\rm Earth}$, $f_{e}=\frac{\left<\delta P_{0g}\right>}{\left<\delta P_{\rm Earth-g}\right>}$. In Figure \ref{fig:expansion_Knudsen} we show this quantity for all samples of all four flights, to summarize and compare the effect of removing gravitational load upon the sample. The value of $f_{e}=1$ is over-plotted and represents the situation in which there is no variation in measured pressure gradient at $0g$ as compared to $g_{\rm Earth}$. 

The only sample material which shows $f_{e}=1$ is the coarse fraction of asteroid regolith simulant, and this is true for all values of $Kn$ (different mass flow rate). In stark contrast, the asteroid regolith samples in smaller size ranges both exhibit extreme variations, of up to more than 80\%. Similarly, the Microgrit exhibits extreme excursions. The cause of this variation cannot be Knudsen number \textit{per se}, since several samples share the same $Kn$ and behave differently. Nevertheless, there is a very clear trend specifically for the three irregularly shaped particle type samples.  To comprehend this, we will look directly at the pressure dependent permeability coefficients in the following section. 

\subsection{Permeability coefficients at variable gas pressure and gravitational load}\label{subsection:coeffs}

\begin{figure}
\includegraphics[width=0.45\textwidth]{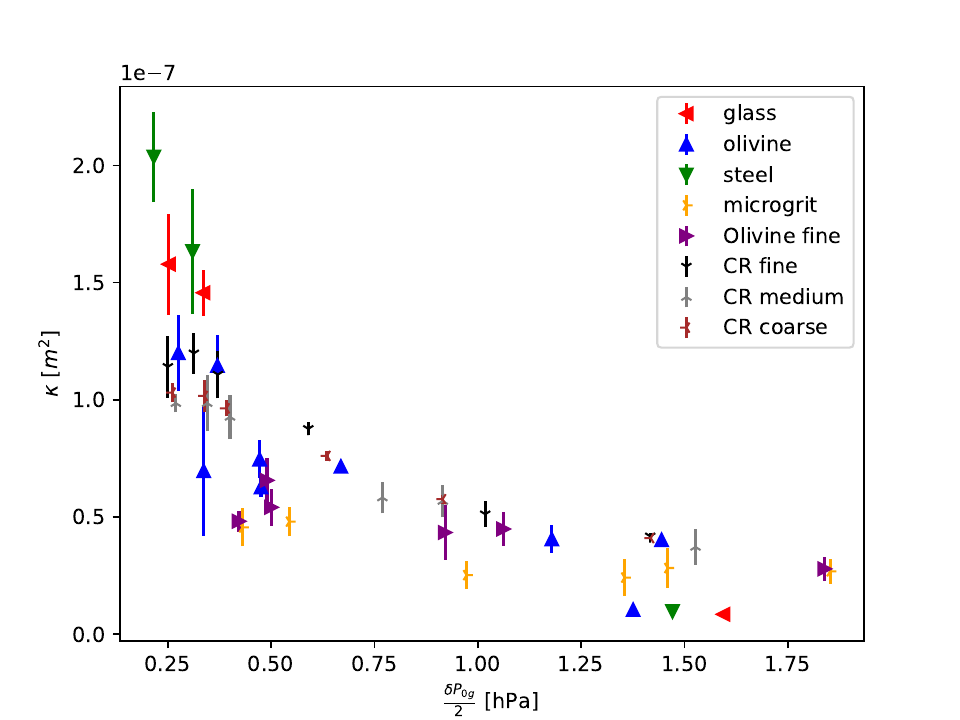}\\
\includegraphics[width=0.45\textwidth]{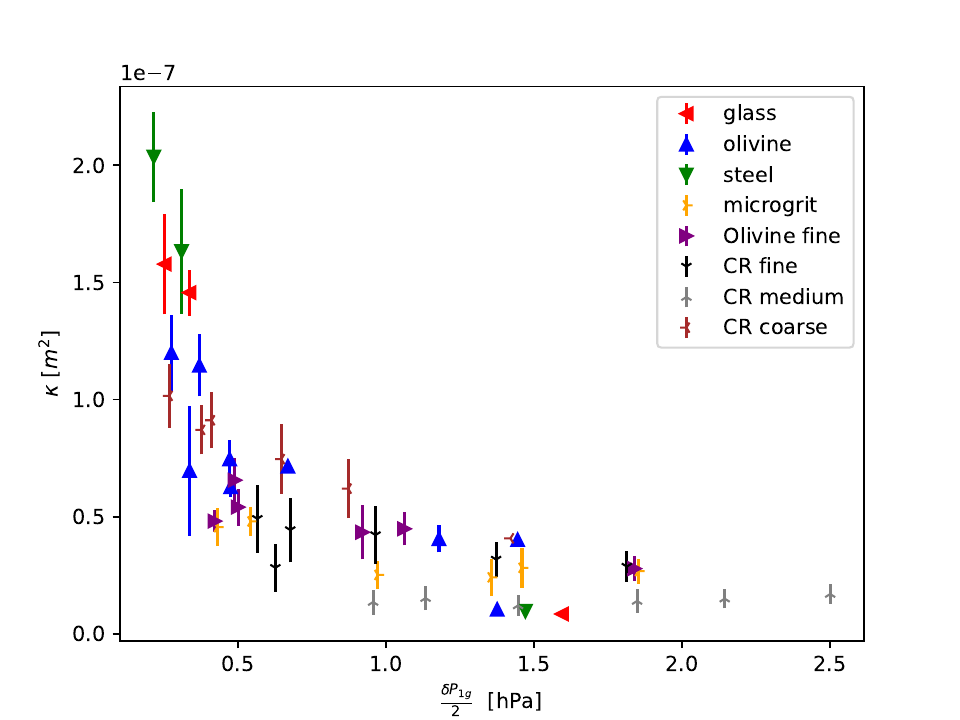}\\
\includegraphics[width=0.45\textwidth]{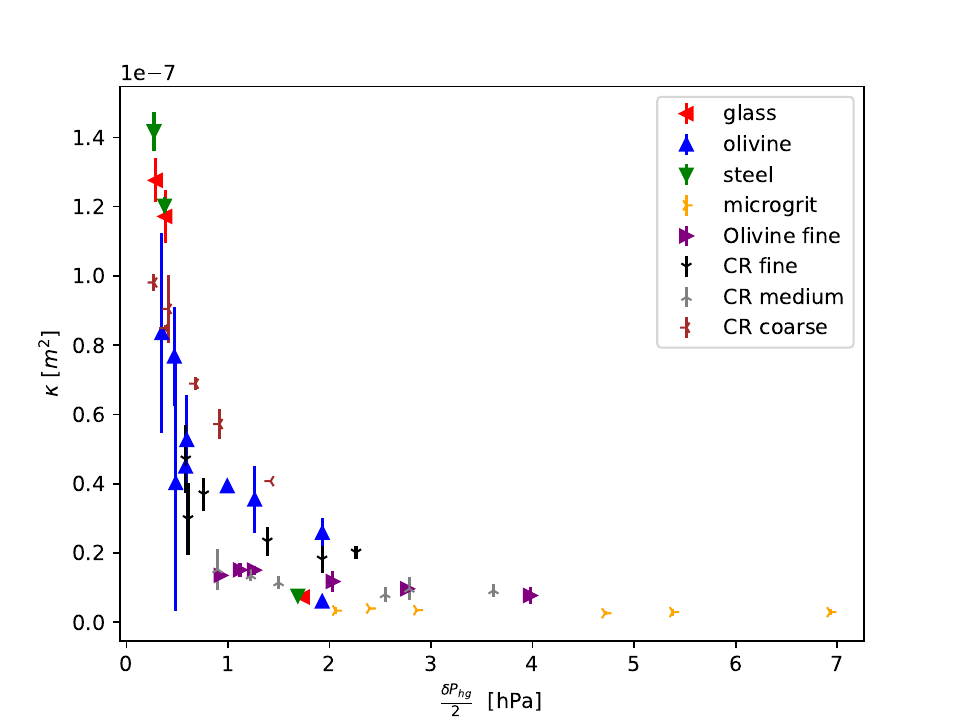}\\
\caption{\label{fig:Klinkenberg_flight_all}Scatter plot of the permeability coefficient $\kappa$, as a function of mean pressure $\frac{\delta P}{2}$, for each chamber and each measurement series. Top panel: $0g$. Middle panel: $g_{\rm Earth}$. Bottom panel: $hg$. Colors and symbols are the same as Figure \ref{fig:compression_Knudsen} and \ref{fig:expansion_Knudsen}.}
\end{figure}

One objective of this work is to extract permeability coefficients for the flow of rarefied gas through granular media. Another is to assess the role of gravitational compression (or lack thereof) on this quantity. Recall the meaning of the permeability coefficient $\kappa$ shown in Equations \ref{equ:pressure_drop} and \ref{equ:kappa}: a reduced kappa has the equivalent measureable effect on the mass flux, for a given pressure gradient, as an enhanced viscosity, and it can be considered a type of resistance, corresponding to the mean open cross-sectional area of the pore space through which the fluid flows. We anticipate that for low $Kn$, viscosity itself is important. But for increasing $Kn$, diffusion plays an increasingly important role \citep[][]{Schweighart:2021}. The correction parameter represented in Equation \ref{klinkenberg} assumes that the variation with gas pressure can be represented by a single parameter $b$, which has a limiting value at infinity, or in other words when the value of $Kn$ is high enough that $\kappa$ can simply be represented with a continuum assumption for the gas properties. 

In Figure \ref{fig:Klinkenberg_flight_all}, we show the extracted values of $\kappa$ as a function of mean pressure in each chamber and for each mass flow rate setting. The labels are the same as in Figure \ref{fig:compression_Knudsen} and \ref{fig:expansion_Knudsen}. 
 The displayed error bars represent values corresponding to the variance in measured $\delta P$, at fixed mass flow rate, and averaged over the particular phase of gravitational loading. As was shown in Figures \ref{flight1_1} - \ref{flight1_4}, there are temporal variations in the pressure differential, corresponding to fluctuations in the gravity level, which contribute to the variance on the mean. We show this separately for the different gravitational loading phase: $g_{\rm Earth}$, $hg$ and $0g$.

For all three gravitational loadings, the data follows the curve that would be expected as molecular diffusivity becomes increasingly relevant at lower mass flow rates (decreasing pressure). No single sample fills in the whole curve, but together, the turnover between the pressure-independent and pressure dependent regimes can be seen nicely. Notably, a qualitatively similar behavior is predicted by equation \ref{klinkenberg}. In appendix \ref{app:permeability_coeffs} we provide fits to this function for a select set of samples.

 Considering the part of the curve where $\kappa$ asymptotes towards the continuum $\kappa_{\infty}$ value, higher average pressures are reached during the compression phase. See especially the x-value range in the top, vs. middle panels of Figure \ref{fig:Klinkenberg_flight_all}. This is consistent with the samples being structurally altered in such a way as to minimize the open pore space, hence increasing the pressure gradient and limiting the flow of gas, and increasing the average over-all pressure. The largest shift in mean pressure occurs for the medium size fraction, followed by the smallest size fraction. 
 The different gravitational loadings produce a transition from a state in which the $\kappa$ values lie close to the constant continuum value -- a relatively flat curve - to a state in which the data starts to diverge significantly from the constant value - a steeper curve. This transition shows us why there are such strong deviations from the extracted pressure differential at $g_{\rm Earth}$, apparent in Figure \ref{fig:compression_Knudsen}. We can also see why there is an apparent pressure dependent effect in Figure \ref{fig:compression_Knudsen}; the change in shape of the curve -- from flat to steep -- while shifting to lower pressures causes such a strong dependence when the ratio is taken and they are compared side by side. This is true for both the compression factor and the comparison with the expanded data, which as shown by Figure \ref{fig:expansion_Knudsen} is even more dramatic.

For the samples in flight 1, the parameters of all three samples are essentially the same, given that the granular bed height was always the same, the particles have the same mean diameter and are either exactly or approximately spherical, and none of the samples showed a strong response to being compressed by accelerations in excess of $g_{\rm Earth}$ (see Figure \ref{flight1_1}). For these materials that are impervious to compression at $\sim 2g$, we can consider them a nearly perfect representation of a randomly packed bed, since one cannot assume that the mass density of the material impacts the porosity of the sample by being heavy and bearing down upon the empty pore space. 

On the other hand, a distinctive feature stands out for the samples of flight 1: under compression at $hg$, the permeability coefficient of the olivine samples shifts significantly more than either glass or coarse steel to lower values. We can attribute this as a particle shape effect. However, we must be also aware that there is a feedback loop between permeability, pressure, and porosity: When the permeability coefficient is low, force builds up inside of the sample, and this works to push it apart to higher porosities. When the confining gravitational pressure changes, then it re-establishes the low permeability and low porosity. This cycle is obviously at play and can explain part of the compression dynamics since some of our samples, particularly the microgrit (see Supplementary Online Material movie 1) are visibly at the limit where they can start to fluidize at $g_{\rm Earth}$. 

Viewed as an ensemble in Figure \ref{fig:Klinkenberg_flight_all}, the data from all of the different samples exhibits a similar behavior. However, this does not mean that all of the data can be represented by a single function, which would imply that the porosity is the same at a given gravitational load, and that the only change is the mass flow rate (and from sample to sample, the particle size). However this highly simplifies the case. First, there is a wide range of scatter about an apparent common relationship, which can be attributed to specific features of the samples themselves. Therefore, in appendix \ref{app:permeability_coeffs}, we present fitting curves overlaid on the measured coefficients, and report fit parameters of the Klinkenberg correction, separately for different samples. Second, we have already seen that the degree of expansion or contraction, and resultant pressure gradient of a given sample can vary wildly. Third, there are additional non-ideal effects from the aircraft - such as jitter and a slightly drifting direction of the gravitational force vector, which cause both strong fluctuations in the location of the sample in both the horizontal and vertical directions. Finally, the concept that all material simply packs into one end of the chamber to the other end, is also a gross over-simplification. 

Despite these complications, it is clear that all samples expand in significant degree during the $0g$ phase and that the $\kappa$ values follow a logical trend with increasing gas pressure, which still yields the expected correction function. In the next section, we will evaluate with more precision the relationship between pore space and measured over-pressure. 

\subsection{Over-pressure and pore space} \label{subsection:overpressure}
\begin{figure}
\includegraphics[width=0.45\textwidth]{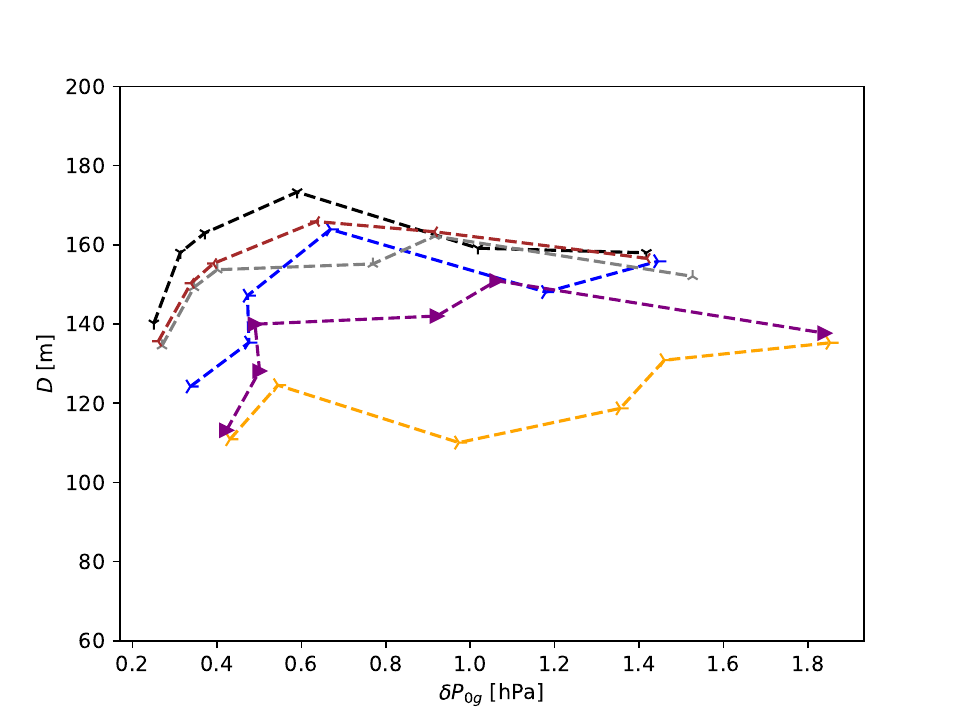}\\
\includegraphics[width=0.45\textwidth]{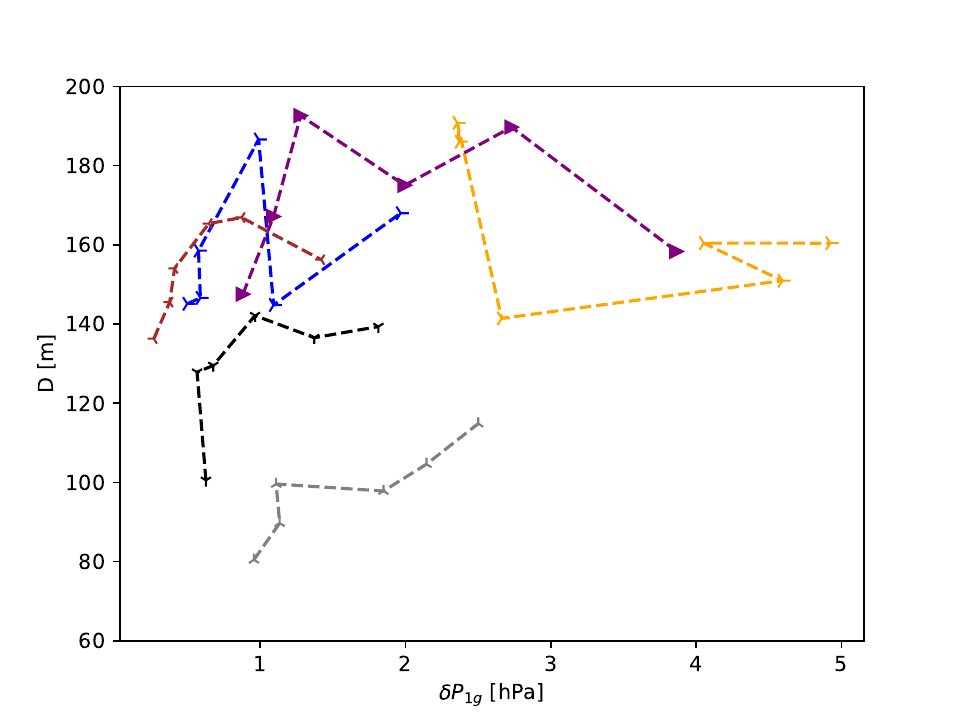}\\
\includegraphics[width=0.45\textwidth]{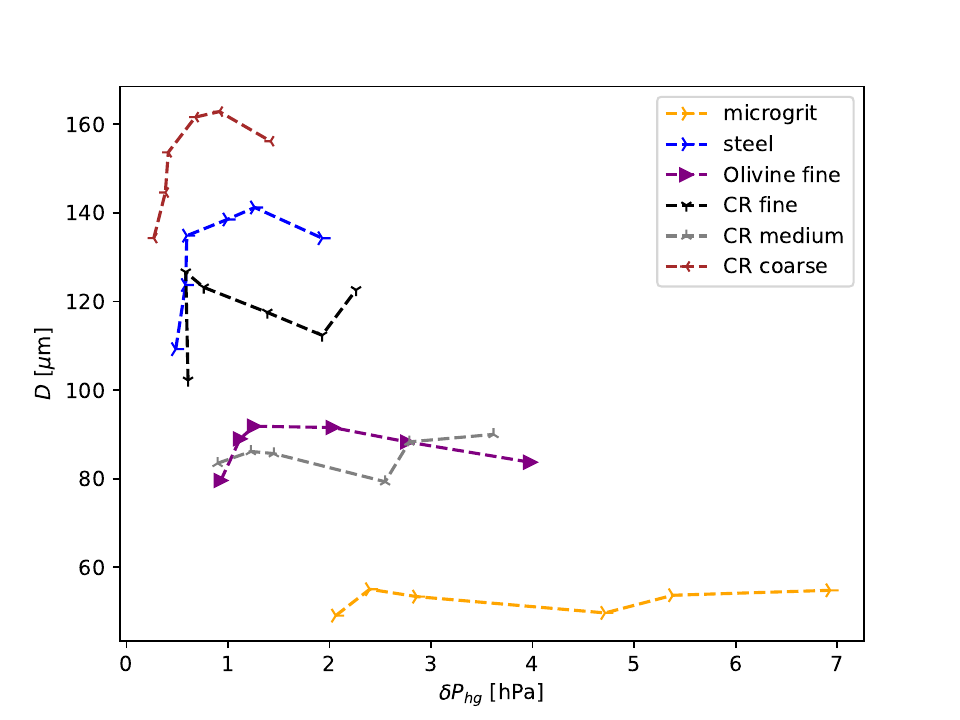}\\
\caption{\label{fig:Interparticle_separation} Mean inter-particle separation vs. pressure gradient, for six samples. Colors and symbols for all three cases are shown in the legend of the lowest panel. Top panel: $0g$. Middle panel: $g_{\rm Earth}$. Bottom panel: $hg$. Inter-particle separations decrease with increasing gravitational load. Separations tend to initially increase, followed by a decrease with increasing gas pressure, demonstrating the role of pressure support.} 
\end{figure}

We would like to determine the relationship between the balance of pressures acting upon the sample and the inter-particle distance. We have already studied the differential gas pressures and derived quantities in the work above; see especially Tables \ref{exp_table1}-\ref{exp_table4} for exact values. We have previously considered the pressure gradient as a drop caused by the flow through granular media. We can now derive the net pressure on the sample by taking the difference with the effective gravitational pressure $P_{grav}=W/a_l$, with $W$ the weight of the sample measured on Earth -- therefore the product of the mass and Earth's gravitational acceleration -- and $a_l$ the surface area of the lower end of the container. We have reported the porosity, 1-$\phi$, and $P_{grav}$ in Table \ref{sample_table}. This was derived simply using the known mass densities and measuring the $W$ and volume occupied by the samples in the laboratory on the ground. The porosity measurements on Earth were also conducted under ambient conditions. One realizes that a pressure gradient bracketing the granular layer will impact the porosity as well. While $\kappa$ is dependent upon porosity, the dependence is non-linear, and so it is more straight-forward to refer to the definition of $\kappa$ as an area corresponding to the cross-section of the average pore in the medium. Since the particles roughly follow the Klinkenberg relation, this assumes that the porosity is unchanging. However, we can still check if small deviations in inter-particle separation occur, once the pressure dependence is accounted for.
We convert the measured permeability coefficients to mean inter-particle separation, using the measured pressure values and dividing off the constants used to determine $\kappa$. In Figure \ref{fig:Interparticle_separation}, we derive the mean inter-particle separation as a diameter D of the area $\kappa=\pi\frac{D^{2}}{4}$. Indeed, each sample follows its own path in this space. 

 When we compare the magnitude of the gravitational pressureat $g_{\rm Earth}$, we see that it is of comparable order of magnitude, but less than, the values of $\delta P$ at $0g$. While there is of course no gravitational pressure at $0g$, we can see from this comparison why the gravitational pressure and gas pressure should be approximately balanced at $g_{\rm Earth}$, with gravity only slightly dominating; hence, the samples are at a threshold where they are poorly confined at $g_{\rm Earth}$ and appear more scattered in $D$.\footnote{We note especially the irregular behavior of microgrit at $g_{\rm Earth}$. We attribute this to the fact that this sample is at the threshold where it begins to fluidize when the gas stream passes through it. This is very well visualized in the supplementary video movie1.avi, where slight bubbling of the sample surface can be seen during steady $g_{\rm Earth}$ flight, corresponding to the beginning and end of the movie.} Finally, at $hg$, the gravitational pressure doubles and gravity is the dominant confining pressure, reducing the interparticle distance noticeably. In both the $hg$ and $0g$ cases the particle separation reaches a plateau, with increasing gas pressure. These small variations in the inter-particle separation, on the order of tens of micrometers, are interesting features which could be investigated further. 

\subsection{Tensile and shear strength of cohesive samples} \label{subsection:cohesion}

The following system attribute would be expected for samples where the grains stick strongly enough to introduce collective dynamical behaviour: all or part of the sample retains its macroscopic shape while free-floating in the presence of the gas flow. Such behavior would indicate that the cohesive strength that binds the material together is in excess of the shear induced by the wind. The supplementary online material movie 2 shows the following scenario:

\begin{itemize}
    \item Material packs into the lower chamber in a conical shape.
    \item The material retains the shape of its container while it levitates and floats to the other end.
    \item Material crushes when it meets the wall of the container and compacts into opposite conical end-segment.
    \item Material sediments and can once again be seen to retain the shape of the chamber.
    \item The aggregate crushes when it reaches the lower end of the chamber.
\end{itemize}

We directly see cohesion effects for only one parabola, that of the lowest pressure setting. Namely, with the asteroid regolith simulant of smallest size range, a bound aggregate can be formed naturally in the system. We point out that such large coherent structures do not form under Earth's gravity at the gas flow rates studied here.  Apparently the operational range of pressures is barely at the limit where cohesion starts to dominate over shear stresses, and only for this sample.  

The next time this chamber is active, at slightly increased gas pressure, we see no such behavior. Instead, we see signs of fluidization of the sample material in response to the gas. Albeit, the lift-off of material, even during $0g$, is feeble. In subsequent parabolas of the set, with increasing gas flow, the gas more and more readily fluidizes and transports the sample across the chamber, to be confined in the other end. 

We can directly determine the tensile strength of this material by calculating the force profile, according to Equation 25 of \cite{Capelo:2022}, and using the experimentally derived permeability coefficient $\kappa$. Since there is a clean break at defined height approximately 2.5 cm below the upper surface of the sample, we can confirm that a simple one-dimensional calculation of the force on a given cross-sectional area is valid. During the $hg$ phase, the pressure above the sample is 0.3 hPa, whereas the pressure below is 1.2 hPa. When we calculate the pressure profile at the location of the break, we retrieve a pressure of 0.6 hPa, which we take to be the tensile strength of the aggregate.  

We adopt the length scale of the longest axis of the aggregate, which is 6.5 cm, resulting in $Re\sim3$. The system was designed to produce a gas flow speed $u$ of 1 ms$^{-1}$. The density of the gas is converted from the pressure measurement using the ideal gas law, and gas constants listed in Table \ref{sample_table}. The application of Eqn. \ref{equ:shear} results in a wind-generated shear stress of $\tau_{w}=3\times 10^{-6}$ hPa. This value is several orders of magnitude less than the derived tensile stress. Since the aggregate appears to remain coherent in the face of a wind exhibiting such pressure (it is destroyed by collision with the container), we can take this value as the lower limit on the cohesive shear strength of the material. The actual shear strength is probably higher, as it is expected to be less than, but similar order of magnitude as the tensile strength \citep{SeizingerEtal:2013c,Scheeres_2018}.

\section{Discussion and Conclusions}\label{section:conclusions}
\subsection{Summary}\label{subsection:summary}
We have studied the development of pressure gradients that form when gas flows through granular media analogous to the dust particles that cover planetesimals in planet-forming discs and which remain preserved in the present day solar system on the surfaces and presumably the interior of comets and asteroids. We consider the application of low pressure to a granular sample to be a generic process which can either be applied literally -- as in the flow of gas through the surface of a comet's regolith layer -- or can be abstracted to represent an arbitrary force applied to a granular medium at scales where significant self-gravity does not apply. 

We apply moderate and high Knudsen numbers $Kn$ on the particle length scale, which are relevant for gas flows through granular aggregates facing a vacuum. Moreover, experimental measurements where the Klinkenberg correction apply are in general scarce in the literature and are usually conducted in settings where the pore space is on the nano-scale, where grain surface effects become dominant \citep{darcy_early, Hutchins:1995,Sharipov,Lasseux:2017}.  
By reducing the pressure of the gas before entry into the chamber, we achieve high $Kn$ even for relatively large pore spaces, namely 10s-100s of micrometers at the most compact configurations and sometimes much greater particle separation during $0g$.

In section \ref{subsubsection:samples} we described the analogue materials studied. Our first flight involved relatively large spherical particles or nearly spherical particles as test material, whereas the last flight used mineral mixtures which reproduce the composition of asteroid matrix with high fidelity. In the two intervening flights, we experimented with particles of differing sizes, size distributions, and shapes. 

In section \ref{subsubsection:pressure_gradient_overview} we presented the time-dependent evolution of pressure gradients for four flights and for samples of increasing complexity. Some notable features from the overview of the flight sequence include: 
\begin{itemize}
\item Pressure differences correlate with the gravitational acceleration. 
\item The pressure measurement signal is somewhat noisy during the reduced gravity phase, however the relevant drop in signal is much greater than the fluctuations.  
\item The acceleration profiles shown in black in Figures \ref{flight1_1} - \ref{flight1_4} are symmetric, reaching similar maximum magnitude, during both the pull-up and pull-out ($hg$ phases preceding and following the $0g$ phase). 
\item The pressure differential values are not necessarily symmetric during the pull-up vs. pull-out phases. 
\item The magnitude of the pressure differential increases with increasing mass flow rate $Q$.
\item For small, irregular particles, the pressure gradient is very responsive to changes in both acceleration and mass flow rate, particularly following an expansion to $0g$ and then transitioning into the pull-out phase of the parabola.
\end{itemize}

Since we found unique behaviors corresponding to the particular phase of gravitational acceleration, the rest of the analysis was conducted upon the average pressure gradient within a fixed range of external force, and for varying mass flow rate, for each sample. 

In Section \ref{section:analysis}, we compared the relative effect on the pressure gradients under either $hg$ or $0g$, as compared to $g_{\rm Earth}$. We first studied simple ratios of the differential pressure at either $hg$ or $0g$, which we denoted the compression factor, or expansion factor, respectively. We then derived from our measurements the permeability coefficients of the samples, and reported best-fit parameters for the typical correction (Klinkenberg) to be applied at decreasing pressures.  We found essentially three classes of behavior: i) particle types that are essentially inert to compression effects at $hg$ and which expand to mildly larger porosities at micro-gravity, and which can be represented by a classical correction parameter at all phases; ii) particle types that respond moderately to both compression and expansion, show no strong dependency of compression or expansion factor to Knudsen number, and which can be represented by a classical correction parameter at all phases; iii) particle types which exhibit extreme response to both compression and expansion, such that the curve for the pressure-dependent correction shifts significantly to show no pressure dependency. 

\subsection{Synthesis}\label{subsection:synthesis} 
While either the compression factor or the expansion factor can summarize the various responses to compaction and expansion among different samples, they do not fully elucidate the differences between the two $hg$ phases of the parabolas, especially for samples exhibiting significant variability.

However, the behavior of $\kappa$ sheds more light on the physical interpretation of the results. We can to some extent comprehend the large variations in $\delta P$ with gravitational load as a transition between the pressure-dependent and pressure-independent regimes of the permeability. Although, this cannot be the whole story, since there is also a remarkable difference between the two different phases of $hg$. 

Significantly, a similar feature of both the pull-out and pull-up phases of the parabola, especially for the compressible samples, is not only a decrease in porosity, but also an increase in relative particle-gas velocity. As the sample compresses in the pull-up phase, this shift is minor, as particles move from a nearly jammed state at $g_{\rm Earth}$, and then pack even more tightly at $hg$. During the pull-out phase however, the particles sediment rapidly from the top of the container to the bottom of the container, as they transition from a diffuse to a compressed state. 

Inspecting Equation \ref{equ:pressure_drop} once again shows that $\frac{\delta P}{\delta z}$ is directly proportional to the flow speed, $u$. We determine and set the fixed linear gas velocity, via the mass flow rate: $Q\propto Pau$. When we increase $Q$ using the mass flow controller, the product $au$ remains fixed, while P increases to adjust to Q.  This quantity $u$ refers to the constant velocity of the gas that is passing through the chamber at low pressure. When the granular bed is static, this also refers to the relative particle-gas velocity. However, if the granular medium is free to move, the relative velocity can be either increased or decreased, depending upon whether the particles are moving with or against the direction of the gas. In this instance, the relative velocity is greatest when the particles are sedimenting against the gas flow during the pull-out phase. Considering the quantities to which our measurements are sensitive, this relative velocity effect is a viable explanation for the strong contribution to the apparently reduced permeability especially during the pull-out phase of the parabola. 

The supplementary online video 1 supports our interpretation. From Parabola 22 microgrit, we see the combined effects quite well: we see for example the nearly fluidized surface at $g_{\rm Earth}$, we see the surface shrink at $hg$, then at the pullout phase we see simultaneous sedimentation and shrinking of the sample. Finally, we see the sample expand and start to levitate at $g_{\rm Earth}$ again. It is somewhat coincidental that the forces acting on the samples were nearly perfectly balanced at $g_{\rm Earth}$ and the highest mass flow rate, so that the sample was just barely starting to fluidize, but remained confined to a relatively small region. 

The possibility that there is a dynamical contribution to the pressure gradient is an intriguing development, the implications for which are discussed below. 

One could speculate that the time dependent evolution is related to the cohesiveness of the sample, and that over several cycles of processing, the material becomes more compacted. But we can rule out this explanation because the samples become more permeable to gas flow at $0g$, as they always expand. See for example Figure \ref{fig:expansion_Knudsen}. We see especially a distinct difference between $g_{\rm Earth}$ and $hg$ that reflects the force balance on the sample, and so simplistic explanations, to the effect that the sample is becoming `more clumpy', do not apply. 

Having said this, while cohesion cannot explain the structural and dynamical effects that we observe, it does not mean that it cannot apply or be studied here. We have shown that the compression state of the granular samples results from a delicate balance between gravitational pressure and pressure build-up within the sample, and the confining pressure above and below the granular bed. For only one sample (CR asteroid regolith simulant), the system naturally produces a bound aggregate that breaks apart due to perpendicular force of the pressure gradient, but does not disrupt in the presence of a flow that is directed in the parallel direction. Significantly, this sample is a material which one expects to have enhanced cohesiveness, due to the presence of many contact points between the particles as well as organic compounds \citep{POCH2016,Bischoff2020}. 

We have characterized the particle types used as sample material and used these average particle properties in our analysis. However, we cannot directly observe and therefore neglect some of the possible features of the granular bed that could result from process such as grain size sorting due to shaking and sedimentation, or aerodynamic focusing of particle orientation.

\subsection{Applications and significance} \label{subsection:significance}

 The ice sublimation pressure below the granular surface on comets is naturally low, while there are mechanisms which can serve to increase the pressure. For example, it is predicted that there is a Knudsen layer formed above a solid surface across which vapour is flowing, which extends over a length-scale of several mean-free paths, such that the pressure does not not automatically drop to zero above the nucleus surface. \citep{davidsson2008}. As our formulation makes apparent, i.e. Equations  \ref{equ:pressure_drop} - \ref{equ:kappa}, the net pressure differential is regulated by the vacuum conditions above the surface. Gas percolation through the granular matrix is another process which can impact the subsurface pressure. According to the model of \cite{Skorov2011}, used also in \cite{Blum_2014}, the pressure under the surface changes with the thickness of the granular dust layer, with static gas pressures resulting from water ice sublimation should be in the range 10$^{-3}$ -- $1$ Pa, depending upon heliocentric distance. \cite{Blum_2014} point out that since dust tensile strengths are of the order of 1 Pa, whereas the gas sublimation pressure is near 10$^{-3}$ Pa, it is an open question how comets produce ejecta. They concluded that there is a characteristic particle size, below which the tensile strengths are too large to be overcome by the viable pressures. Our results provide insight into this set of issues. For example, we have demonstrated that the smaller particles serve to increase the subsurface pressure more than larger particles do, because of their lower permeability coefficients and propensity to pack more tightly. Small particles within the matrix on comets may also increase the pressure and help overcome the tensile strength of the regolith. It should also be pointed out that the sublimation of other volatile species, such as CO2, occur at higher pressures than water ice. Moreover, it was observed that the ejection of dust from the day side of  Comet 67P and subsequent fallback can lead to dust layers with meter-scale depths \citep{keller_2015, keller_2017, hu_2017, davidsson_2021}. As discussed by \cite{Skorov2011} and also supported by the methodology presented in this work (see Eqs. \ref{equ:pressure_drop},\ref{equ:kappa}, and \ref{equ:profile}), increasing the thickness of the dust layer results in a greater pressure differential.

Our experimental findings have implications for understanding the compaction mechanisms on comets, particularly in relation to the SESAME-PP measurements from the Rosetta mission to Comet 67P. Figure \ref{fig:expansion_Knudsen} shows the pressure difference between the top and bottom of grain aggregates under micro-gravity, compared to that of Earth gravity. If the grains were entirely free-floating, the effective filling factor would be relatively low and the pressure difference relatively small, as is the case for some of the samples (such as the microgrit). However, for other samples (e.g., large CR, glass, coarse olivine), the pressure difference is $\gtrapprox 80\%$ of that obtained under Earth gravity. We also discussed in \ref{subsection:overpressure} that the gravitational pressure is of similar magnitude to the pressure gradient. This indicates that the gas pressure gradient is sufficient to cause almost the same level of grain aggregate compaction as achieved with Earth gravity itself.  The corresponding values in tables \ref{exp_table1} - \ref{exp_table4} show that in cases where the expansion factor (\(f_{\rm e}\)) is greater than 0.8, the pressure differences observed range from 50 to 300 Pa. These levels are consistent with the pressures that can be reached during CO2 sublimation. This set of observations align with the hypothesis proposed by \cite{Davidsson_2022}, who suggested that CO2 sublimation pressure gradients below the sublimation front could induce compaction of the dust/water-ice solid matrix underneath.

It is an important issue, at which scale one considers the permeability. For example, \cite{Blum_2014} work with mm particles that are themselves aggregates composed of smaller $\mu m$ units. It is an open question: what is the permeability of a granular bed composed of granular aggregates. In particular, one can see that the sets of gas pressures and particle sizes considered in the model of \cite{Blum_2014} correspond to either transitional or free molecular flow, depending on the location of the comet and whether one considers the scale of the aggregate or the compositional dust. While neither previous authors nor this work have addressed this level of complexity across scales, we have shown that small particle sizes generate high flow impedance, which overcomes the effect of molecular diffusion upon the permeability. An implication is that the hierarchical structure of dust aggregates could in fact contribute to an increase in the sub-surface pressures on comets as well. However, we have studied the limiting behavior of the permeability and find that as molecular diffusion starts to dominate, the values of $\kappa$ increase.

\cite{Skorov2011} lays out the argument for why the calculation of the gas mass flow rate in a low-pressure, gradient driven flow is challenging, mainly due to the difficulty in appropriately describing the properties of the porous medium. \cite{Skorov2022} addresses the influence of the regolith microstructure upon the gas production rate and delineates important parameters such as the average pore size, the effective porosity, and degree of particle filling. Here we provide measurement-derived values on the average pore space. We have provided direct measurements of flow through media of various types, which produce different impact on the generated pressure gradient, for fixed and controlled mass flow rate. These measurements should therefore be useful to discriminate between theoretical models used to understand gas flow through the surfaces of comets.

\cite{CHRISTOU2018} and \cite{CHRISTOU2020} performed simulations of outgassing through porous surface layers with structural properties derived from terrestrial rocks, which can result in flows in a non-radial direction from the comet nucleus surface. However, our results challenge the assumption that the porous structure of the medium remains rigid during outgassing on a low-gravity object, since the pressure support from the gas serves to increase the average pore size of the material. The effect of pressure support reducing the porosity might be overcome, if the cohesiveness of the granular material is significant. We have shown that instead of fluidizing, a cohesive dust-layer forms an aggregate and simply breaks off from the granular bed. The experiments presented in \cite{POCH2016} also show that the formation of cohesive surface layers in mixtures of sublimating ice and dust correspond to explosive ejection events, most likely due to the build-up of sub-surface pressures.

The compactness state of dust particle aggregates, and the bodies comprised of them, are directly dependent upon their assembly history. Dust aggregates, themselves composed of primordial grains in the nm-$\mu$m range, assemble through collisions, and the compactness depends upon the distribution of impact velocities and size ratios. Additionally, when aggregates face a headwind in a protoplanetary disc, they should become compacted due to the dynamical pressure of the gas \citep{Kataoka_2013}. That is, if they are not destroyed by shear stress \citep{Demirci:2020,Schaffer:2020}. 

There are many instances where the structural stability and cohesiveness of dust aggregates play a critical role in understanding the formation pathways of planetesimals and therefore of planets. While the particles that we study here are much larger than the sub-micron grains originally comprising the Solar nebula, these grains are expected to aggregate to form pebbles of up to cm size. As regards the gas-particle interaction, previous works \citep{lambrechts, Capelo:2019, Schaffer:2020, Capelo:2022} have provided useful dimensionless scaling relations for particle aerodynamic behavior based upon the parameters $Kn$, friction time (equivalently Stokes number, which depends upon drag force), and $Re$ . For example, it was shown that in the midplane of a Solar-mass disc, $Kn=1$ occurs around 1 Astronomical Unit for $\sim 6$ cm aggregates. In this work, the particle-scale $Kn$ ranges from $\sim 0.1-2.0$, and so the individual particles in our samples (particularly the particles of 10s of micrometers) can be considered as analogs to the dust found in the habitable region of proto-planetary discs. 

Here, we have discovered the limit where a compact aggregate remains cohesive in the presence of a wind or applied force, and where it is destroyed by pressures resulting from a gas flow. The tensile strength that we report falls within the range of values given by \cite{Bischoff2020}, for a suite of Brazilian disc tests upon organic materials. However, we cannot yet conclude that the carbonaceous material is the explanation for the cohesion. \cite{Murdoch2013} performed shearing experiments on granular media under microgravity conditions and find that the force contact-network between monomer grains is reduced in microgravity. Our results could also be partially explained by what \cite{Murdoch2013} proposes, which is that irregularly shaped particles of diverse sizes should increase the strength of the granular matrix. Regardless, both the derived shear and tensile strength of the aggregates is low, in the 0.1-1 hPa range, as is expected.

It is known that small particles are in the cohesive regime more readily than larger particles, and our results are consistent with this prior understanding. For a similar set of confining pressures, we find a much different apparent cohesive strength of our aggregates, which depends upon how we choose to represent the dust. It is interesting that the sample material which exhibits particular cohesiveness is also the one that represents asteroid composition with the highest fidelity. This is however a complex material and therefore it is challenging to discern whether it is the peculiar shape, the various surface energies of the diverse mineral sub-components, or the presence of organic molecules which are responsible. Given the dramatic effect, as well as the important but poorly understood role of cohesive forces between grains, further work to disentangle the effect of these properties is recommended.  

Shape matters. We have shown that the samples most responsive to external forces are those which cannot be classified as simple monomers. In an additional separate work (Bodenan et al. in preparation), we evaluate in detail the shapes of real dust aggregates, as preserved in meteorites and comets. Indeed, the more realistic case is that the smallest constituents of dust aggregates are not themselves spherical. 
\cite{Ivanovski2017} showed that in situ measurements of the dust ejection speeds from comet 67P/Churyumov-Gerasimenko were inconsistent with a spherical assumption for the grain shape when calculating the drag law on the particles. Understanding how complex grain structure impacts the applicable drag laws is important therefore in modeling dust comae. 

When one considers the balance of forces at play on the surface of small bodies, cohesion comes to mind as an important, but often poorly understood, term in the force balance. Here we showed that our setup is suitable to study the cohesive properties of a sample, and that the current operational parameters are close to the transition between cohesion-dominated regime to a regime in which aggregates are easily compressed or broken apart by moderate pressures in the 50-Pa regime. With small modification to the mass flow settings and sample choices, we could explore the range of cohesive strengths in the $\sim$1 Pa range, which is of great interest for comparison to the strengths of minor bodies in our solar system \citep{Hirabayashi_2022,Raducan_2022,Ferrari_2022,Perry2022,Attree:2018,Steckloff:2015}. 

Some features of our data could best be explained by a form of dynamical pressure as additional contributor to the primary cause of pressure gradients, which is the tortuous flow of fluid through the granular bed. This is a significant finding as it is the first instance where direct measurements of flow symmetry breaking by granular media have been reported. Typically, the development of pressure gradients due to objects in a flow are set by the Reynolds number. We have shown that even in low Reynolds number flow, rapidly sedimenting particles can also cause an asymmetry in the flow properties up and downstream of the granular mass. A similar phenomenon was predicted to occur in sedimenting fluid in the so-called drafting instability \citep{lambrechts}.  An important distinction, however is that the drafting instability is driven by mass over densities, rather than by particle filling over density. Indeed \cite{Capelo:2019} found that even for extremely dilute particle concentrations, massive particles could act collectively to accelerate their sedimentation speeds. In the present work, we cannot conclude any drag-regime dependence of such effect, because there is also a bias introduced by the particular force balance in this system: at low pressure, the samples are not pressure supported, and so they are less susceptible to either expansion by forces within, or to compression by gravitational force.  
 
 In general, this work opens up the possibility to equate drag forces on swarms of particles to a pressure differential. This is the original idea of \cite{Brinkman}, which has never before been explicitly tested in the Knudsen regime. While such dynamical pressure gradients were generated with the particles falling in the direction of gravity, we point out that even in the case of particles lifted radially from the surface of a comet, there is naturally a large relative velocity between the particles and the gas due to non-slip boundary conditions and poor coupling. The implication would therefore be that a form of collective drag effect could be at play in comet tails, particularly close to the nucleus. Collective drag effects, which are ultimately responsible for fluid-drag instabilities \citep{You_good2005} have already gained traction as being very important for dust-gas interaction in pressure-supported proto-planetary discs and a number of contexts in galactic astronomy \citep{Hopkins_Squire:2018,Squire_Hopkins:2018a,Squire_Hopkins:2018b}. But the relevance of collective drag effect has in general not ever been considered in the pressure-driven flows derived from gas sublimation in comets. 


Above, we addressed the relatively densely packed regime, analogous to a regolith bed or dust aggregate. However, the samples prepared for flight 3 used much less material and allow us to directly address extremely low filling factors, which is more applicable to the early development stages of dust in orbit and drifting in a protoplanetary disc. The effects that we find in the dense-packing regime presumably all depend upon the compactness of the particle configuration. In the extremely loose-packing regime, we will put to test that mass itself can serve to cause a pressure gradient in granular two-phase flow. These results are presented in paper II of this series.

We have shown values of the mean pressure gradients for a few parabolas conducted at partial gravity: two at Martian gravity levels and one at Lunar gravity. Incidentally, the mean pressure in the chambers was corresponding to the mean atmospheric pressure on Mars. We found very little difference in the behavior of the sample when comparing Martian and Earth gravity, whereas the same mass flow at Lunar gravity caused apparent reduction in the porosity. We have not exploited this data further because there are relatively few measurements and because the samples were idealized spheres, and so not necessarily representative of the soil types on either of these lower gravity objects. We therefore suggest that a dedicated partial gravity campaign could help to address topics such as Martian and Lunar subsurface ice stability and surface feature current activity that are potentially attributable to subsurface flows \citep{KRAEMER2019131,KRUSS2020113438,SIZEMORE2008606,BRYSON2008446}.

\section*{Acknowledgements}

H.L.C. and M.J. acknowledge support from the Swiss National Science Foundation (project number 200021\_207359).
This work has been carried out within the framework of the NCCR PlanetS supported by the Swiss National Science Foundation under grants 51NF40\_182901 and 51NF40\_205606. We acknowledge contributions from the CoPhyLab project funded by the D-A-CH program.

\section*{Data Availability}

The data will be made available on Zenodo and the doi given simultaneously with the publication of this manuscript.



\bibliographystyle{mnras}
\bibliography{mnras_bib} 

\begin{thebibliography}{}
\makeatletter
\relax
\def\mn@urlcharsother{\let\do\@makeother \do\$\do\&\do\#\do\^\do\_\do\%\do\~}
\def\mn@doi{\begingroup\mn@urlcharsother \@ifnextchar [ {\mn@doi@}
  {\mn@doi@[]}}
\def\mn@doi@[#1]#2{\def\@tempa{#1}\ifx\@tempa\@empty \href
  {http://dx.doi.org/#2} {doi:#2}\else \href {http://dx.doi.org/#2} {#1}\fi
  \endgroup}
\def\mn@eprint#1#2{\mn@eprint@#1:#2::\@nil}
\def\mn@eprint@arXiv#1{\href {http://arxiv.org/abs/#1} {{\tt arXiv:#1}}}
\def\mn@eprint@dblp#1{\href {http://dblp.uni-trier.de/rec/bibtex/#1.xml}
  {dblp:#1}}
\def\mn@eprint@#1:#2:#3:#4\@nil{\def\@tempa {#1}\def\@tempb {#2}\def\@tempc
  {#3}\ifx \@tempc \@empty \let \@tempc \@tempb \let \@tempb \@tempa \fi \ifx
  \@tempb \@empty \def\@tempb {arXiv}\fi \@ifundefined
  {mn@eprint@\@tempb}{\@tempb:\@tempc}{\expandafter \expandafter \csname
  mn@eprint@\@tempb\endcsname \expandafter{\@tempc}}}

\bibitem[\protect\citeauthoryear{Arakawa et~al.,}{Arakawa
  et~al.}{2020}]{Arakawa2020}
Arakawa M.,  et~al., 2020, \mn@doi [Science] {10.1126/science.aaz1701}, 368, 67

\bibitem[\protect\citeauthoryear{{Attree} et~al.,}{{Attree}
  et~al.}{2018}]{Attree:2018}
{Attree} N.,  et~al., 2018, \mn@doi [Astronomy \& Astrophysics]
  {10.1051/0004-6361/201732155}, \href
  {https://ui.adsabs.harvard.edu/abs/2018A&A...611A..33A} {611, A33}

\bibitem[\protect\citeauthoryear{Bischoff, Kreuzig, Haack, Gundlach  \&
  Blum}{Bischoff et~al.}{2020}]{Bischoff2020}
Bischoff D.,  Kreuzig C.,  Haack D.,  Gundlach B.,   Blum J.,  2020, \mn@doi
  [Monthly Notices of the Royal Astronomical Society] {10.1093/mnras/staa2126},
  497, 2517

\bibitem[\protect\citeauthoryear{Blum, Gundlach, Mühle  \&
  Trigo-Rodriguez}{Blum et~al.}{2014}]{Blum_2014}
Blum J.,  Gundlach B.,  Mühle S.,   Trigo-Rodriguez J.,  2014, \mn@doi
  [Icarus] {10.1016/j.icarus.2014.03.016}, 235, 156–169

\bibitem[\protect\citeauthoryear{{Brinkman}}{{Brinkman}}{1949}]{Brinkman}
{Brinkman} H.~C.,  1949, \mn@doi [Applied Scientific Research]
  {https://doi.org/10.1007/BF02120313}, 1, 27

\bibitem[\protect\citeauthoryear{Britt et~al.,}{Britt
  et~al.}{2019}]{Britt:2019}
Britt D.~T.,  et~al., 2019, \mn@doi [Meteoritics {\&} Planetary Science]
  {10.1111/maps.13345}, 54, 2067

\bibitem[\protect\citeauthoryear{{Brouet, Y.}, {Levasseur-Regourd, A. C.},
  {Sabouroux, P.}, {Encrenaz, P.}, {Thomas, N.}, {Heggy, E.}  \& {Kofman,
  W.}}{{Brouet, Y.} et~al.}{2015}]{Brouet_2015}
{Brouet, Y.} {Levasseur-Regourd, A. C.} {Sabouroux, P.} {Encrenaz, P.} {Thomas,
  N.} {Heggy, E.}  {Kofman, W.} 2015, \mn@doi [A&A]
  {10.1051/0004-6361/201526099}, 583, A39

\bibitem[\protect\citeauthoryear{Bryson, Chevrier, Sears  \& Ulrich}{Bryson
  et~al.}{2008}]{BRYSON2008446}
Bryson K.~L.,  Chevrier V.,  Sears D.~W.,   Ulrich R.,  2008, \mn@doi [Icarus]
  {https://doi.org/10.1016/j.icarus.2008.02.011}, 196, 446

\bibitem[\protect\citeauthoryear{{Bukhari Syed}, {Blum}, {Wahlberg Jansson}  \&
  {Johansen}}{{Bukhari Syed} et~al.}{2017}]{mohtashim}
{Bukhari Syed} M.,  {Blum} J.,  {Wahlberg Jansson} K.,   {Johansen} A.,  2017,
  \mn@doi [The Astrophysical Journal] {10.3847/1538-4357/834/2/145}, \href
  {http://adsabs.harvard.edu/abs/2017ApJ...834..145B} {834, 145}

\bibitem[\protect\citeauthoryear{{Burn}, {Marboeuf}, {Alibert}  \&
  {Benz}}{{Burn} et~al.}{2019}]{Burn:2019}
{Burn} R.,  {Marboeuf} U.,  {Alibert} Y.,   {Benz} W.,  2019, \mn@doi [\aap]
  {10.1051/0004-6361/201935780}, \href
  {https://ui.adsabs.harvard.edu/abs/2019A&A...629A..64B} {629, A64}

\bibitem[\protect\citeauthoryear{{Capelo}, {Mol{\'a}{\v{c}}ek}, {Lambrechts},
  {Lawson}, {Johansen}, {Blum}, {Bodenschatz}  \& {Xu}}{{Capelo}
  et~al.}{2019}]{Capelo:2019}
{Capelo} H.~L.,  {Mol{\'a}{\v{c}}ek} J.,  {Lambrechts} M.,  {Lawson} J.,
  {Johansen} A.,  {Blum} J.,  {Bodenschatz} E.,   {Xu} H.,  2019, \mn@doi
  [Astronomy \& Astrophysics] {10.1051/0004-6361/201833702}, \href
  {https://ui.adsabs.harvard.edu/\#abs/2019A&A...622A.151C} {622, A151}

\bibitem[\protect\citeauthoryear{{Capelo} et~al.,}{{Capelo}
  et~al.}{2022}]{Capelo:2022}
{Capelo} H.~L.,  et~al., 2022, \mn@doi [Review of Scientific Instruments]
  {10.1063/5.0087030}, \href
  {https://ui.adsabs.harvard.edu/abs/2022RScI...93j4502C} {93, 104502}

\bibitem[\protect\citeauthoryear{Carman}{Carman}{1937}]{CARMAN}
Carman P.,  1937, \mn@doi [Chemical Engineering Research and Design]
  {https://doi.org/10.1016/S0263-8762(97)80003-2}, 75, S32

\bibitem[\protect\citeauthoryear{Christou et~al.,}{Christou
  et~al.}{2018}]{CHRISTOU2018}
Christou C.,  et~al., 2018, \mn@doi [Planetary and Space Science]
  {https://doi.org/10.1016/j.pss.2018.06.009}, 161, 57

\bibitem[\protect\citeauthoryear{Christou, Dadzie, Marschall  \&
  Thomas}{Christou et~al.}{2020}]{CHRISTOU2020}
Christou C.,  Dadzie S.~K.,  Marschall R.,   Thomas N.,  2020, \mn@doi
  [Planetary and Space Science] {https://doi.org/10.1016/j.pss.2019.104752},
  180, 104752

\bibitem[\protect\citeauthoryear{{Darcy}}{{Darcy}}{1856}]{darcy_early}
{Darcy} H.,  1856, {Les fontaines publiques de la ville de Dijon}.
Libraire des corps imperiaux des ponts et chaussees et des mines

\bibitem[\protect\citeauthoryear{Davidsson}{Davidsson}{2008}]{davidsson2008}
Davidsson 2008, \mn@doi [Space Science Reviews] {10.1007/s11214-008-9305-8},
  138, 207

\bibitem[\protect\citeauthoryear{Davidsson et~al.,}{Davidsson
  et~al.}{2021}]{davidsson_2021}
Davidsson B.~J.,  et~al., 2021, \mn@doi [Icarus]
  {https://doi.org/10.1016/j.icarus.2020.114004}, 354, 114004

\bibitem[\protect\citeauthoryear{Davidsson et~al.,}{Davidsson
  et~al.}{2022}]{Davidsson_2022}
Davidsson B. J.~R.,  et~al., 2022, \mn@doi [Monthly Notices of the Royal
  Astronomical Society] {10.1093/mnras/stac2560}, 516, 6009

\bibitem[\protect\citeauthoryear{Demirci, Kruss, Teiser, Bogdan, Jungmann,
  Schneider  \& Wurm}{Demirci et~al.}{2019}]{Demirci:2019}
Demirci T.,  Kruss M.,  Teiser J.,  Bogdan T.,  Jungmann F.,  Schneider N.,
  Wurm G.,  2019, \mn@doi [Monthly Notices of the Royal Astronomical Society]
  {10.1093/mnras/stz107}, 484, 2779

\bibitem[\protect\citeauthoryear{{Demirci}, {Schneider}, {Teiser}  \&
  {Wurm}}{{Demirci} et~al.}{2020}]{Demirci:2020}
{Demirci} T.,  {Schneider} N.,  {Teiser} J.,   {Wurm} G.,  2020, \mn@doi [\aap]
  {10.1051/0004-6361/202039312}, \href
  {https://ui.adsabs.harvard.edu/abs/2020A&A...644A..20D} {644, A20}

\bibitem[\protect\citeauthoryear{Donev, Cisse, Sachs, Variano, Stillinger,
  Connelly, Torquato  \& Chaikin}{Donev et~al.}{2004}]{Donev_2004}
Donev A.,  Cisse I.,  Sachs D.,  Variano E.~A.,  Stillinger F.~H.,  Connelly
  R.,  Torquato S.,   Chaikin P.~M.,  2004, \mn@doi [Science]
  {10.1126/science.1093010}, 303, 990

\bibitem[\protect\citeauthoryear{{Epstein}}{{Epstein}}{1924}]{epstein}
{Epstein} P.~S.,  1924, \mn@doi [Physical Review] {10.1103/PhysRev.23.710},
  \href {http://adsabs.harvard.edu/abs/1924PhRv...23..710E} {23, 710}

\bibitem[\protect\citeauthoryear{Feder, Flekkøy  \& Hansen}{Feder
  et~al.}{2022}]{feder_flekkøy_hansen_2022}
Feder J.,  Flekkøy E.~G.,   Hansen A.,  2022, Physics of Flow in Porous Media.
Cambridge University Press, \mn@doi{10.1017/9781009100717}

\bibitem[\protect\citeauthoryear{Ferrari, Raducan, Soldini  \& Jutzi}{Ferrari
  et~al.}{2022}]{Ferrari_2022}
Ferrari F.,  Raducan S.~D.,  Soldini S.,   Jutzi M.,  2022, \mn@doi [The
  Planetary Science Journal] {10.3847/PSJ/ac7cf0}, 3, 177

\bibitem[\protect\citeauthoryear{{Groussin} et~al.,}{{Groussin}
  et~al.}{2015}]{groussin:2015}
{Groussin} O.,  et~al., 2015, \mn@doi [\aap] {10.1051/0004-6361/201527020},
  \href {https://ui.adsabs.harvard.edu/abs/2015A&A...583A..36G} {583, A36}

\bibitem[\protect\citeauthoryear{Hirabayashi et~al.,}{Hirabayashi
  et~al.}{2022}]{Hirabayashi_2022}
Hirabayashi M.,  et~al., 2022, \mn@doi [The Planetary Science Journal]
  {10.3847/PSJ/ac6eff}, 3, 140

\bibitem[\protect\citeauthoryear{{Hopkins} \& {Squire}}{{Hopkins} \&
  {Squire}}{2018}]{Hopkins_Squire:2018}
{Hopkins} P.~F.,  {Squire} J.,  2018, \mn@doi [Monthly Notices of the Royal
  Astronomical Society] {10.1093/mnras/sty1982}, \href
  {http://adsabs.harvard.edu/abs/2018MNRAS.tmp.1886H} {}

\bibitem[\protect\citeauthoryear{{Hu} et~al.,}{{Hu} et~al.}{2017}]{hu_2017}
{Hu} X.,  et~al., 2017, \mn@doi [\aap] {10.1051/0004-6361/201629910}, \href
  {https://ui.adsabs.harvard.edu/abs/2017A&A...604A.114H} {604, A114}

\bibitem[\protect\citeauthoryear{Hutchins, Harper  \& Felder}{Hutchins
  et~al.}{1995}]{Hutchins:1995}
Hutchins D.~K.,  Harper M.~H.,   Felder R.~L.,  1995, \mn@doi [Aerosol Science
  and Technology] {10.1080/02786829408959741}, 22, 202

\bibitem[\protect\citeauthoryear{Ivanovski et~al.,}{Ivanovski
  et~al.}{2017}]{Ivanovski2017}
Ivanovski S.~L.,  et~al., 2017, \mn@doi [Monthly Notices of the Royal
  Astronomical Society] {10.1093/mnras/stx3008}, 469, S774

\bibitem[\protect\citeauthoryear{Johansen \& Youdin}{Johansen \&
  Youdin}{2007}]{Jo_you2007j}
Johansen A.,  Youdin A.,  2007, The Astrophysical Journal, 662, 627

\bibitem[\protect\citeauthoryear{{Johansen}, {Oishi}, {Mac Low}, {Klahr},
  {Henning}  \& {Youdin}}{{Johansen} et~al.}{2007}]{Johansen2007}
{Johansen} A.,  {Oishi} J.~S.,  {Mac Low} M.-M.,  {Klahr} H.,  {Henning} T.,
  {Youdin} A.,  2007, \mn@doi [Nature] {10.1038/nature06086}, \href
  {http://adsabs.harvard.edu/abs/2007Natur.448.1022J} {448, 1022}

\bibitem[\protect\citeauthoryear{Jutzi}{Jutzi}{2015}]{JUTZI20153}
Jutzi M.,  2015, \mn@doi [Planetary and Space Science]
  {https://doi.org/10.1016/j.pss.2014.09.012}, 107, 3

\bibitem[\protect\citeauthoryear{{Jutzi}, {Holsapple}, {W{\"u}nneman}  \&
  {Michel}}{{Jutzi} et~al.}{2015}]{Jutzi:2015}
{Jutzi} M.,  {Holsapple} K.~A.,  {W{\"u}nneman} K.,   {Michel} P.,  2015, in ,
  Asteroids IV.
University of Arizona Press, pp 679--699,
  \mn@doi{10.2458/azu_uapress_9780816532131-ch035}

\bibitem[\protect\citeauthoryear{{Jutzi}, {Benz}, {Toliou}, {Morbidelli}  \&
  {Brasser}}{{Jutzi} et~al.}{2017}]{Jutzi:2017}
{Jutzi} M.,  {Benz} W.,  {Toliou} A.,  {Morbidelli} A.,   {Brasser} R.,  2017,
  \mn@doi [Astronomy \& Astrophysics] {10.1051/0004-6361/201628963}, 597, A61

\bibitem[\protect\citeauthoryear{Jutzi, Raducan, Zhang, Michel  \&
  Arakawa}{Jutzi et~al.}{2022}]{Jutzi:2022}
Jutzi M.,  Raducan S.~D.,  Zhang Y.,  Michel P.,   Arakawa M.,  2022, \mn@doi
  [Nature Communications] {10.1038/s41467-022-34540-x}, 13, 7134

\bibitem[\protect\citeauthoryear{{Kataoka}, {Tanaka}, {Okuzumi}  \&
  {Wada}}{{Kataoka} et~al.}{2013}]{Kataoka_2013}
{Kataoka} A.,  {Tanaka} H.,  {Okuzumi} S.,   {Wada} K.,  2013, \mn@doi
  [Astronomy \& Astrophysics] {10.1051/0004-6361/201322151}, \href
  {http://adsabs.harvard.edu/abs/2013A%26A...557L...4K} {557, L4}

\bibitem[\protect\citeauthoryear{{Keller} et~al.,}{{Keller}
  et~al.}{2015}]{keller_2015}
{Keller} H.~U.,  et~al., 2015, \mn@doi [\aap] {10.1051/0004-6361/201525964},
  \href {https://ui.adsabs.harvard.edu/abs/2015A&A...583A..34K} {583, A34}

\bibitem[\protect\citeauthoryear{Keller et~al.,}{Keller
  et~al.}{2017}]{keller_2017}
Keller H.~U.,  et~al., 2017, \mn@doi [Monthly Notices of the Royal Astronomical
  Society] {10.1093/mnras/stx1726}, 469, S357

\bibitem[\protect\citeauthoryear{Klinkenberg}{Klinkenberg}{2012}]{Klinkenberg2012ThePO}
Klinkenberg L.~J.,  2012, Drilling and Production Practice, pp 200--213

\bibitem[\protect\citeauthoryear{Kozeny}{Kozeny}{1924}]{kozeny1924kapillaren}
Kozeny J.,  1924, Kulturtechniker, 27, 11

\bibitem[\protect\citeauthoryear{Kraemer, Teiser, Steinpilz, Koester  \&
  Wurm}{Kraemer et~al.}{2019}]{KRAEMER2019131}
Kraemer A.,  Teiser J.,  Steinpilz T.,  Koester M.,   Wurm G.,  2019, \mn@doi
  [Planetary and Space Science] {https://doi.org/10.1016/j.pss.2018.09.004},
  166, 131

\bibitem[\protect\citeauthoryear{{Krijt}, {Ormel}, {Dominik}  \&
  {Tielens}}{{Krijt} et~al.}{2015}]{krijt}
{Krijt} S.,  {Ormel} C.~W.,  {Dominik} C.,   {Tielens} A.~G.~G.~M.,  2015,
  \mn@doi [\aap] {10.1051/0004-6361/201425222}, \href
  {https://ui.adsabs.harvard.edu/abs/2015A&A...574A..83K} {574, A83}

\bibitem[\protect\citeauthoryear{Kruss, Musiolik, Demirci, Wurm  \&
  Teiser}{Kruss et~al.}{2020}]{KRUSS2020113438}
Kruss M.,  Musiolik G.,  Demirci T.,  Wurm G.,   Teiser J.,  2020, \mn@doi
  [Icarus] {https://doi.org/10.1016/j.icarus.2019.113438}, 337, 113438

\bibitem[\protect\citeauthoryear{{Lambrechts} \& {Johansen}}{{Lambrechts} \&
  {Johansen}}{2014}]{Lambrechts_2014}
{Lambrechts} M.,  {Johansen} A.,  2014, \mn@doi [Astronomy \& Astrophysics]
  {10.1051/0004-6361/201424343}, 572, A107

\bibitem[\protect\citeauthoryear{{Lambrechts}, {Johansen}, {Capelo}, {Blum}  \&
  {Bodenschatz}}{{Lambrechts} et~al.}{2016}]{lambrechts}
{Lambrechts} M.,  {Johansen} A.,  {Capelo} H.~L.,  {Blum} J.,   {Bodenschatz}
  E.,  2016, \mn@doi [Astronomy \& Astrophysics] {10.1051/0004-6361/201526272},
  \href {http://adsabs.harvard.edu/abs/2016A%26A...591A.133L} {591, A133}

\bibitem[\protect\citeauthoryear{Lasseux \& Valdés-Parada}{Lasseux \&
  Valdés-Parada}{2017}]{Lasseux:2017}
Lasseux D.,  Valdés-Parada F.~J.,  2017, \mn@doi [Comptes Rendus Mécanique]
  {https://doi.org/10.1016/j.crme.2017.06.005}, 345, 660

\bibitem[\protect\citeauthoryear{Mannel, Bentley, Schmied, Jeszenszky,
  Levasseur-Regourd, Romstedt  \& Torkar}{Mannel et~al.}{2016}]{Mannel:2016}
Mannel T.,  Bentley M.~S.,  Schmied R.,  Jeszenszky H.,  Levasseur-Regourd
  A.~C.,  Romstedt J.,   Torkar K.,  2016, \mn@doi [Monthly Notices of the
  Royal Astronomical Society] {10.1093/mnras/stw2898}, 462, S304

\bibitem[\protect\citeauthoryear{Murdoch, Rozitis, Green, Michel, de Lophem  \&
  Losert}{Murdoch et~al.}{2013}]{Murdoch2013}
Murdoch N.,  Rozitis B.,  Green S.~F.,  Michel P.,  de Lophem T.-L.,   Losert
  W.,  2013, \mn@doi [Monthly Notices of the Royal Astronomical Society]
  {10.1093/mnras/stt742}, 433, 506

\bibitem[\protect\citeauthoryear{Nakagawa, Sekiya  \& Hayashi}{Nakagawa
  et~al.}{1986}]{Nakagawa1986}
Nakagawa Y.,  Sekiya M.,   Hayashi C.,  1986, \mn@doi [Icarus]
  {http://dx.doi.org/10.1016/0019-1035(86)90121-1}, 67, 375

\bibitem[\protect\citeauthoryear{{Paszun} \& {Dominik}}{{Paszun} \&
  {Dominik}}{2009}]{Paszun}
{Paszun} D.,  {Dominik} C.,  2009, \mn@doi [\aap]
  {10.1051/0004-6361/200810682}, \href
  {https://ui.adsabs.harvard.edu/abs/2009A&A...507.1023P} {507, 1023}

\bibitem[\protect\citeauthoryear{P{\"a}tzold et~al.,}{P{\"a}tzold
  et~al.}{2016}]{patzold2016}
P{\"a}tzold M.,  et~al., 2016, \mn@doi [Nature] {10.1038/nature16535}, 530, 63

\bibitem[\protect\citeauthoryear{Perry et~al.,}{Perry et~al.}{2022}]{Perry2022}
Perry M.~E.,  et~al., 2022, \mn@doi [Nature Geoscience]
  {10.1038/s41561-022-00937-y}, pp~1--6

\bibitem[\protect\citeauthoryear{Poch, Pommerol, Jost, Carrasco, Szopa  \&
  Thomas}{Poch et~al.}{2016}]{POCH2016}
Poch O.,  Pommerol A.,  Jost B.,  Carrasco N.,  Szopa C.,   Thomas N.,  2016,
  \mn@doi [Icarus] {https://doi.org/10.1016/j.icarus.2015.11.006}, 266, 288

\bibitem[\protect\citeauthoryear{{Pollack}, {Hubickyj}, {Bodenheimer},
  {Lissauer}, {Podolak}  \& {Greenzweig}}{{Pollack}
  et~al.}{1996}]{PollackEtal:1996}
{Pollack} J.~B.,  {Hubickyj} O.,  {Bodenheimer} P.,  {Lissauer} J.~J.,
  {Podolak} M.,   {Greenzweig} Y.,  1996, \mn@doi [Icarus]
  {10.1006/icar.1996.0190}, \href
  {http://adsabs.harvard.edu/abs/1996Icar..124...62P} {124, 62}

\bibitem[\protect\citeauthoryear{Raducan \& Jutzi}{Raducan \&
  Jutzi}{2022}]{Raducan_2022}
Raducan S.~D.,  Jutzi M.,  2022, \mn@doi [The Planetary Science Journal]
  {10.3847/PSJ/ac67a7}, 3, 128

\bibitem[\protect\citeauthoryear{Richardson, Harker  \& Backhurst}{Richardson
  et~al.}{2002}]{richardson}
Richardson J.,  Harker J.,   Backhurst J.,  2002, in Richardson J.,  Harker J.,
    Backhurst J.,  eds, Chemical Engineering Series, Chemical Engineering
  (Fifth Edition), fifth edition edn, Butterworth-Heinemann, Oxford, pp
  191--236, \mn@doi{https://doi.org/10.1016/B978-0-08-049064-9.50015-1}, \url
  {https://www.sciencedirect.com/science/article/pii/B9780080490649500151}

\bibitem[\protect\citeauthoryear{Sakatani et~al.,}{Sakatani
  et~al.}{2021}]{Sakatani2021}
Sakatani N.,  et~al., 2021, \mn@doi [Nature Astronomy]
  {10.1038/s41550-021-01371-7}, 5, 766

\bibitem[\protect\citeauthoryear{{Schaffer}, {Johansen}, {Cedenblad}, {Mehling}
   \& {Mitra}}{{Schaffer} et~al.}{2020}]{Schaffer:2020}
{Schaffer} N.,  {Johansen} A.,  {Cedenblad} L.,  {Mehling} B.,   {Mitra} D.,
  2020, \mn@doi [\aap] {10.1051/0004-6361/201935763}, \href
  {https://ui.adsabs.harvard.edu/abs/2020A&A...639A..39S} {639, A39}

\bibitem[\protect\citeauthoryear{{Scheeres}, {Hartzell}, {S{\'a}nchez}  \&
  {Swift}}{{Scheeres} et~al.}{2010}]{Scheeres_2018}
{Scheeres} D.~J.,  {Hartzell} C.~M.,  {S{\'a}nchez} P.,   {Swift} M.,  2010,
  \mn@doi [\icarus] {10.1016/j.icarus.2010.07.009}, \href
  {http://adsabs.harvard.edu/abs/2010Icar..210..968S} {210, 968}

\bibitem[\protect\citeauthoryear{{Schr{\"a}pler} \& {Blum}}{{Schr{\"a}pler} \&
  {Blum}}{2011}]{SchraeplerBulm:2011}
{Schr{\"a}pler} R.,  {Blum} J.,  2011, \mn@doi [The Astrophysical Journal]
  {10.1088/0004-637X/734/2/108}, \href
  {http://adsabs.harvard.edu/abs/2011ApJ...734..108S} {734, 108}

\bibitem[\protect\citeauthoryear{Schwartz, Michel, Jutzi, Marchi, Zhang  \&
  Richardson}{Schwartz et~al.}{2018}]{Schwartz}
Schwartz S.~R.,  Michel P.,  Jutzi M.,  Marchi S.,  Zhang Y.,   Richardson
  D.~C.,  2018, \mn@doi [Nature Astronomy] {10.1038/s41550-018-0395-2}, 2, 379

\bibitem[\protect\citeauthoryear{{Schweighart}, {Macher}, {Kargl}, {Gundlach}
  \& {Capelo}}{{Schweighart} et~al.}{2021}]{Schweighart:2021}
{Schweighart} M.,  {Macher} W.,  {Kargl} G.,  {Gundlach} B.,   {Capelo} H.~L.,
  2021, \mn@doi [Monthly Notices of the Royal Astronomical Society]
  {10.1093/mnras/stab934}, \href
  {https://ui.adsabs.harvard.edu/abs/2021MNRAS.504.5513S} {504, 5513}

\bibitem[\protect\citeauthoryear{Scott \& Kilgour}{Scott \&
  Kilgour}{1969}]{Scott_1969}
Scott G.~D.,  Kilgour D.~M.,  1969, \mn@doi [Journal of Physics D: Applied
  Physics] {10.1088/0022-3727/2/6/311}, 2, 863

\bibitem[\protect\citeauthoryear{{Seizinger}, {Speith}  \& {Kley}}{{Seizinger}
  et~al.}{2013}]{SeizingerEtal:2013c}
{Seizinger} A.,  {Speith} R.,   {Kley} W.,  2013, \mn@doi [Astronomy \&
  Astrophysics] {10.1051/0004-6361/201322046}, 559, A19

\bibitem[\protect\citeauthoryear{{Sharipov}}{{Sharipov}}{2011}]{Sharipov}
{Sharipov} F.,  2011, \mn@doi [Journal of Physical and Chemical Reference Data]
  {10.1063/1.3580290}, \href
  {http://adsabs.harvard.edu/abs/2011JPCRD..40b3101S} {40, 023101}

\bibitem[\protect\citeauthoryear{{Simon}, {Blum}, {Birnstiel}  \&
  {Nesvorn{\'y}}}{{Simon} et~al.}{2022}]{Simon:2022}
{Simon} J.~B.,  {Blum} J.,  {Birnstiel} T.,   {Nesvorn{\'y}} D.,  2022, arXiv
  e-prints, \href {https://ui.adsabs.harvard.edu/abs/2022arXiv221204509S} {p.
  arXiv:2212.04509}

\bibitem[\protect\citeauthoryear{Sizemore \& Mellon}{Sizemore \&
  Mellon}{2008}]{SIZEMORE2008606}
Sizemore H.~G.,  Mellon M.~T.,  2008, \mn@doi [Icarus]
  {https://doi.org/10.1016/j.icarus.2008.05.013}, 197, 606

\bibitem[\protect\citeauthoryear{Skorov, van Lieshout, Blum  \& Keller}{Skorov
  et~al.}{2011}]{SKOROV2011}
Skorov Y.~V.,  van Lieshout R.,  Blum J.,   Keller H.~U.,  2011, \mn@doi
  [Icarus] {https://doi.org/10.1016/j.icarus.2011.01.018}, 212, 867

\bibitem[\protect\citeauthoryear{Skorov, Reshetnyk, Bentley, Rezac, Hartogh  \&
  Blum}{Skorov et~al.}{2022}]{Skorov2022}
Skorov Y.,  Reshetnyk V.,  Bentley M.~S.,  Rezac L.,  Hartogh P.,   Blum J.,
  2022, \mn@doi [Monthly Notices of the Royal Astronomical Society]
  {10.1093/mnras/stab3760}, 510, 5520

\bibitem[\protect\citeauthoryear{Spadaccia, Capelo, Pommerol, Schuetz, Alibert,
  Ros  \& Thomas}{Spadaccia et~al.}{2021}]{Spadaccia:2021}
Spadaccia S.,  Capelo H.~L.,  Pommerol A.,  Schuetz P.,  Alibert Y.,  Ros K.,
  Thomas N.,  2021, \mn@doi [Monthly Notices of the Royal Astronomical Society]
  {10.1093/mnras/stab3196}, 509, 2825

\bibitem[\protect\citeauthoryear{{Squire} \& {Hopkins}}{{Squire} \&
  {Hopkins}}{2018a}]{Squire_Hopkins:2018b}
{Squire} J.,  {Hopkins} P.~F.,  2018a, \mn@doi [Monthly Notices of the Royal
  Astronomical Society] {10.1093/mnras/sty854}, \href
  {http://adsabs.harvard.edu/abs/2018MNRAS.477.5011S} {477, 5011}

\bibitem[\protect\citeauthoryear{{Squire} \& {Hopkins}}{{Squire} \&
  {Hopkins}}{2018b}]{Squire_Hopkins:2018a}
{Squire} J.,  {Hopkins} P.~F.,  2018b, \mn@doi [Astrophysical Journal Letters]
  {10.3847/2041-8213/aab54d}, \href
  {http://adsabs.harvard.edu/abs/2018ApJ...856L..15S} {856, L15}

\bibitem[\protect\citeauthoryear{{Steckloff}, {Johnson}, {Bowling}, {Jay
  Melosh}, {Minton}, {Lisse}  \& {Battams}}{{Steckloff}
  et~al.}{2015}]{Steckloff:2015}
{Steckloff} J.~K.,  {Johnson} B.~C.,  {Bowling} T.,  {Jay Melosh} H.,  {Minton}
  D.,  {Lisse} C.~M.,   {Battams} K.,  2015, \mn@doi [\icarus]
  {10.1016/j.icarus.2015.06.032}, \href
  {https://ui.adsabs.harvard.edu/abs/2015Icar..258..430S} {258, 430}

\bibitem[\protect\citeauthoryear{{Thomas}, {Alexander}  \& {Keller}}{{Thomas}
  et~al.}{2008}]{thomas_2008}
{Thomas} N.,  {Alexander} C.,   {Keller} H.~U.,  2008, \mn@doi [\ssr]
  {10.1007/s11214-008-9332-5}, \href
  {https://ui.adsabs.harvard.edu/abs/2008SSRv..138..165T} {138, 165}

\bibitem[\protect\citeauthoryear{{Thomas} et~al.,}{{Thomas}
  et~al.}{2015}]{Thomas:2015}
{Thomas} N.,  et~al., 2015, \mn@doi [Astronomy \& Astrophysics]
  {10.1051/0004-6361/201526049}, 583, A17

\bibitem[\protect\citeauthoryear{Wada et~al.,}{Wada et~al.}{2018}]{wada}
Wada K.,  et~al., 2018, \mn@doi [Progress in Earth and Planetary Science]
  {10.1186/s40645-018-0237-y}, 5, 82

\bibitem[\protect\citeauthoryear{{Wahlberg Jansson} \& {Johansen}}{{Wahlberg
  Jansson} \& {Johansen}}{2014}]{WahlbergI}
{Wahlberg Jansson} K.,  {Johansen} A.,  2014, \mn@doi [Astronomy \&
  Astrophysics] {10.1051/0004-6361/201424369}, \href
  {http://adsabs.harvard.edu/abs/2014A%26A...570A..47W} {570, A47}

\bibitem[\protect\citeauthoryear{{Wahlberg Jansson}, {Johansen}, {Bukhari Syed}
   \& {Blum}}{{Wahlberg Jansson} et~al.}{2017}]{WahlbergII}
{Wahlberg Jansson} K.,  {Johansen} A.,  {Bukhari Syed} M.,   {Blum} J.,  2017,
  \mn@doi [The Astrophysical Journal] {10.3847/1538-4357/835/1/109}, \href
  {http://adsabs.harvard.edu/abs/2017ApJ...835..109W} {835, 109}

\bibitem[\protect\citeauthoryear{{Weidenschilling}}{{Weidenschilling}}{1977}]{Weidenschilling:1977a}
{Weidenschilling} S.~J.,  1977, Monthly Notices of the Royal Astronomical
  Society, \href {http://adsabs.harvard.edu/abs/1977MNRAS.180...57W} {180, 57}

\bibitem[\protect\citeauthoryear{Wooden}{Wooden}{2008}]{Wooden:2008}
Wooden D.~H.,  2008, \mn@doi [Space Science Reviews]
  {10.1007/s11214-008-9424-2}, 138, 75

\bibitem[\protect\citeauthoryear{{Youdin} \& {Goodman}}{{Youdin} \&
  {Goodman}}{2005}]{You_good2005}
{Youdin} A.~N.,  {Goodman} J.,  2005, \mn@doi [The Astrophysical Journal]
  {10.1086/426895}, \href {http://adsabs.harvard.edu/abs/2005ApJ...620..459Y}
  {620, 459}

\bibitem[\protect\citeauthoryear{Youdin \& Johansen}{Youdin \&
  Johansen}{2007}]{You_jo2007j}
Youdin A.,  Johansen A.,  2007, The Astrophysical Journal, 662, 613

\bibitem[\protect\citeauthoryear{Zaccone}{Zaccone}{2022}]{zaccone_2022}
Zaccone A.,  2022, \mn@doi [Phys. Rev. Lett.] {10.1103/PhysRevLett.128.028002},
  128, 028002

\bibitem[\protect\citeauthoryear{van Dyke}{van Dyke}{1982}]{vanDyke:1982}
van Dyke M.,  1982, An album of fluid motion.
Parabolic Press, Stanford, CA, USA

\makeatother
\end{thebibliography}



\clearpage
\newpage
\appendix
\section{Sample properties and sources}\label{app:sample_tab}
In Tables \ref{exp_table1}-\ref{exp_table2}, we have already listed the samples, their sizes, which chambers they were used in, and which mass flow rates they were subject to. In Table \ref{sample_table} we provide more information about the particle properties and their source.   
\begin{table}
\label{sample_table}
\footnotesize
\begin{tabular}{c c c c c c c c c }  
\hline\hline  
Sample name& provider & nature/composition &  P$_{grav}$ [hPa] & solid density [g cm$^{-3}$] & porosity on Earth & particle shape \\
\hline
glass beads &Sigmund Linder & silicon dioxide &0.1 & 2.5&0.4 &   Spherical  \\
olivine coarse &Microbeads &magnesium iron silicate & 0.1& 3.3 & 0.5&   Grains  \\
steel ball-bearings & Simply Bearings Ltd. & chrome steel& 0.34 & 7.8  & 0.4 &   Spherical \\
microgrit & Micro Abrasives Corp &aluminum oxide &0.1 &4.0  & 0.4-0.6 &   platelet  \\
olivine fine &Microbeads &magnesium iron silicate & 0.1& 3.3 & 0.5&   Grains  \\
steel beads & Thermo Scientific& Stainless steel & 0.029& 8.0 & 0.4 &   Spheres   \\
olivine finest &Microbeads &magnesium iron silicate & 0.03/0.01 & 3.3 & 0.5 &   Grains  \\
CR regolith simulant & Exolith Labs & Average of 5 CR chondrites &0.02/0.03/0.07  &1.10 & & irregular grains  \\
\hline \hline
\end{tabular}
\end{table}

\clearpage
\newpage
\section{list of parameters used in this paper}\label{app:parameter_tab}
We summarise the parameters, their meaning, and if applicable quantities adopted, in Table \ref{parameter_table}.   

\begin{table*}
\caption{summary of the quantities studied and constants employed for calculations.}
\label{parameter_table}
\footnotesize
\begin{tabular}{c c c c}  
\hline\hline  
symbol & meaning & value &units  \\
\hline
$\phi$& filling factor & -&-\\
g$_{\rm Earth}$& Earth's gravitational acceleration &9.81 &ms$^{-2}$\\
P & pressure & - &hPa\\
$\delta P$&differential pressure & - &hPa\\
$\delta P_{\rm min}$&mean $\delta P$ at zero or partial gravity & - &hPa\\
$\delta P_{\rm Earth}$& mean $\delta P$ at Earth's gravity & - &hPa\\
$\delta P_{\rm hg}$& mean $\delta P$ at hyper gravity & - &hPa\\
$\delta z$& granular bed height&- & cm\\
$\kappa$& permeability coefficient & -& m$^{2}$\\
c& porosity & -&-\\
$d_{\rm p}$& particle diameter& - &$\mu$m\\
D& average pore diameter& -& $\mu$m\\
${K}_{\rm Koz}$& Kozeny constant&2/90&-\\
Re& Reynolds number&-&-\\
$\rho$ &gas density &- &kg m$^{-3}$ \\
u & gas velocity& 1.0 & ms$^{-1}$\\
$\mu$ & dynamic viscosity &$1.8\times 10^{-5}$ &Pa s\\
$\eta$ & kinematic viscosity & & \\
R & universal gas constant & 8.314& m$^{3}$ Pa K${^-1}$ mol${-1}$\\
T & temperature & 293 & K\\
M & molar molecular mass & .02891& \\
$\beta$& conical opening angle &15 &$\circ$ \\
b & Klinkenberg correction parameter &- & Pa\\
Kn &Knudsen number & -& -\\
Q & mass flow rate & & kg s$^{-1}$ \\
f$_{c}$ & compression factor &- & -\\
f$_{e}$ & expansion factor& -&- \\

\hline
\end{tabular}
\end{table*}
\section{Analytical fits to Permeability coefficients}\label{app:permeability_coeffs}
\begin{figure}
\centering
\includegraphics[width=0.45\textwidth]{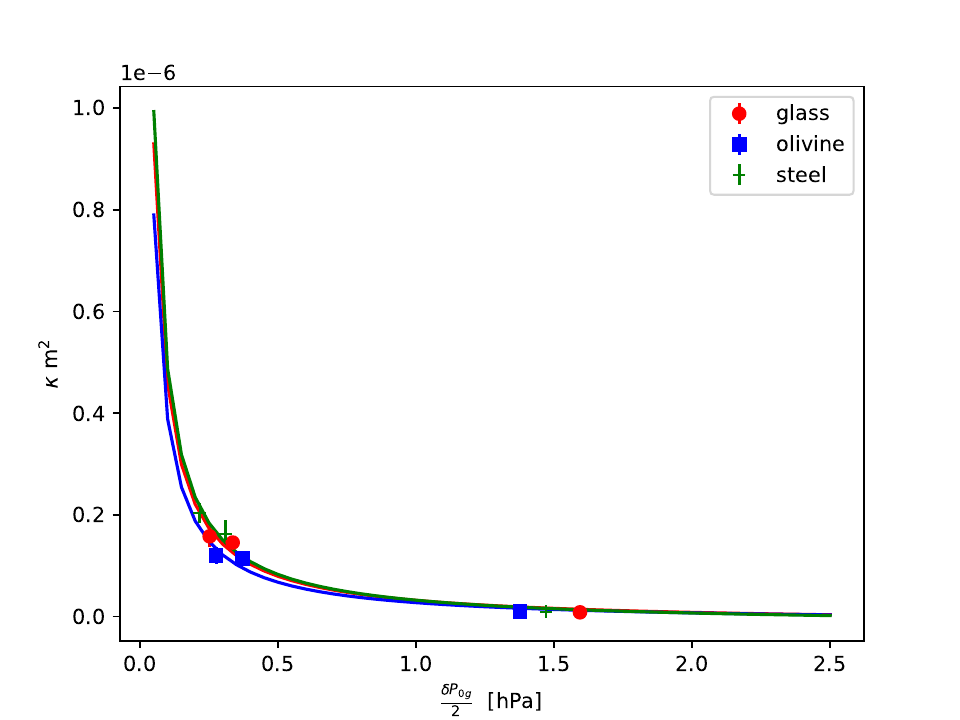}\\
\includegraphics[width=0.45\textwidth]{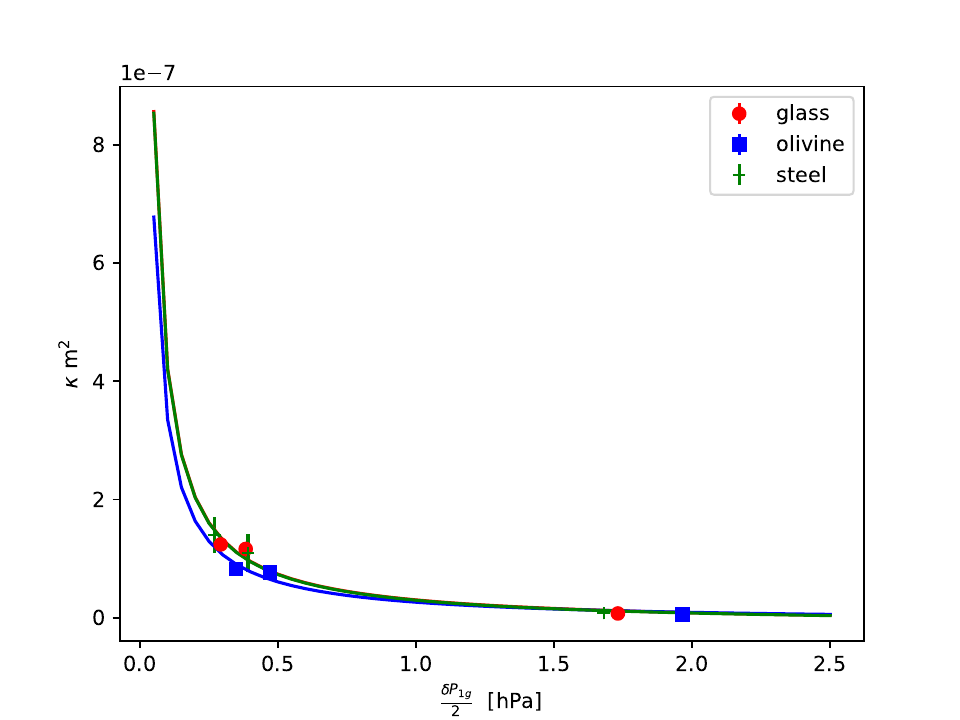}\\
\includegraphics[width=0.45\textwidth]{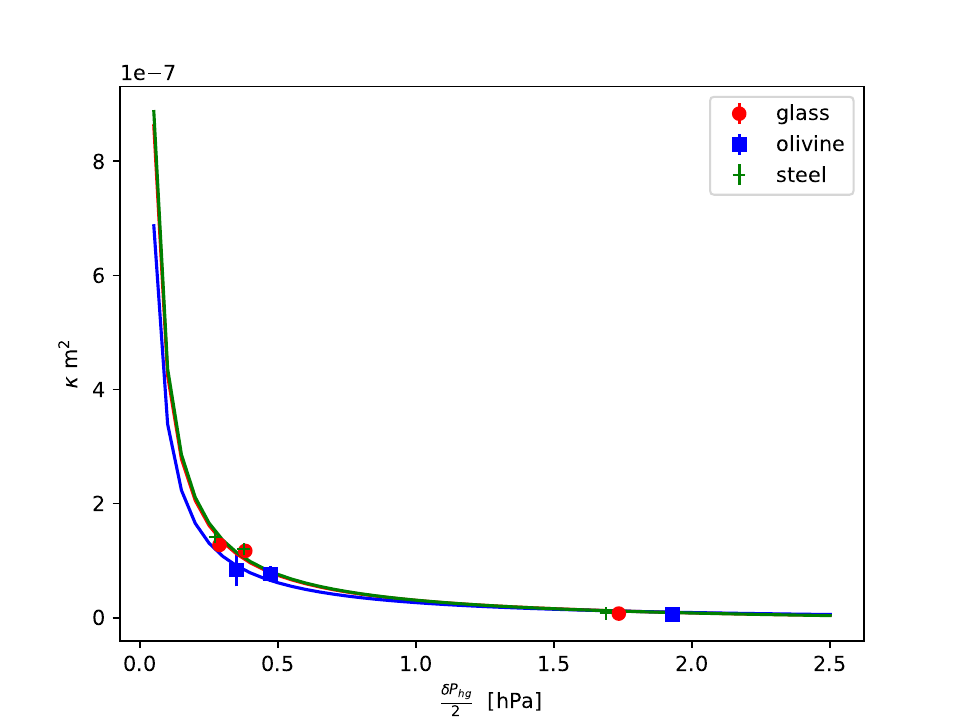}\\

\caption{\label{fig:Klinkenberg_flight1}Scatter plot of the permeability coefficient $\kappa$, as a function of mean pressure $\frac{\delta P}{2}$, for each chamber and each measurement series, of flight 4. Red circles correspond to chamber 1; blue squares correspond to chamber 2; green plus marks correspond to chamber 3. Error bars are similar in scale to the size of the markers. Top: coefficients were extracted from the data obtained at $g_{\rm Earth}$; Middle: coefficients were extracted from the data obtained during $hg$; Bottom: coefficients were extracted from the data obtained during $0g$. A functional fit of the Klinkenberg correction is shown for comparison. }

\end{figure}

\begin{figure}
\centering
\includegraphics[width=0.45\textwidth]{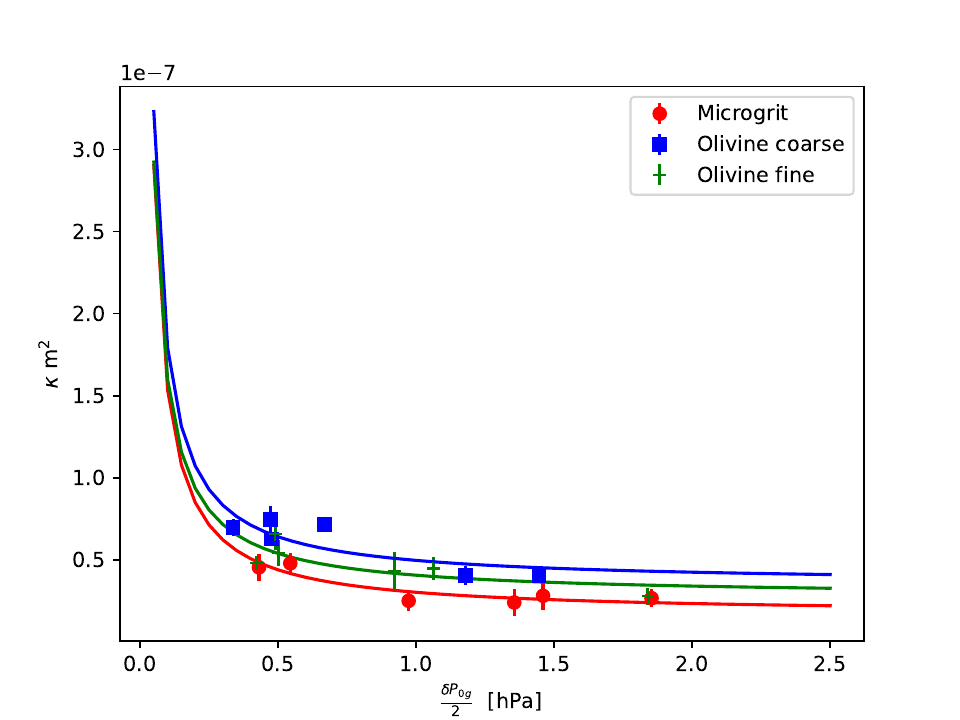}\\
\includegraphics[width=0.45\textwidth]{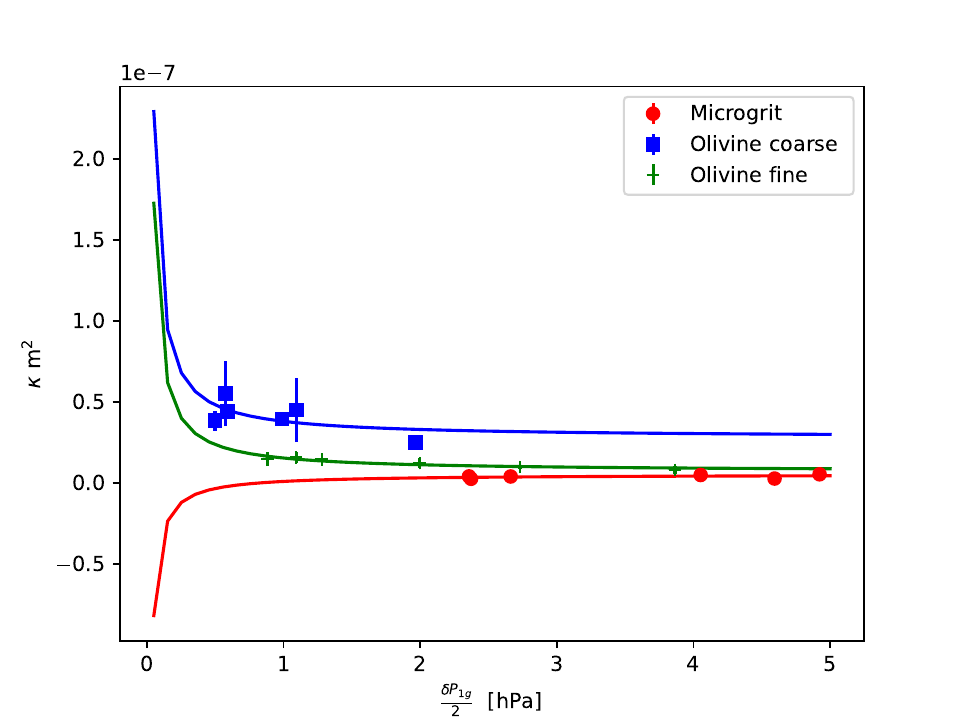}\\
\includegraphics[width=0.45\textwidth]{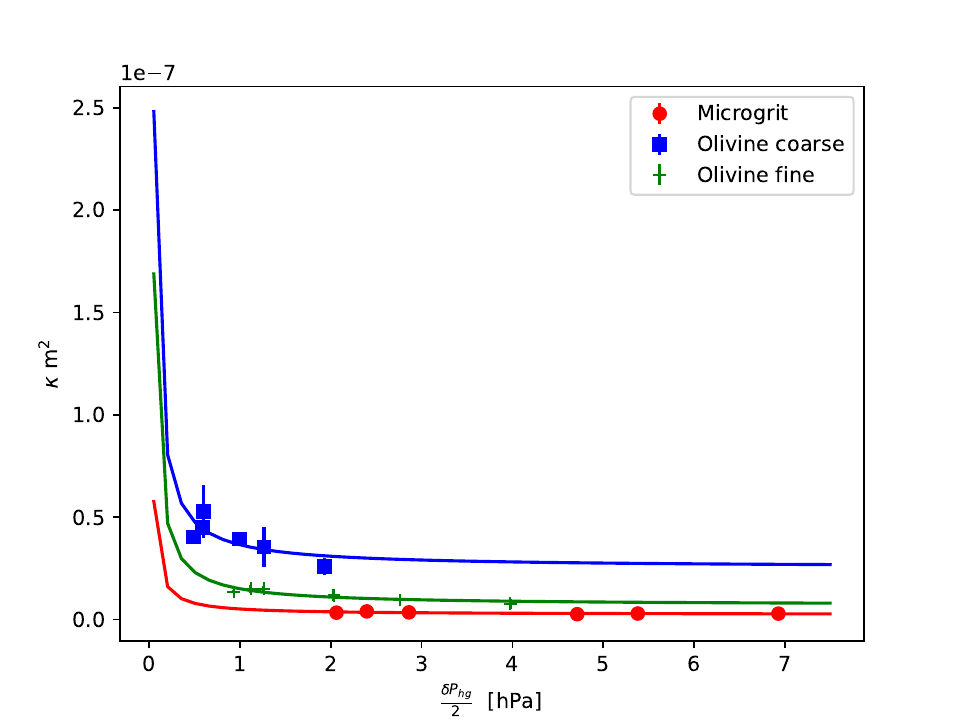}\\

\caption{\label{fig:Klinkenberg_flight2} Same as for Figure \ref{fig:Klinkenberg_flight1}, with flight 2 data.}

\end{figure}

\begin{figure}
\centering
\includegraphics[width=0.45\textwidth]{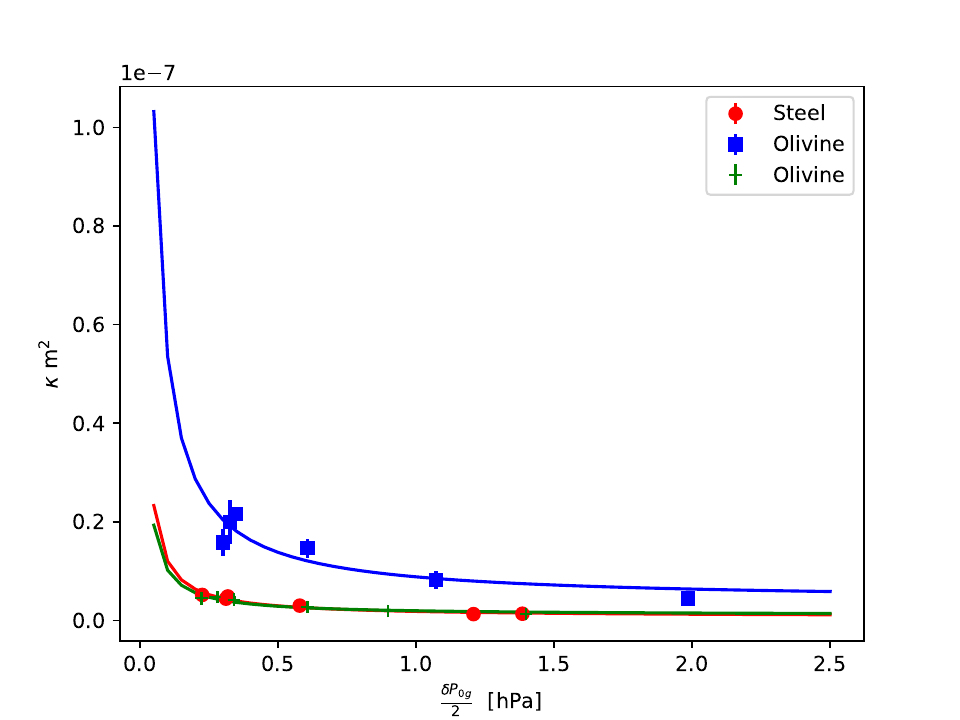}\\
\includegraphics[width=0.45\textwidth]{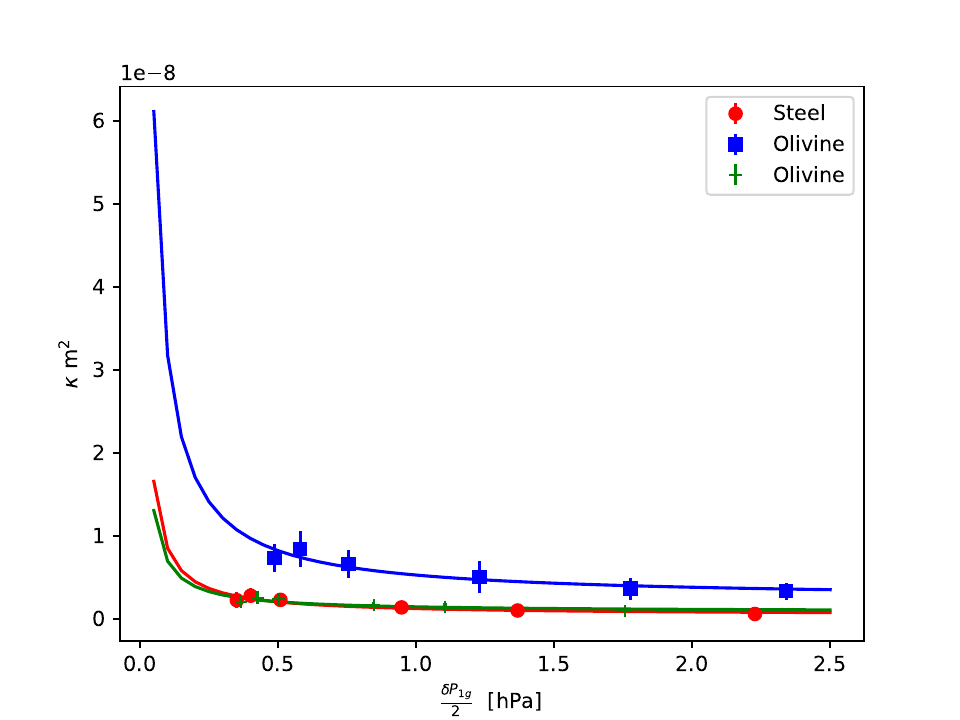}\\
\includegraphics[width=0.45\textwidth]{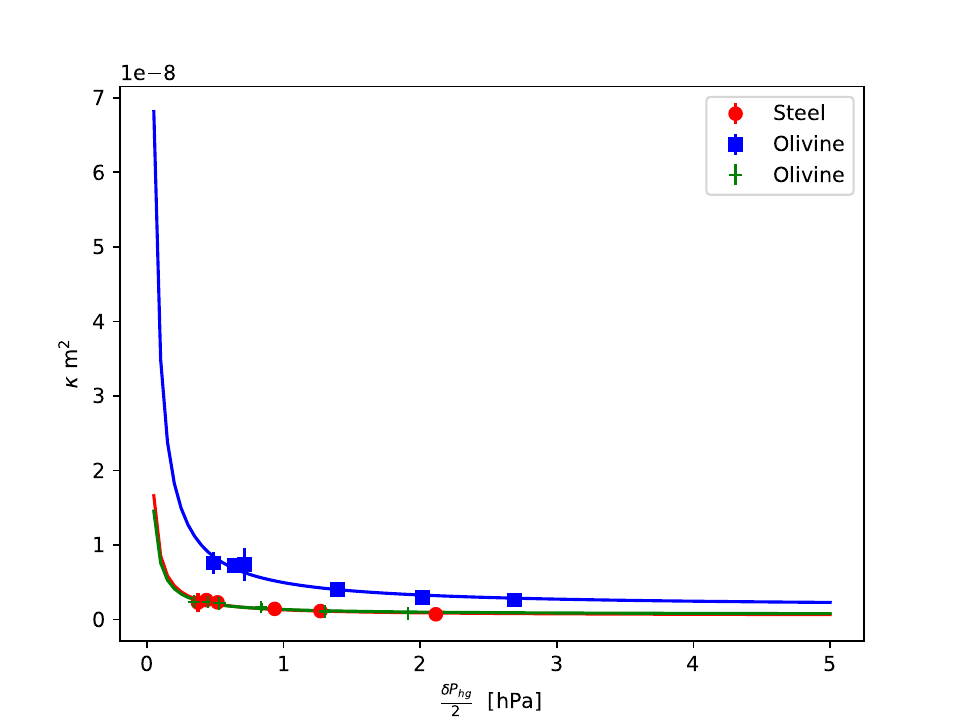}\\

\caption{\label{fig:Klinkenberg_flight3} Same as for Figure \ref{fig:Klinkenberg_flight1}, with flight 3 data.}
\end{figure}

\begin{figure}
\centering
\includegraphics[width=0.45\textwidth]{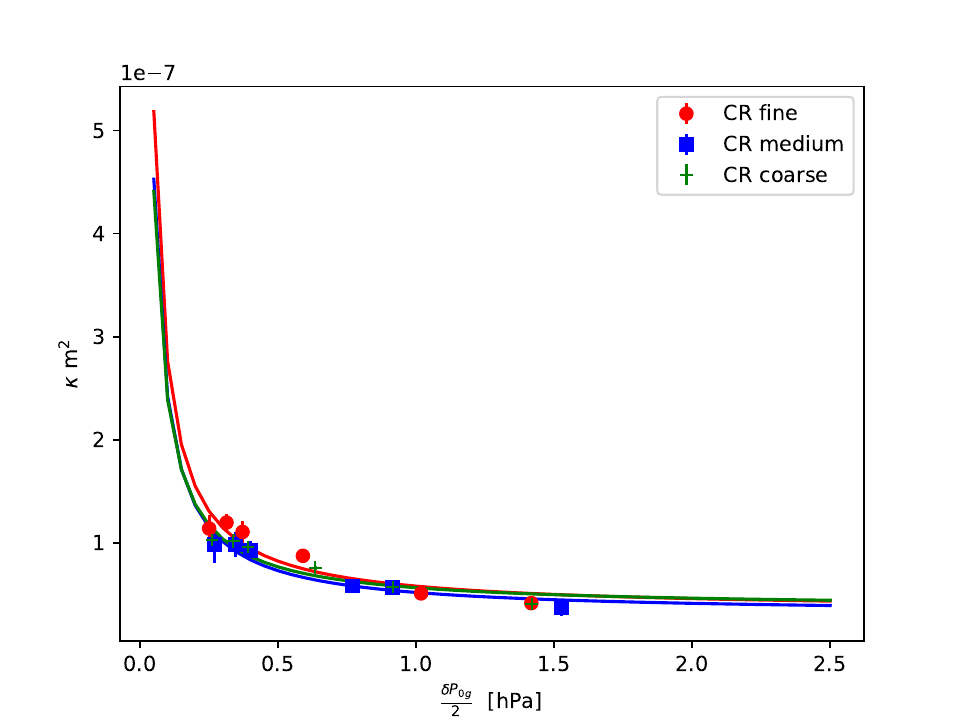}\\
\includegraphics[width=0.45\textwidth]{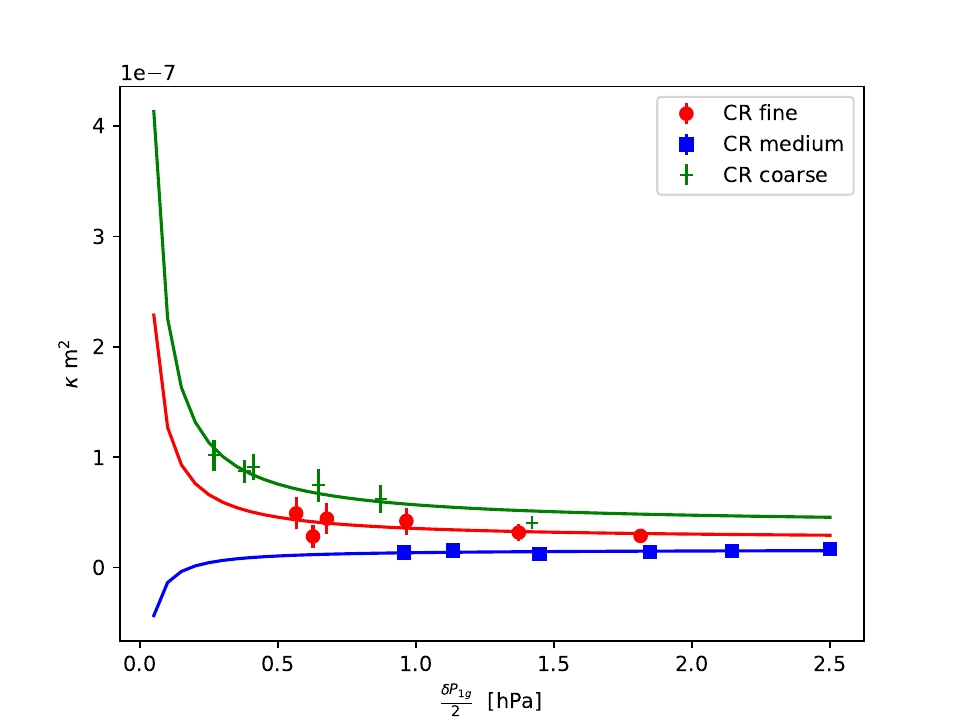}\\
\includegraphics[width=0.45\textwidth]{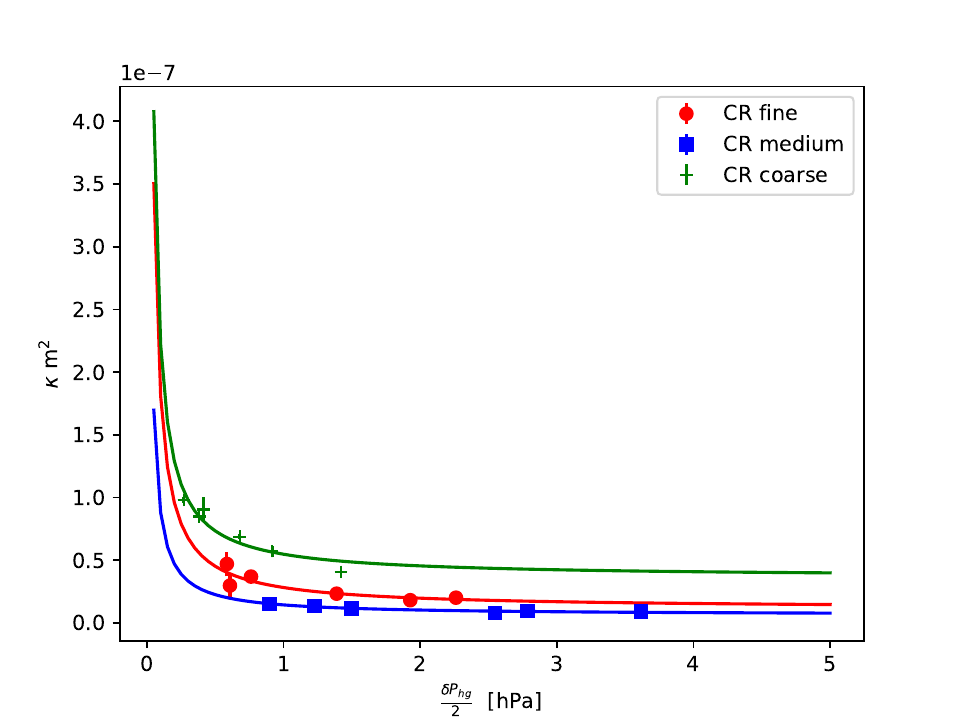}\\

\caption{\label{fig:Klinkenberg_flight4} Same as for Figure \ref{fig:Klinkenberg_flight1}, with flight 4 data.}
\end{figure}

We fit Equation \ref{klinkenberg} to the calculated permeability coefficients at a particular gravitational load for all samples and all flights. We report the best-fit parameters for $b$ and $\kappa_{\rm inf}$ in Table \ref{fit_table}. We also illustrate the resulting best-fit functions compared to the data-derived permeability coefficients, for all flights, in figures \ref{fig:Klinkenberg_flight1} - \ref{fig:Klinkenberg_flight4}. We emphasize that the interpretation of the best fit parameters should be treated carefully. In particular, the data from each sample type does not span the extrema of the functions (both extremely high and extremely low pressures), and so it is possible that different curves might result if there were a wider range of pressures covered. Because the pressure values do not necessarily expand into the continuum flow regime, where the permeability coefficients should be constant with pressure, we left $\kappa_{\rm infty}$ as a free parameter. In table \ref{fit_table}, some combinations of $b$ and $\kappa_{\rm inf}$ are both negative (particularly for flight 1, where only three points are fit for each sequence); when the permeability values are close to zero, the resultant value given by equation \ref{klinkenberg} shall still be positive. It can also be the case that all of the data lies in the regime where the permeability is constant, as is the case for the highly compressible samples at high compression, and so fitting a curve to nearly flat data does not produce sensible results. This is illustrated especially for the Microgrit sample in Figure \ref{fig:Klinkenberg_flight2}, see also the relatively constant values of steel and Olivine of similar volume in figure \ref{fig:Klinkenberg_flight3}. 

Despite the few caveats just mentioned, comparing the data to the best-fit curves is useful to verify that our derivation of the permeability coefficients -- using mass and momentum conservation -- produces the basic expectation that the data should have a form similar to equation \ref{klinkenberg}. The results from flight one provide a clear example that is illustrated in figure \ref{fig:Klinkenberg_flight1}. We find that for flight 1, the three very similar samples produce similar results, except perhaps the irregularly shaped olivine responding more, however subtly, to changes in gravitational load. With regards to fight 2, for which all three chambers contained irregularly-shaped particles, the effect of increasing the average gas pressure with increasing compression can clearly be observed.

For flight three, the two samples with similar bed heights have similar curves, whereas the sample with same mass, but dissimilar  follows a different relationship. This is because the mean pressure, $\frac{\delta z}{2}$ is lower for this sample, and so the system is pushed away from the continuum flow limit.  We speculate that a distinction may be that the mass density of the material, in addition to pore space, also plays a part in generating the pressure gradient, a possibility that we investigate further in future work.

\begin{table*}

\caption{Best-fit parameters for the Klinkenberg correction to the permeability coefficient. }
\label{fit_table}
\footnotesize
\begin{tabular}{c c c c c c c c }  
\hline\hline  
flight & sample & 1g & & hg & & 0g&\\
\hline
 &  & $b$ &$\kappa_{\rm inf}$ & $b$ &$\kappa_{\rm inf}$ & $b$ &$\kappa_{\rm inf}$ \\
\hline
1 & Glass beads &-3.3 &-1.3 $\times10^{-8}$ &-3.3 &-1.3 $\times10^{-8}$ & 0.8& 1.7 $\times 10^{-8}$\\
1 & Olivine coarse &-4.3 &-7.9 $\times 10^{-9}$ &-4.2 &-8.3 $\times 10^{-9}$ &0.4&3.4 $\times 10^{-8}$\\
1 & Steel beads &-3.2 &-1.4 $\times 10^{-8}$ &-3.2 &-1.4 $\times 10^{-8}$ &0.5 &2.8 $\times 10^{-8}$\\
 2 & Microgrit WCA 40 &-0.8 &5.3 $\times 10^{-9}$ &1.2 &2.4$\times 10^{-9}$&0.8 &1.7$\times 10^{-8}$\\
2 & Olivine coarse &0.4 &2.8 $\times 10^{-8}$& 0.4 &2.5$ \times 10^{-8}$ &0.4&3.5 $\times 10^{-8}$\\
 2& Olivine fine &1.2 &7.1 $\times 10^{-9}$ &1.2 &6.9 $\times 10^{-9}$&0.5 &2.8 $\times 10^{-8}$\\
3 & Steel (m$_{\rm s}$,V$_{\rm s}$) &0.4&2.5 $\times 10^{-8}$&1.2 &2.4 $\times 10^{-9}$ &0.8 &1.7 $\times 10^{-8}$\\
3& Olivine (m$_{\rm s}$) &-0.2&1.7 $\times 10^{-8}$&0.4 &2.5 $\times 10^{-8}$&0.4 &3.5 $\times 10^{-8}$\\
3 & Olvine (V$_{\rm s}$) &0.5 &3.8 $\times 10^{-8}$&1.7 &6.9 $\times 10^{-9}$&0.5 &2.8 $\times 10^{-8}$\\
4& CR fine &1.9&4.3 $\times 10^{-10}$&1.6 &5.0 $\times 10^{-10}$&1.9 &4.3 $\times 10^{-10}$\\
4 & CR medium&1.3 &2.3 $\times 10^{-9}$ &2.0 &1.6 $\times 10^{-9}$ &1.3 & 2.3 $\times 10^{-9}$\\
4 & CR coarse & 0.7 & 8.4 $\times 10^{-10}$ &1.1 &6.5 $\times 10^{-10}$ &0.7 &8.4 $\times 10^{-10}$\\
\hline
\end{tabular}
\end{table*}

We note that the values at $0g$ are a limit, rather than absolute parameter, since the height of the sample is poorly defined when the particles are fluidized and since the porosity increases, obviously the height increases as well. However, increasing the bed height can only serve to increase $\kappa$, and so the reported values at $0g$ are a lower limit upon $\kappa$. It is true that the uncertainty in the sample height could affect the derived values and also in turn the best fit parameters.

We make similar calculations for the samples of flight 4 (CR asteroid regolith simulant, see Table \ref{exp_table4}, and Figure \ref{flight1_4}). We derive $\kappa$ from measured pressure gradients, and compare to a fit of the Klinkenberg correction.  Figure \ref{fig:Klinkenberg_flight4} shows the results for each consecutive phase of the parabolas: $g_{\rm Earth}$, $hg$ and $0g$, in the top, middle and bottom panels. During the two compression stages, the data have been collectively fit by a single function and the data lie either close to the knee of the function, or close to the limit of the function. However, on closer inspection, the points representing the fine and medium size fractions in the $g_{\rm Earth}$ and $hg$ cases, tend rather to be flat and lie under the curve. This indicates that the three data sets cannot necessarily be represented by the same function, and actually might have their own track in this phase space.

Remarkably, during $0g$, the measured values for $\kappa$ collapse on top of one another. Here, a fit to the data sits above the measured points, because the mean pressure has shifted so dramatically to lower values, that there is no appropriate choice to set the anchor value of $\kappa_{\infty}$: all of the data clusters near the knee and at higher values of the permeability. Striking, however, is that the trend for the coarse material has not changed, while the curves for the fine and medium fractions alter themselves so as to behave the same as the coarse material.


\bsp	
\label{lastpage}
\end{document}